\documentclass[11pt, a4paper]{article}
\pdfoutput=1
\usepackage{jheppub}
\usepackage{amssymb,amsmath}
\usepackage{graphicx}
\usepackage{booktabs}
\usepackage{subfigure}
\usepackage{enumerate}
\usepackage{xspace}

\newcommand{\beq}{\begin{equation}}
\newcommand{\eeq}{\end{equation}}
\newcommand{\beqa}{\begin{eqnarray}}
\newcommand{\eeqa}{\end{eqnarray}}
\newcommand{\col}{\mathrm{col}}
\newcommand{\calM}{{\cal M}}
\newcommand{\calN}{{\cal N}}
\newcommand{\calA}{{\cal A}}
\newcommand{\calC}{{\cal C}}
\newcommand{\calK}{{\cal K}}
\newcommand{\cm}{{\cal M}^0}

\newcommand{\comb}{{\cal M}}
\newcommand{\cmb}{{\cal M}}
\newcommand{\ds}{{\rm d}\hat{\sigma}}
\newcommand{\dphi}{{\rm d}\Phi}
\newcommand{\gev}{{\rm GeV}}
\newcommand{\wt}{\widetilde}
\newcommand{\re}{{\rm{Re}}}
\newcommand{\norm}{{\cal N}}
\newcommand{\ssoft}[3]{{\cal S}_{#1 #2 #3}}
\newcommand{\soft}[4]{{\cal S}_{#1 #2 #3 #4}}
\newcommand{\order}[1]{{\cal O}(#1)}
\newcommand{\q}[1]{#1_q}

\newcommand{\Q}[1]{#1_Q}
\newcommand{\Qb}[1]{#1_{\bar{Q}}}
\newcommand{\qi}[1]{\hat{#1}_q}
\newcommand{\qbi}[1]{\hat{#1}_{\bar{q}}}

\newcommand{\gl}[1]{#1_g}
\newcommand{\gli}[1]{\hat{#1}_g}
\newcommand{\ph}[1]{#1_{\gamma}}

\newcommand{\ione}[2]{{\bf I}^{(1)}_{#1 #2}}
\newcommand{\Dzero}{{\cal D}_0}

\newcommand{\cep}{C(\epsilon)}
\newcommand{\cepb}{\bar{C}(\epsilon)}
\newcommand{\asmu}{\alpha_s(\mu)}
\newcommand{\poles}{{\cal P}oles}

\newcommand{\GG}{{\bf \Gamma}}
\newcommand{\gaussf}[4]{\ensuremath{\, _2F_1 \left(#1,#2,#3;#4\right)}}

\def\e{\epsilon}

\newcommand{\LO}{\mathrm{LO}}
\newcommand{\NLO}{\mathrm{NLO}}
\newcommand{\NNLO}{\mathrm{NNLO}}
\newcommand{\RR}{\mathrm{RR}}
\newcommand{\MF}{\mathrm{MF}}
\newcommand{\RV}{\mathrm{RV}}
\newcommand{\VV}{\mathrm{VV}}
\newcommand{\VS}{\mathrm{VS}}
\newcommand{\rR}{\mathrm{R}}
\newcommand{\rV}{\mathrm{V}}
\newcommand{\rS}{\mathrm{S}}
\newcommand{\rT}{\mathrm{T}}
\newcommand{\rd}{\mathrm{d}}
\newcommand{\ri}{\mathrm{i}}
\newcommand{\rU}{\mathrm{U}}
\newcommand{\ren}{\mathrm{ren}}
\newcommand{\bare}{\mathrm{bare}}

\newcommand{\OpenLoops}{{\sc OpenLoops}\xspace}
\newcommand{\Collier}{{\sc Collier}\xspace}
\newcommand{\Cuttools}{{\sc Cuttools}\xspace}

\title{NNLO QCD subtraction for  top-antitop production in the $q\bar{q}$ channel
}

\author[a]{Gabriel Abelof}
\author[b,c]{Aude Gehrmann-De Ridder}
\author[c]{Philipp Maierh\"ofer}
\author[c]{Stefano Pozzorini}

\affiliation[a]{Department of Physics \& Astronomy, Northwestern University, Evanston, IL 60208, USA}
\affiliation[b]{Institute for Theoretical Physics, ETH, CH-8093 Z\"urich, Switzerland}
\affiliation[c]{
Physics Institute,
University of Z\"urich, Winterthurerstrasse 190, CH-8057, Z\"urich}

\emailAdd{gabriel.abelof@northwestern.edu}
\emailAdd{gehra@itp.phys.ethz.ch}
\emailAdd{philipp@physik.uzh.ch}
\emailAdd{pozzorin@physik.uzh.ch}


\abstract{ 
We present the computation of the double real and real-virtual contributions
to top-antitop pair production in the quark-antiquark channel at leading
colour. The $q \bar q \to t \bar{t} g$  amplitudes contributing to
the real-virtual part are computed with \OpenLoops, and their numerical
stability in the soft and collinear regions is found to be sufficiently
high to perform a realistic NNLO calculation in double precision. 
The subtraction terms required at
real-real and real-virtual levels are constructed within the antenna
subtraction formalism extended to deal with the presence of coloured massive
final state particles.  We show that those subtraction terms approximate the
real-real and real-virtual matrix elements in all their singular limits.  }
  
\preprint{
\hfill\begin{minipage}[t]{8em}
\today\\    
      ZU-TH 17/14\\
      LPN14-067\\
  \end{minipage}
}

\begin{document}
\bibliographystyle{JHEP-2}
\allowdisplaybreaks
\maketitle


\section{Introduction}\label{sec.intro} 

Top quark physics has become precision physics at the LHC.  Some
observables, like the total cross section for $t\bar{t}$ production, are
expected to be measured with accuracies at the percent level.  In addition,
the ATLAS and CMS collaborations at CERN have reported first measurements of
differential observables in top-quark pair production, such as the
transverse momentum and rapidity of the $t\bar t$ system \cite{Aad:2012hg},
its invariant mass \cite{CMS:fxa}, and the top quark transverse
momentum~\cite{CMS:cxa}.  Those measurements will allow for a much more
detailed probe of the top quark production mechanism than what can obtained
from the total cross section.  To reliably interpret these data, these
precise measurements have to be matched onto equally accurate theoretical
predictions.  Those can be obtained by computing these hadron collider
observables at the next-to-next-to leading order (NNLO) in perturbative QCD.
At present, a fully differential NNLO calculation of the cross section for
top pair production including all partonic channels is still missing. 
Intermediate results have recently become available in
\cite{Abelof:2011ap,Anastasiou:2008vd,Baernreuther:2012ws,Bierenbaum:2011gg,Bonciani:2008az,Bonciani:2009nb,Bonciani:2010mn,Czakon:2008zk,Czakon:2011ve,Czakon:2012zr,Kniehl:2008fd,Korner:2008bn}.


Most notably, the inclusive total hadronic $t\bar{t}$
production cross section has been presented in \cite{Czakon:2013goa}.

At NNLO, perturbative calculations of collider observables, like jet or heavy quark 
cross sections and associated kinematical distributions, 
are typically carried out using parton-level event generators.
These programs generate events for all
parton-level subprocesses relevant to a given final state configuration up
to NNLO accuracy and provide full kinematical information 
on an event-by-event basis. 
Towards this ultimate goal for top-pair production observables, in this
paper we consider the quark-antiquark initiated channel at leading
colour and compute two essential contributions to the NNLO top pair
production cross section, namely the double real and real-virtual parts.

An NNLO event generator for observables with $n$ final-state 
particles or jets involves three main building blocks:
the two-loop corrections to the $n$-parton production process, 
denoted as double-virtual contributions ${\rm d}\sigma^{VV}$,
the one-loop corrections to the $(n+1)$-parton production process,
called real-virtual contribution ${\rm d}\sigma^{RV}$,
and the tree-level $(n+2)$-parton double real contribution,
${\rm d}\sigma^{RR}$.
These three building blocks involve infrared divergences
that arise from the exchange or emission of soft and collinear partons 
and cancel only in their sum.
In addition, the real-virtual and virtual-virtual contributions to hadron
collider observables involve initial-state collinear singularities
that must be absorbed into mass factorisation counter
terms. Those are labelled as ${\rm{d}}\sigma^{MF,1}$ and ${\rm {d}}\sigma^{MF,2}$,
respectively.

The combination of subprocesses of different particle multiplicity and the
consistent cancellation of the respective infrared singularities is one of
the major challenges in the construction of NNLO parton-level event
generators.  In each subprocess, infrared singularities assume a different
form: in the virtual corrections they are explicit, while in the real
contributions they are implicit and become explicit only after phase space
integration.  To compute an observable beyond leading order, a
regularization procedure is therefore required to extract and cancel the
infrared singularities among different partonic channels before those can be
implemented in the parton-level event generator.
This goal is typically achieved by means of subtraction methods,
where all relevant singularities of the matrix elements are subtracted 
by means of universal auxiliary terms, which 
are sufficiently simple to be added back after 
analytic integration over the unresolved phase space.
In the past, this approach was successfully applied to 
various NNLO calculations using 
sector decomposition~\cite{Binoth:2000ps,Anastasiou:2003gr,
Binoth:2004jv,Anastasiou:2010pw},  $q_T$-subtraction \cite{Catani:2007vq}, 
antenna subtraction~\cite{GehrmannDeRidder:2005cm} and  most recently 
with an approach based on sector-improved residue subtraction 
\cite{Czakon:2010td,Czakon:2011ve}.

Two of these methods have been extended to treat massive final state fermions and
applied to top pair hadro-production.  In \cite{Czakon:2013goa} the total
cross section for inclusive $t\bar{t}$ production was obtained with 
the {\sc Stripper} method \cite{Czakon:2010td,Boughezal:2011jf}, which combines
the FKS subtraction method \cite{Frixione:1997np} and sector decomposition 
\cite{Anastasiou:2003gr,Binoth:2004jv}.  Moreover, the antenna subtraction formalism
with massive fermions has been applied to the evaluation of the double real
contributions to $t\bar{t}$ production for the pure fermionic processes
\cite{Abelof:2011ap} and for the gluon initiated process $ g g \rightarrow t
\bar{t} q \bar{q}$ \cite{Abelof:2012rv}.
In this paper, we  shall employ the massive extension of 
antenna subtraction to extract the infrared behaviour of
double real and real-virtual NNLO contributions 
to the $q \bar q \to t \bar{t}$ channel at leading colour. 

While the computation of NNLO corrections to observables involving massive
particles require the same kind of ingredients as for massless observables,
namely real-real, real-virtual and virtual-virtual contributions, the
presence of massive fermions in the final state introduces a few
simplifications as well as new complications.  First,
due to the presence of massive final states, the ultraviolet renormalisation
procedure of one and two loop amplitudes is more involved than for their
massless counterparts.  Not only couplings but also mass and wave function
ultraviolet renormalisations are required.  For all loop amplitudes
encountered in this paper, we shall use the ultraviolet regularisation
procedure described in \cite{Bonciani:2008az,Bonciani:2010mn}.
Concerning infrared singularities, massive quarks do not give rise to
final-state collinear singularities, and 
the quasi-collinear effects described in \cite{Abelof:2011jv,Catani:2002hc} 
can be safely ignored for $t\bar t$ production at the LHC.  
Thus only divergencies associated with soft radiation 
and with collinear emissions off massless partons 
require explicit subtraction terms.
On the other hand, the non-vanishing parton masses
introduce a new scale,
which represents a considerable 
source of complexity both for 
the final-state kinematics and for the 
integration of the subtraction terms.\\

Employing a subtraction method, the NNLO partonic cross section for top-pair production in a given partonic channel (and proportional to a specific colour factor) has the 
general structure \cite{GehrmannDeRidder:2005cm}
\beqa\label{eq.sigNNLO}
{\rm d}\hat\sigma_{\NNLO}&=&\int_{\Phi_{4}}\left({\rm{d}}\hat\sigma_{\NNLO}^{\RR}-{\rm{d}}\hat\sigma_{\NNLO}^{\rS}\right)+\int_{\Phi_{4}}{\rm{d}}\hat\sigma_{\NNLO}^{\rS}\nonumber\\
&+&\int_{\Phi_{3}}\left({\rm{d}}\hat\sigma_{\NNLO}^{\RV}-{\rm{d}}\hat\sigma_{\NNLO}^{\VS}\right)+\int_{\Phi_{3}}{\rm{d}}\hat\sigma_{\NNLO}^{\VS}+\int_{\Phi_{3}}{\rm{d}}\hat\sigma_{\NNLO}^{\MF,1}\nonumber\\
&+&\int_{\Phi_2}{\rm{d}}\hat\sigma_{\NNLO}^{\VV}+\int_{\Phi_2}{\rm{d}}\hat\sigma_{\NNLO}^{\MF,2}.
\eeqa
Two types of subtraction terms are introduced: ${\rm d} \hat\sigma^{\rS}_{\NNLO}$ for the $4$-parton final state, and ${\rm d} \hat\sigma^{\VS}_{\NNLO}$ for the $3$-parton final state. The former approximates the behaviour of the double real contributions ${\rm d} \hat\sigma^{\RR}_{\NNLO}$ in their single and double unresolved limits, whereas the latter reproduces the single unresolved behaviour of the mixed real-virtual contributions ${\rm d} \hat\sigma^{\RV}_{\NNLO}$.

In the context of the antenna subtraction framework employed in this paper, 
we decompose further the double real subtraction term $\ds_{\NNLO}^{\rS}$. This term contains distinct pieces
corresponding to different limits and different colour-ordered
configurations.  Some of these pieces ought to be integrated analytically
over the unresolved phase space of one particle and combined with the
$3$-parton final state, while the remaining terms are to be integrated
over the unresolved phase space of two particles and combined with the
$2$-parton contributions.  This separation amounts to splitting the
integrated form of $\ds_{\NNLO}^{\rS}$ as
\cite{Currie:2013dwa,Currie:2013vh,GehrmannDeRidder:2011aa,GehrmannDeRidder:2012dg}
\beq
\int_{\Phi_{4}}\ds_{\NNLO}^{\rS}= \int_{\Phi_{3}} \int_1 \ds_{\NNLO}^{\rS,1}+\int_{\Phi_{3}} \int_2 \ds_{\NNLO}^{\rS,2},
\eeq
which allows us to rearrange the different terms in eq.(\ref{eq.sigNNLO}) into the more convenient form
\beqa\label{eq.subnnlo}
\ds_{\NNLO}&=&\int_{\Phi_{4}}\left[\ds_{\NNLO}^{\RR}-\ds_{\NNLO}^{\rS}\right]\nonumber \\
&+& \int_{\Phi_{3}}\left[\ds_{\NNLO}^{\RV}-\ds_{\NNLO}^{\rT}\right] \nonumber \\
&+&\int_{\Phi_{2}}\left[\ds_{\NNLO}^{\VV}-\ds_{\NNLO}^{\rU}\right],
\eeqa
with 
\beqa
\label{eq.Tdef}  \ds_{\NNLO}^{\rT} &=& \phantom{ -\int_1 }\ds_{\NNLO}^{\VS}- \int_1 \ds_{\NNLO}^{\rS,1} - \ds_{\NNLO}^{\MF,1},  \\
\label{eq.Udef}  \ds_{\NNLO}^{\rU} &=& -\int_1 \ds_{\NNLO}^{\VS}-\int_2 \ds_{\NNLO}^{\rS,2}-\ds_{\NNLO}^{\MF,2}. 
\eeqa

In this paper, we shall explicitly construct  the antenna subtraction terms $\ds_{\NNLO}^{\rS}$ and $\ds_{\NNLO}^{\rT} $ entering at the 
four- and three-parton contributions to the NNLO top pair 
production cross section (\ref{eq.subnnlo}) for the quark-antiquark 
channel at leading colour. The virtual-virtual subtraction term,
 $\ds_{\NNLO}^{\rU}$, will be derived elsewhere.

Based on the universal factorisation properties of QCD colour-ordered
amplitudes,
the antenna formalism
\cite{Abelof:2011ap,Abelof:2011jv,Abelof:2012rv,Bernreuther:2011jt,
Bernreuther:2013uma,Boughezal:2010mc, Daleo:2006xa,Daleo:2009yj,Gehrmann:2011wi,
GehrmannDeRidder:2005cm,GehrmannDeRidder:2007jk,GehrmannDeRidder:2009fz,
GehrmannDeRidder:2011aa,GehrmannDeRidder:2012ja,Glover:2010im}
provides a general framework for the construction of subtraction terms that
reproduce the singular behaviour of the double real and mixed real-virtual
NNLO corrections.
Subtraction terms are constructed as products of antenna functions and
reduced matrix elements squared with remapped momenta, and the subtraction 
procedure is based on the colour ordering representation.

The antenna functions capture all the unresolved
radiation emitted between a pair of hard partons, referred to as hard
radiators.  In hadronic collisions, the hard radiators can be initial or
final state partons, and therefore three types of antennae must be
distinguished: final-final (f-f), initial-final (i-f) and initial-initial
(i-i).  While NLO subtraction terms only involve three-parton tree-level
antennae, at NNLO four-parton tree-level antennae and three-parton 
antennae are also needed in the double real and real-virtual contributions,
respectively.  In addition, $3\to 2$ and $4\to 2$ phase space mappings are
required for the reduced matrix elements multiplying the antenna functions
in the subtraction terms. Moreover, the analytic integration of the
subtraction terms over the appropriate unresolved patch of the phase space 
requires an exact and Lorentz invariant factorisation
of the phase space.
Both the mappings and the factorisations are different in f-f, i-f, and i-i
configurations.  They can all be found for the massive case in
\cite{Abelof:2011ap,Abelof:2011jv}.
   
The framework outlined above for the construction of NNLO antenna subtraction terms was set up in \cite{GehrmannDeRidder:2011aa,GehrmannDeRidder:2012dg,Glover:2010im} in the context of a proof-of-principle implementation of the purely gluonic leading-colour NNLO contributions to di-jet production at hadron colliders. In \cite{GehrmannDeRidder:2011aa,GehrmannDeRidder:2012dg}, the correctness of the method was checked by showing a complete cancellation of all explicit and implicit infrared divergences that arise in the intermediate steps of the calculation. These results were numerically implemented in the {\tt NNLOJET} parton-level event generator \cite{GehrmannDeRidder:2013mf}, producing the first NNLO results for hadronic di-jet production. 
A considerable reduction of the theoretical scale uncertainty was observed, and for the first time double differential distributions in $p_T$  and $\eta$  for inclusive-jet and di-jet NNLO cross sections were presented. Recently, these results have been upgraded to include the full colour dependence in \cite{Currie:2013dwa}.

As outlined above, the goal of this paper is to employ the antenna subtraction method in its extension to the massive case to compute the double real and real-virtual corrections to $t\bar{t}$ hadro-production in the $q\bar{q}$ channel. In particular we shall focus on the leading-colour pieces of processes $q\bar{q} \to t\bar{t}gg$ at tree-level and $q\bar{q} \to t \bar{t}g$ at  one-loop and their corresponding antenna subtraction terms denoted as $\ds_{\NNLO}^{\rS}$ and $\ds_{\NNLO}^{\rT}$ in eq.(\ref{eq.subnnlo}) . This will require the use of known phase space factorisations and mappings \cite{Abelof:2011ap,Abelof:2011jv,Abelof:2012he}  and several massive antennae.  
From those, the three-parton massive antennae are known. New tree-level four-parton and 
 three-parton antennae will be derived here for the first time. 
The general structure of the subtraction terms remains unchanged with respect to the massless case \cite{GehrmannDeRidder:2005cm,GehrmannDeRidder:2011aa,GehrmannDeRidder:2012dg,Glover:2010im} though, and we shall not repeat it here. We shall instead restrict our presentation to the new elements that are relevant for the present calculation and are related to the presence of massive final states.

Besides antenna subtraction, also the calculation of the $2\to 3$ one-loop
amplitudes represent a nontrivial ingredient of the $2\to 2$ NNLO
calculation at hand.  Thanks to the recent advent of fully automated NLO
tools, such contributions can be in principle computed on a routinely basis. 
However, the application of NLO tools in the framework of NNLO calculations
poses new and still poorly explored challenges.  First of all, depending on
the employed tool, the numerical character of the new one-loop algorithms
might imply a serious CPU speed penalty as compared to analytic approaches. 
Moreover, the integration of the (subtracted) one-loop contributions over
the soft and collinear regions of phase space can lead to serious numerical
instabilities.  In particular, the well known spurious singularities related
to small Gram determinants are inevitably enhanced in the infrared phase
space regions, and the resulting loss of numerical accuracy can be strongly
enhanced by the large cancellations between matrix elements and subtraction
terms.  It is thus a priori not clear if automated one-loop generators can
guarantee an adequate level of numerical stability and speed for NNLO
applications.  In this paper we address these issues using the
\OpenLoops~\cite{Cascioli:2011va} one-loop generator in combination with the
\Cuttools~\cite{Ossola:2007ax} reduction library, which allows us to study
the behaviour of one-loop matrix elements in the deep infrared regime using
quadruple precision.  As we will show, in spite of the presence of severe
instabilities associated with very soft gluon emissions, in the antenna
subtraction framework the employed tools turn out to be sufficiently stable
to perform a realistic NNLO calculation in double precision.  Given the high
speed of \OpenLoops and the fact that quadruple precision can be avoided
almost completely, this guarantees a highly efficient integration of the
real-virtual NNLO contributions.
 
The paper is organised as follows: In section \ref{sec.ttbNLO}, we shall
present the cross section for the top-antitop production up to the NLO
level.  This will enable us to set up the normalisation and present the NLO
ingredients required for the computation of the top-pair production cross
section at NNLO.  The new tree-level four-parton antenna and the 
three-parton antenna functions required at the double real and real-virtual
level of this computation will be presented in sections \ref{sec.ant4} and
\ref{sec:A31loop}, respectively.  The double real contributions and their
subtraction terms are derived in section \ref{sec.RR}.  Sections
\ref{sec.structureRV} and \ref{sec.openloops} contain the real-virtual
contributions.  Their general structure is presented in section
\ref{sec.structureRV} while their computation with \OpenLoops is described 
in section \ref{sec.openloops}. In section \ref{sec.RVsub}, we explicitly construct the 
real-virtual subtraction terms which cancel the explicit infrared poles of the
real-virtual contributions and approximate these contributions in all their
single unresolved limits.  Sections \ref{sec.tests} 
and \ref{sec.integration}
present various detailed checks on the consistency and numerical 
stability of the double real and real-virtual subtractions.  Finally,
section \ref{sec.conclusions} contains our conclusions. In appendix \ref{sec.unresolvedfactors},
the universal single unresolved soft and collinear factors are presented, in
appendix \ref{sec.iones} the colour-ordered infrared singularity operators
are included, while in appendix \ref{sec.a04} the full expression of the
antenna presented in section \ref{sec.ant4} is given.


\section{Top-antitop production in the $q\bar{q}$ channel at NLO}\label{sec.ttbNLO}
In this section we shall present the main ingredients that are required in the computation of the NLO cross section for $t\bar{t}$ hadronic production in the $q\bar{q}$ channel.  Besides setting up the notation and the general framework that we will follow throughout the present paper, those NLO contributions will be an essential input for the NNLO mass factorisation counter term $\ds_{q\bar{q},\NNLO}^{\MF,1}$ which shall be derived explicitly in section~\ref{sec.RVsub}.


\subsection{Notation and conventions}
To facilitate the reading of our expressions, we shall closely follow the notation in~\cite{Abelof:2011ap,Abelof:2011jv} for the matrix elements and subtractions terms. In order to identify the colour-ordered sub-amplitudes with the colour factors that multiply them in the colour decomposition of the full amplitude, we use the following conventions: Different colour strings are separated with double semicolons. A colour string $(T^{a_1}\ldots T^{a_n})_{ij}$ corresponds to $\ldots;;i,a_1,\ldots,a_n,j;;\ldots$ in the argument of the corresponding sub-amplitude. Adjacent partons within one colour string are colour-connected. An antiquark (or an initial state quark) at the end of a colour chain and a like-flavour quark (or initial state antiquark) at the beginning of a different colour chain are also colour-connected, since the two chains merge in the collinear limit where the $q\bar{q} $ pair clusters into a gluon. When decoupling identities are used, we denote the gluons which are photon-like and only couple to quark lines with the index $\gamma$ instead of $g$. In sub-amplitudes where all gluons are photon-like no semicolons are used, since the concept of colour connection in not meaningful. Finally, a hat over the label of a certain parton indicates that it is an initial state particle (for example, $\qi{1}$ is an initial state quark with momentum $p_1$).

Concerning the kinematics, we will use the following definition of invariants: 
\beq
\label{invariants}
s_{ij}=2 p_i  \cdot p_j,
\qquad s_{ijk}=2 p_i  \cdot p_j + 2 p_i \cdot p_k + 2 p_j  \cdot p_k,
\eeq
both for massless and massive momenta to make the mass-dependent terms explicitly proportional to $m_{Q}$. The momenta $p_{i,j,k}$ in eq.(\ref{invariants}) have to be understood as physical incoming or outgoing momenta with $p_{i,j,k}^0>0$. The above invariants are thus always positive, and crossing transformations have to be accompanied by sign-flips $s_{ij}\to -s_{ij}$,  $s_{ijk}\to -s_{ijk}$ whenever appropriate.

 
\subsection{$t\bar{t}$ production at LO}
The hadronic cross section for $t\bar{t}$ production at leading order involves two partonic channels, with either a $q\bar{q}$ pair or a pair of gluons in the initial state. It is given by  
\beqa
&&\hspace{-0.5in}{\rm d}\sigma_{\LO}(H_1,H_2)=\int\frac{{\rm d}\xi_1}{\xi_1}\frac{{\rm d}\xi_2}{\xi_2}\bigg( f_g(\xi_1,\mu)f_g(\xi_2,\mu)\:\ds_{gg,\LO}(p_1,p_2)\nonumber\\
&&\hspace{0.965in}+\sum_q f_q(\xi_1,\mu)f_{\bar{q}}(\xi_2,\mu)\:\ds_{q\bar{q},\LO}(p_1,p_2) \bigg),
\eeqa
where $H_1$ and $H_2$ are the momenta of the incoming hadrons, $p_i=\xi_i H_i$, and the sum runs over all quark flavours. Restricting ourselves to the $q \bar{q}$ initiated process, the leading order partonic cross section takes the form: 
\beq\label{eq.qqblo}
\ds_{q\bar{q},\LO}=\norm_{\LO}^{\:q\bar{q}}\int\dphi_2(p_3,p_4;p_1,p_2)\:|\cm_4(\Q{3},\Qb{4},\qbi{2},\qi{1})|^2 J^{(2)}_2(p_3,p_4),
\eeq
where, $\dphi_2(p_3,p_4;p_1,p_2)$ is the $2 \to 2$ partonic phase space, $J^{(2)}_2(p_3,p_4)$ is a so-called measurement function, which ensures that a pair of final state massive quarks of momenta $p_3$ and $p_4$ are observed.  $\cm_4(...)$ is the colour-ordered and coupling-stripped tree-level amplitude.  It is related to the full amplitude through the (trivial) colour decomposition
\beq\label{eq.coldecqqblo}
M_4^0(q_1\bar{q}_2\rightarrow Q_3 \bar{Q}_4)=g_s^2\left( \delta_{i_3i_1}\delta_{i_2i_4}-\frac{1}{N_c}\delta_{i_3i_4}\delta_{i_2i_1}\right)\cm_4(\Q{3},\Qb{4},\qbi{2},\qi{1}).
\eeq
The normalisation factor is
\beq\label{eq.normlo}
\norm_{\LO}^{\:q\bar{q}}=\frac{1}{2s}\:\left( \frac{\asmu}{2\pi}\right)^2\:\frac{\cepb^2}{\cep^2}\:\frac{(N_c^2-1)}{4N_c^2},
\eeq
where $s$ is the energy squared in the hadronic center-of-mass frame. Included in this normalisation factor are the flux factor, as well as the sum and average over colour and spin.

The constants $\cep$ and $\cepb$  are defined as:
\beq
\cep=\frac{(4\pi)^{\e}}{8\pi^2}e^{-\e \gamma_E} \hspace{1.5in}   
\cepb=(4\pi)^{\e} e^{-\e \gamma_E},
\eeq
providing the useful relation
\beq
g_s^2=4\pi\alpha_s=\left( \frac{\alpha_s}{2\pi}\right)\frac{\cepb}{\cep}.
\eeq


\subsection{$t\bar{t}$ production at NLO}
At the next-to-leading order, three different partonic channels enter: the $q\bar{q}$, the $gg$ and the $qg$ channels. The hadronic cross section for $t\bar{t}$ production at this order is therefore given by
\beqa
&&\hspace{-0.25in}{\rm d}\sigma_{\NLO}(P_1,P_2)=\int\frac{{\rm d}\xi_1}{\xi_1}\frac{{\rm d}\xi_2}{\xi_2}\bigg[ f_g(\xi_1,\mu)f_g(\xi_2,\mu)\:\ds_{gg,\NLO}(p_1,p_2)\phantom{\sum_q}\nonumber\\
&&\hspace{-0.25in}\phantom{{\rm d}\sigma_{\NLO}(P_1,P_2)}+\sum_q\bigg( f_q(\xi_1,\mu)f_{\bar{q}}(\xi_2,\mu)\:\ds_{q\bar{q},\NLO}(p_1,p_2)\nonumber\\
&&\hspace{-0.25in}\phantom{{\rm d}\sigma_{\NLO}(P_1,P_2)}+ \Big(f_q(\xi_1,\mu)+ f_{\bar{q}}(\xi_1,\mu)\Big)f_{g}(\xi_2,\mu)\ds_{qg,\NLO}(p_1,p_2)
\nonumber\\
&&\hspace{-0.25in}\phantom{{\rm d}\sigma_{\NLO}(P_1,P_2)}+ f_{g}(\xi_1,\mu)\Big(f_q(\xi_2,\mu)+ f_{\bar{q}}(\xi_2,\mu)\Big)\ds_{gq,\NLO}(p_1,p_2)
\bigg)\bigg],
\eeqa
where we have used the fact that the partonic cross sections for the $qg$ and the $\bar{q}g$ are identical due to their invariance under charge conjugation.

Restricting ourselves to the $q\bar{q}$ channel and employing a subtraction method at NLO, the partonic cross section takes the form, 
\beq\label{eq.qqbnlo}
\ds_{q\bar{q},\NLO}=\int_{{\rm d}\Phi_3}\left( \ds_{q\bar{q},\NLO}^{\rR}-\ds_{q\bar{q},\NLO}^{\rS} \right) +  \int_{{\rm d}\Phi_2}\left(
\ds_{q\bar{q},\NLO}^{\rV}
+\int_1 \ds_{q\bar{q},\NLO}^{\rS} +\ds_{q\bar{q},\NLO}^{\MF}\right).
\eeq
The three-parton final state contains the real radiation contributions  $\ds_{q\bar{q},\NLO}^{\rR}$ and their corresponding subtraction term $\ds_{q\bar{q},\NLO}^{\rS}$, whereas the two-parton final state includes the virtual contributions $\ds_{q\bar{q},\NLO}^{\rV}$, the integrated subtraction term $\int_1 \ds_{q\bar{q},\NLO}^{\rS}$ and the NLO mass factorisation counter term $\ds_{q\bar{q},\NLO}^{\MF}$. The latter is related to the leading order partonic cross section and will be given below. 

By grouping the different contributions to the NLO partonic cross section as in eq.(\ref{eq.qqbnlo}), the difference $(\ds_{q\bar{q},\NLO}^{\rR}-\ds_{q\bar{q},\NLO}^{\rS})$ is numerically well behaved in all regions of the $2\rightarrow 3$ phase space. It can be integrated numerically in four dimensions. Furthermore, the two-parton contributions $(\ds_{q\bar{q},\NLO}^{\rV}+\int_1 \ds_{q\bar{q},\NLO}^{\rS} +\ds_{q\bar{q},\NLO}^{\MF})$ are free of poles in the dimensional regulator $\e$ as we shall demonstrate below.


\subsubsection{Real radiation contributions}
The real radiation corrections to the $q\bar{q}$ channel for $t\bar{t} $ hadronic production are due to the process $q\bar{q}\rightarrow t\bar{t} g$. The colour decomposition of the corresponding tree-level amplitude is
\beqa\label{eq.colourdecqqbR}
&&\hspace{-0.1in}M_5^0(q_1\bar{q}_2\rightarrow Q_3\bar{Q}_4g_5)=\nonumber\\
&&\hspace{0.215in}g_s^3\sqrt{2}\bigg\{\bigg[ (T^{a_5})_{i_3i_1}\delta_{i_2i_4}\cm_5(\Q{3},\gl{5},\qi{1};;\qbi{2},\Qb{4})+(T^{a_5})_{i_2i_4}\delta_{i_3i_1}\cm_5(\Q{3},\qi{1};;\qbi{2},\gl{5},\Qb{4})\bigg]\nonumber\\
&&\hspace{0.35in}-\frac{1}{N_c}\bigg[(T^{a_5})_{i_3i_4}\delta_{i_2i_1}\cm_5(\Q{3},\gl{5},\Qb{4};;\qbi{2},\qi{1})+(T^{a_5})_{i_2i_1}\delta_{i_3i_4}\cm_5(\Q{3},\Qb{4};;\qbi{2},\gl{5},\qi{1})\bigg]\bigg\}.\nonumber\\
\eeqa
Squaring this expression and combining it with the $2 \to 3$ phase space, the appropriate overall factors and the measurement function, we can write the real radiation contributions as 
\beqa\label{eq.qqbR}
\ds_{q\bar{q},\NLO}^{\rR}&=&\norm_{\NLO}^{\rR,q\bar{q}}\:\dphi_3(p_3,p_4,p_5;p_1,p_2)\nonumber\\
&& \times\bigg\{ N_c\bigg[ |\cm_5(\Q{3},\gl{5},\qi{1};;\qbi{2},\Qb{4})|^2 + |\cm_5(\Q{3},\qi{1};;\qbi{2},\gl{5},\Qb{4})|^2 \bigg]\nonumber\\
&& \hspace{0.085in}+\frac{1}{N_c}\bigg[ |\cm_5(\Q{3},\gl{5},\Qb{4};;\qbi{2},\qi{1})|^2 + |\cm_5(\Q{3},\Qb{4};;\qbi{2},\gl{5},\qi{1})|^2\nonumber\\
&& \hspace{0.3in}-2|\cm_5(\Q{3},\Qb{4},\qbi{2},\qi{1},\ph{5})|^2\bigg]\bigg\}J^{(2)}_3(p_3,p_4,p_5)
\eeqa
where we have defined 
\beqa
\cm_5(\Q{3},\Qb{4},\qbi{2},\qi{1},\ph{5})&=&\cm_5(\Q{3},\gl{5},\qi{1};;\qbi{2},\Qb{4})+\cm_5(\Q{3},\qi{1};;\qbi{2},\gl{5},\Qb{4})\nonumber\\
&=&\cm_5(\Q{3},\gl{5},\Qb{4};;\qbi{2},\qi{1})+\cm_5(\Q{3},\Qb{4};;\qbi{2},\gl{5},\qi{1})
\eeqa
in which the gluon is $U(1)$-like. The normalisation factor $\norm_{\NLO}^{\rR,q\bar{q}}$ is given by
\beq
\norm_{\NLO}^{\rR,q\bar{q}}=\norm_{\LO}^{\:q\bar{q}}\:\frac{\asmu}{2\pi}\:\frac{\cepb}{\cep},
\eeq
and the measurement or jet function denoted by $J^{(2)}_3(p_3,p_4,p_5)$ guarantees that out of three-parton with momenta $p_3, p_4$ and $p_5$ a final state with a massive heavy quark pair is formed.

The matrix elements squared in eq.(\ref{eq.qqbR}) can become singular when the gluon, whose momentum is denoted by $p_5$ in the above equation, is either soft or collinear to either of the incoming partons. The antenna subtraction term that reproduces the behaviour of $\ds_{q\bar{q},\NLO}^{\rR}$ in those limits is known \cite{Abelof:2011jv}.  It is constructed entirely with products of A-type antennae and reduced matrix elements in final-final, initial-final and initial-initial kinematical configurations
\beqa\label{eq.qqbS}
\ds_{q\bar{q},\NLO}^{\rS}&=&\norm_{\NLO}^{\rR,q\bar{q}}\:\dphi_3(p_3,p_4,p_5;p_1,p_2)\nonumber\\
&&\times\bigg\{ N_c\bigg[ A_3^0(\Q{3},\gl{5},\qi{1})|\cm_4(\Q{(\wt{35})},\Qb{4},\qbi{2},\qi{\bar{1}})|^2 J^{(2)}_2(\wt{p_{35}},p_4)\nonumber\\
&&\hspace{0.35in}+A_3^0(\Qb{4},\gl{5},\qbi{2})|\cm_4(\Q{3},\Qb{(\wt{45})},\qbi{\bar{2}},\qi{1})|^2 J^{(2)}_2(p_3,\wt{p_{45}})\bigg]\nonumber\\
&&\hspace{0.075in}+\frac{1}{N_c}\bigg[ 2 A_3^0(\Q{3},\gl{5},\qbi{2})|\cm_4(\Q{(\wt{35})},\Qb{4},\qbi{\bar{2}},\qi{1})|^2 J^{(2)}_2(\wt{p_{35}},p_4)\nonumber\\
&&\hspace{0.36in}+2 A_3^0(\Qb{4},\gl{5},\qi{1})|\cm_4(\Q{3},\Qb{(\wt{45})},\qbi{2},\qi{\bar{1}})|^2 J^{(2)}_2(p_3,\wt{p_{45}})\phantom{\bigg[}\nonumber\\
&&\hspace{0.36in} -2A_3^0(\Q{3},\gl{5},\qi{1})|\cm_4(\Q{(\wt{35})},\Qb{4},\qbi{2},\qi{\bar{1}})|^2 J^{(2)}_2(\wt{p_{35}},p_4)\phantom{\bigg[}\nonumber\\
&&\hspace{0.36in}-2A_3^0(\Qb{4},\gl{5},\qbi{2})|\cm_4(\Q{3},\Qb{(\wt{45})},\qbi{\bar{2}},\qi{1})|^2 J^{(2)}_2(p_3,\wt{p_{45}})\phantom{\bigg[}\nonumber\\
&&\hspace{0.36in}- A_3^0(\Q{3},\gl{5},\Qb{4})|\cm_4(\Q{(\wt{35})},\Qb{(\wt{45})},\qbi{2},\qi{1})|^2 J^{(2)}_2(\wt{p_{35}},\wt{p_{45}})\phantom{\bigg[}\nonumber\\
&&\hspace{0.36in}- A_3^0(\qi{1},\gl{5},\qbi{2})|\cm_4(\Q{\tilde{3}},\Qb{\tilde{4}},\qbi{\bar{2}},\qi{\bar{1}})|^2 J^{(2)}_2(\wt{p_3},\wt{p_4}) \bigg]\bigg\}.
\eeqa

The reduced matrix elements and measurement functions in the equation above
contain redefined momenta that are obtained from the original ones through
Lorentz invariant on-shell mappings, whose form is different in subtraction
terms involving final-final, initial-final and initial-initial antennae. 
Final state and initial state remapped momenta are denoted with tildes (e.g. 
$\wt{p_{35}}$) and bars (e.g.  $\bar{p}_1$), respectively.  In final-final
subtraction terms the mappings employed are of the form
$\{p_i,p_j,p_k\}\to\{\wt{p_{ij}},\wt{p_{jk}}\}$ and both redefined momenta
are obtained from all three original momenta in the antenna system. 
Initial-final mappings are of the form
$\{p_i,p_j,p_k\}\to\{\bar{p}_i,\wt{p_{jk}}\}$, where $p_i$ is an initial
state momentum which the mapping rescales, and $\wt{p_{jk}}$ is obtained
from all three momenta in the antenna system.  For subtraction terms in
initial-initial configurations, the mapping rescales both initial state
momenta and performs a Lorentz boost on all remaining final state particles
in order to preserve momentum conservation in the reduced matrix elements. 
The precise definitions of all these mappings can be found, for example, in
\cite{Abelof:2011jv}.

The construction of the subleading colour pieces ($1/N_c$) of eq.(\ref{eq.qqbS}) requires a special procedure, which was explained in \cite{Abelof:2011ap,Abelof:2011jv}.

The integrated form of the NLO subtraction term $\ds_{q\bar{q},\NLO}^{\rS}$ is obtained by factorising the $2\to 3$ phase space into an antenna phase space and a reduced $2\to 2$ phase space, and integrating the antenna functions $A_3^0$ in eq.(\ref{eq.qqbS}) inclusively over the antenna phase space. This factorisation, as well as the specific form of the antenna phase space is different for final-final (f-f),initial-final( i-f) and initial-initial (i-i) configurations. It has been derived in the massless case in \cite{Daleo:2006xa,GehrmannDeRidder:2005cm} and in the massive case in \cite{Abelof:2011jv,GehrmannDeRidder:2009fz}. The integrated forms of the A-type antennae in eq.(\ref{eq.qqbS}) are denoted as ${\cal A}$.  We shall only make explicit use of their pole parts, which can be entirely written in terms of universal splitting kernels and infrared singularity operators as
\beqa
&&\hspace{-0.3in}\poles\left({\cal A}^0_{QgQ}(\e,s_{ij},x_1,x_2)\right)=-2\ione{Q}{Q}(\e,s_{ij})\delta(1-x_1)\delta(1-x_2)\phantom{\frac{1}{2}}\nonumber\\
&&\hspace{-0.3in}\poles\left({\cal A}^0_{q,Qg}(\e,s_{ij},x_1,x_2)\right)=-2\ione{Q}{q}(\e,s_{ij})\delta(1-x_1)\delta(1-x_2)+\frac{1}{2}\Gamma^{(1)}_{qq}(x_1)\delta(1-x_2)\label{eq.A03flint}\nonumber\\
&&\hspace{-0.3in}\poles\left({\cal A}^0_{q\bar{q},Q}(\e,s_{ij},x_1,x_2)\right)=-2\ione{q}{\bar{q}}(\e,s_{ij})\delta(1-x_1)\delta(1-x_2)\phantom{\frac{1}{2}}\nonumber\\
&&\hspace{-0.3in}\phantom{\poles\left({\cal A}_{q\bar{q},g}(\e,s_{ij},x_1,x_2)\right)=}+\frac{1}{2}\Gamma^{(1)}_{qq}(x_1)\delta(1-x_2)+\frac{1}{2}\Gamma^{(1)}_{qq}(x_2)\delta(1-x_1).
\eeqa
The colour-ordered infrared singularity operators of the form $\ione{i}{j}$ appearing in the above equation are given in appendix \ref{sec.iones}. The splitting kernel $\Gamma_{qq}^{(1)}(x)$ in $D=4-2\e$ dimensions is given by
\beq\label{eq.kernelqq}
\Gamma_{qq}^{(1)}(x)=-\frac{1}{\e}\left(\frac{3}{2}\delta(1-x)+2\Dzero(x)-1-x\right),
\eeq
with
\beq
\Dzero(x)=\bigg(\frac{1}{1-x}\bigg)_+.
\eeq

From these equations, we can express the pole part of the integrated form of eq.(\ref{eq.qqbS}) compactly as
\beqa\label{eq.intsubtermnlo}
&&\hspace{-0.35in}{\cal P}oles\left( \int_1 \ds_{q\bar{q},\NLO}^{\rS}  \right) = \norm_{\NLO}^{\rV,q\bar{q}}\int\frac{{\rm d}x_1}{x_1}\frac{{\rm d}x_2}{x_2} \dphi_2(p_3,p_4;x_1 p_1,x_2 p_2)|\cm_4(\Q{3},\Qb{4},\qbi{\bar{2}},\qi{\bar{1}})|^2\nonumber\\
&&\hspace{0.05in}\times\bigg\{ \delta(1-x_1)\delta(1-x_2)\bigg[ N_c\bigg(-2\ione{Q}{\bar{q}}(\e,s_{13})-2\ione{Q}{\bar{q}}(\e,s_{24}) \bigg)-\frac{1}{N_c}\bigg( 4\ione{Q}{\bar{q}}(\e,s_{14})\nonumber\\
&&\hspace{0.4in}   + 4\ione{Q}{\bar{q}}(\e,s_{23})- 4\ione{Q}{\bar{q}}(\e,s_{13}) - 4\ione{Q}{\bar{q}}(\e,s_{24}) - 2\ione{Q}{\bar{Q}}(\e,s_{34})- 2\ione{q}{\bar{q}}(\e,s_{12})\bigg)\bigg]\nonumber\\
&&\hspace{0.125in}+\bigg( \frac{N_c^2-1}{2N_c}\bigg)\bigg[\Gamma_{qq}^{(1)}(x_1)\delta(1-x_2)+\Gamma_{qq}^{(1)}(x_2)\delta(1-x_1)\bigg]\bigg\}J^{(2)}_2(p_3,p_4).\nonumber\\
\eeqa
with the normalisation factor $\norm_{\NLO}^{\rV,q\bar{q}}$ given by
\beq\label{eq.NV_NLO}
\norm_{\NLO}^{\rV,q\bar{q}}=\norm_{\LO}^{\:q\bar{q}}\:\frac{\asmu}{2\pi}\:\cepb=\norm_{\NLO}^{\rR,q\bar{q}}\:\cep.
\eeq


\subsubsection{Virtual contributions}\label{sec.virtualnlo}
The virtual contributions denoted by $\ds_{q\bar{q},\NLO}^{\rV}$ in eq.(\ref{eq.qqbnlo}) are due to the process $q \bar{q} \to t\bar{t}$ at . The colour decomposition of the relevant one loop amplitude is
\beqa\label{eq.coldecqqbV}
&&\hspace{-0.25in}M_4^1(q_1\bar{q}_2\rightarrow Q_3 \bar{Q}_4)=\nonumber\\
&&\hspace{0.15in}g_s^4\:\cep\bigg[ \delta_{i_3i_1}\delta_{i_2i_4}\cmb_{4,1}^{1}(\Q{3},\Qb{4},\qbi{2},\qi{1})+\delta_{i_3i_4}\delta_{i_2i_1}\cmb_{4,2}^{1}(\Q{3},\Qb{4},\qbi{2},\qi{1})\bigg].
\eeqa
Each of the partial amplitudes can be still decomposed into primitives as
\beqa\label{eq.partialdecqqbV}
\comb_{4,i}^{1}(\Q{3},\Qb{4},\qbi{2},\qi{1})&=&
N_c\:\comb_{4,i}^{[lc]}(\Q{3},\Qb{4},\qbi{2},\qi{1})+N_l\:\comb_{4,i}^{[l]}(\Q{3},\Qb{4},\qbi{2},\qi{1})\nonumber\\
&+&N_h\:\comb_{4,i}^{[h]}(\Q{3},\Qb{4},\qbi{2},\qi{1})-\frac{1}{N_c}\:\comb_{4,i}^{[slc]}(\Q{3},\Qb{4},\qbi{2},\qi{1}),
\eeqa
where  $N_l$ and $N_h$ are respectively the number of light and heavy flavours. Using eqs.(\ref{eq.coldecqqbV}) and (\ref{eq.partialdecqqbV}) together with the colour decomposition given in eq.(\ref{eq.coldecqqblo}) for the corresponding tree-level amplitude, we can write the virtual contributions to the $t \bar{t}$ production cross section in the $q\bar{q}$ channel as
\beqa\label{eq.qqbvirt}
&&\hspace{-0.2in}\ds_{q\bar{q},\NLO}^{\rV}=\norm_{\NLO}^{\rV,q\bar{q}}\:{\rm d}\Phi_2(p_3,p_4;p_1,p_2)
\nonumber\\
&&\times\bigg\{ N_c \big|\cmb_{4,1}^{[lc]}(\Q{3},\Qb{4},\qbi{2},\qi{1})\big|_{\NLO}^2 + N_l \big|\cmb_{4,1}^{[l]}(\Q{3},\Qb{4},\qbi{2},\qi{1})\big|_{\NLO}^2\nonumber\\
&&\hspace{0.1in}+N_h \big|\cmb_{4,1}^{[h]}(\Q{3},\Qb{4},\qbi{2},\qi{1})\big|_{\NLO}^2 - \frac{1}{N_c} \big|\cmb_{4,1}^{[slc]}(\Q{3},\Qb{4},\qbi{2},\qi{1})\big|_{\NLO}^2
\bigg\}J_2^{(2)}(p_3,p_4),\nonumber\\
\eeqa
where we have introduced the following compact notation
\beq
\label{eq:treeloopint}
\big|\cmb_{4,1}^{X}(\Q{3},\Qb{4},\qbi{2},\qi{1})\big|_{\NLO}^2=2{\rm Re}\bigg( \cmb_{4,1}^{X}(\Q{3},\Qb{4},\qbi{2},\qi{1})\cm_{4}(\Q{3},\Qb{4},\qbi{2},\qi{1})^{\dagger}\bigg).
\eeq
Interestingly, the partial amplitude $\cmb_{4,2}^{1}(\Q{3},\gli{1},\gli{2},\Qb{4})$ present in eq.(\ref{eq.coldecqqbV}) vanishes when interfered with the tree-level amplitude of eq.(\ref{eq.coldecqqblo}) and it drops out of $\ds_{q\bar{q},\NLO}^{\rV}$. These virtual contributions have been computed in \cite{Badger:2011yu,Beenakker:1988bq,Nason:1989zy}. Our expressions are in full agreement with those known results.

The  matrix elements in eq.(\ref{eq.qqbvirt}) contain ultraviolet as well as infrared divergences.  While the infrared divergences cancel when added to the integrated subtraction terms and mass factorisation terms, the UV poles are removed by renormalisation.  For all loop amplitudes throughout this paper we shall follow the renormalisation scheme described in \cite{Bonciani:2008az,Bonciani:2009nb}, in which the heavy quark mass and wave function are renormalised on shell, while the strong coupling constant is renormalised in the $\overline{\rm{MS}}$ scheme.  In the particular case of the  amplitude for the process $q\bar{q}\to Q\bar{Q}$ no mass renormalisation is needed since the corresponding tree-level process does not contain any internal massive propagators.  With this simplification, the amplitude is renormalised as
\beq
M^{(1)}_{\ren} = M_{\bare}^{(1)}+\bigg(\delta Z_{WF,Q}^{(1)}+2\delta Z_{\alpha_s}^{(1)}\bigg)M_{\bare}^{(0)},
\eeq
where the subscripts ($\bare$) and ($\ren$) stand for bare and renormalised respectively, and the renormalisation constants are given by
\beqa\label{eq:Zcst}
&&\hspace{-0.3in}\delta Z_{\alpha_s}^{(1)}=\cepb\left( -\frac{\beta_0}{\e}\right)\\
&&\hspace{-0.3in}\delta Z_{WF,Q}^{(1)}=\delta Z_{m_Q}^{(1)}=(4\pi)^{\e}\Gamma(1+\e)\left(\frac{\mu^2}{m_Q^2}\right)^{\e}C_F\left( -\frac{3}{2\e}-\frac{2}{1-2\e}\right)
\eeqa
with
\beq
\beta_0=\frac{11}{6}C_A-\frac{1}{3}(N_h+N_l).
\eeq
$C_A$ and $C_F$ are the $SU(N_c)$ Casimir operators, given by $C_A=N_c$, and $C_F=(N_c^2-1)/2N_c$. 

The explicit infrared pole structure of the UV-renormalised virtual contributions \linebreak $\ds_{q\bar{q},\NLO}^{\rV}$ can be casted in terms of massless and massive colour-ordered infrared singularity operators $\ione{i}{j}$ as,
\beqa\label{eq.polesvirtnlo}
&&\hspace{-0.2in}\poles\left(  \ds_{q\bar{q},\NLO}^{\rV}  \right) =\norm_{\NLO}^{\rV,q\bar{q}}\:\dphi_2(p_3,p_4;p_1,p_2)|\cm_4(\Q{3},\Qb{4},\qbi{2},\qi{1})|^2 \nonumber\\
&&\hspace{0.05in}\times \bigg[ N_c\, \bigg( 2\ione{Q}{\bar{q}}(\e,s_{13})+2\ione{Q}{\bar{q}}(\e,s_{24})\bigg)+\frac{1}{N_c}\bigg(  4 \ione{Q}{\bar{q}}(\e,s_{23})+4 \ione{Q}{\bar{q}}(\e,s_{14})\nonumber\\
&&\hspace{0.27in}- 4\ione{Q}{\bar{q}}(\e,s_{13})- 4\ione{Q}{\bar{q}}(\e,s_{24}) - 2\ione{Q}{\bar{Q}}(\e,s_{34})- 2\ione{q}{\bar{q}}(\e,s_{12},0)\bigg)\bigg]J^{(2)}_2(p_3,p_4).\nonumber\\
\eeqa
As can be seen in the equation above, after UV renormalisation, the remaining infrared poles of the virtual contributions are proportional to the colour factors $N_c$ and $1/N_c$.  The absence of infrared poles in the closed-fermion-loop contributions, that is, the contributions proportional to $N_l$ and $N_h$, is expected, since the real radiation contributions in eq.(\ref{eq.qqbR}) have no terms proportional to $N_l$ or $N_h$.

We have cross checked eq.(\ref{eq.polesvirtnlo}) against the known universal pole structure of  QCD amplitudes with massive fermions \cite{Catani:2000ef}, and found complete agreement.


\subsubsection{The mass factorisation counter term at NLO}
The general form of the NLO mass factorisation counter term is related to the leading order cross section through
\beqa\label{eq:MFNLO}
\ds_{ij,\NLO}^{\MF}(p_1,p_2) = -\int \frac{{\rm d}x_1}{x_1} \frac{{\rm d}x_2}{x_2}\,\GG^{(1)}_{ij;kl}(x_1,x_2)\,\ds_{kl,\LO}(x_1 p_1,x_2 p_2),
\eeqa
with the kernel $\GG^{(1)}_{ij;kl}$ defined as 
\beq\label{eq.oneloopkernelgeneral}
\GG^{(1)}_{ij;kl}(x_1,x_2)= \delta(1-x_2) \,\delta_{lj}\,\GG^{(1)}_{ki}(x_1)   +\delta(1-x_1)\,\delta_{ki}\,\GG^{(1)}_{lj}(x_2), 
\eeq
and $\GG^{(1)}_{ab}(z)$ are  Altarelli-Parisi kernels \cite{Altarelli:1977zs}.

Applying this expression to the $q\bar{q}$ initiated channel we find
\beqa\label{eq.mfnloqqb}
&&\hspace{-0.2in}\ds_{q\bar{q},\NLO}^{\MF} (p_1,p_2) = -\norm_{\NLO}^{\rV,q\bar{q}}\bigg( \frac{N_c^2-1}{2N_c}\bigg)\int \frac{{\rm d}x_1}{x_1}\frac{{\rm d}x_2}{x_2}\dphi_2(p_3,p_4;x_1p_1,x_2p_2)\nonumber\\
&&\hspace{0.5in}\times\bigg( \Gamma^{(1)}_{qq}(x_1)\delta(1-x_2)+\Gamma^{(1)}_{qq}(x_2)\delta(1-x_1)\bigg)|\cm_4(\Q{3},\Qb{4},\qbi{\bar{2}},\qi{\bar{1}})|^2J^{(2)}_2(p_3,p_4),\nonumber\\
\eeqa
where $\qi{\bar{1}}$ and $\qbi{\bar{2}}$ have momenta $x_1p_1$ and $x_2p_2$ respectively and $\Gamma^{(1)}_{qq}(x)$ was given in eq.(\ref{eq.kernelqq}).

Combining eqs.(\ref{eq.intsubtermnlo}), (\ref{eq.polesvirtnlo}) and (\ref{eq.mfnloqqb}), it is straightforward to see that 
\beq
{\cal P}oles\left(  \ds_{q\bar{q},\NLO}^{\rV} +  \int_1 \ds_{q\bar{q},\NLO}^{\rS} + \ds_{q\bar{q},\NLO}^{\MF} \right) =0.
\eeq


\section{The massive initial-final antenna $A_4^0(\Q{1},\gl{3},\gl{4},\qi{2})$}\label{sec.ant4}
Within the antenna formalism \cite{GehrmannDeRidder:2005cm}, the singular limits of the double real contributions that occur when a pair of colour-connected partons become simultaneously unresolved are captured by tree-level four-parton antenna functions. In general, these four-parton antenna functions are denoted as $X^0_4(i,j,k,l)$, and depend on the parton momenta $p_i,p_j,p_k,p_l$  and the masses of the hard radiators $m_i$ and $m_l$ in the massive case. They are obtained from ratios of colour-ordered matrix elements squared as 
\beq\label{eq.fourpantennae}
X_4^0(i,j,k,l)=S_{ijkl,IL}\,\frac{|\cm_4(i,j,k,l)|^2}{|\cm_2(I,L)|^2},
\eeq
where $S_{ijkl,IL}$ denotes a symmetry factor associated with the antenna which accounts both for potential identical particle symmetries and for the presence of more than one antenna in the basic two-parton process. This factor is fixed by demanding that the antennae collapse exactly into the unresolved factors appropriate to each unresolved limit. The flavours of the partons $I$ and $L$ in the two-parton matrix element are determined by the flavour of the two particles that the matrix elements $\cm_4(i,j,k,l)$ collapses onto when $j$ and $k$ become unresolved. According to the species of partons $I$ and $L$, antennae can be classified as quark-antiquark, quark-gluon, and gluon-gluon antennae, and depending on whether the hard radiators $i$ and $l$ are in the initial or in the final state, we distinguish between final-final (f-f), initial-final (i-f)and initial-initial (i-i) antennae.

In the context of this paper, one new massive tree-level four-parton antenna is needed to capture the double unresolved behaviour of the real matrix elements squared associated to the partonic channel $q\bar{q} \to t\bar{t}gg$ in the leading colour component. It is an A-type initial-final flavour-violating antenna which is denoted as $A_4^0(\Q{1},\gl{3},\gl{4},\qi{2})$. It is evaluated from the flavour violating tree-level process $\gamma^{*}q\rightarrow Qgg$ through the ratio
\beq
\frac{|\cm_4(\Q{1},\gl{3},\gl{4},\qi{2})|^2}{|\cm_2(\Q{(\wt{134})},\qi{\bar{2}})|^2}.
\eeq
The full expression of this antenna is rather lengthy and it will be left for appendix \ref{sec.a04}.  In the remaining part of this section, we shall present the single and double unresolved limits of this antenna $A_4^0(\Q{1},\gl{3},\gl{4},\qi{2})$.  We will start by presenting the double unresolved factors related to its double unresolved limits. The single unresolved factors are well known and are collected in appendix \ref{sec.unresolvedfactors} for completeness. The integrated form of $A_4^0(\Q{1},\gl{3},\gl{4},\qi{2})$ is presently unknown, it will be obtained by integrating the antenna over the appropriate initial-final antenna phase space, using the techniques developed in \cite{Abelof:2012he}. 
This integrated form will be part of  $\ds_{\NNLO}^{\rU}$ given in eq. \ref{eq.Udef} which will be derived elsewhere.


\subsection{Universal double unresolved factors} 
When a pair of massless particles becomes simultaneously unresolved, colour-ordered amplitudes squared factorise into a product of a double unresolved factor and a reduced matrix element with two particles less. The form of the double unresolved factor depends crucially on the colour-connection of the unresolved particles: when these are colour-connected a genuine double unresolved factor is obtained, whereas when they are colour-unconnected a product of two single unresolved factors is obtained. In the following we shall present the genuine colour-connected double unresolved factors that we encounter in the unresolved limits of the double real matrix elements squared associated with the partonic process $ q \bar{q} \to t \bar{t} g g $. These are a massless triple collinear factor corresponding to the triple collinear limit of both final state gluons and one of the initial state fermions, a massive double soft factor, and a massive soft-collinear factor. 

   
\subsubsection{Double soft factor of two colour-connected gluons}
When two colour-connected gluons $j$ and $k$ become soft between their neighbours $i$ and $l$ an $m$-particle colour-ordered matrix element factorises as
\beq\label{eq.factampgg}
\cm_m(\ldots,i,j,k,l,\ldots) \stackrel{^{p_j,p_k \rightarrow 0}}{\longrightarrow} \e^{\mu_1}(p_j,\lambda_1) \e^{\mu_2}(p_k,\lambda_2)J_{\mu_1\mu_2}(p_i,p_j,p_k,p_l)\cm_{m-2}(\ldots,i,l,\ldots),
\eeq
with the double soft current given by \cite{Berends:1988zn}
\beqa
&&\hspace{-0.3in}J^{\mu_1\mu_2}(p_i,p_j,p_k,p_l)=\frac{1}{2}\bigg[\frac{g^{\mu_1\mu_2}}{p_j\cdot p_k}\left(1-\frac{p_i\cdot p_j}{p_i\cdot(p_j+p_k)}-\frac{p_k\cdot p_l}{p_l\cdot(p_j+p_k)}\right)-\frac{p_i^{\mu_1}p_l^{\mu_2}}{\left(p_i\cdot p_j\right)\,\left(p_k\cdot p_l\right)}\nonumber\\
&&+\frac{p_i^{\mu_1}p_i^{\mu_2}}{\left(p_i\cdot p_j\right)\,\left(p_i\cdot(p_j+p_k)\right)}+\frac{p_l^{\mu_1}p_l^{\mu_2}}{\left(p_k\cdot p_l\right)\,\left(p_l\cdot(p_j+p_k)\right)}+\frac{p_i^{\mu_1}p_j^{\mu_2}}{\left( p_j\cdot p_k\right)\left( p_i\cdot(p_j+p_k) \right)}\nonumber\\
&&+\frac{p_k^{\mu_1}p_l^{\mu_2}}{\left(p_j\cdot p_k \right)\left( p_l\cdot(p_j+p_k)\right)}-\frac{p_i^{\mu_2}p_k^{\mu_1}}{\left(p_j\cdot p_k \right)\left(p_i\cdot (p_j+p_k) \right)}-\frac{p_j^{\mu_2}p_k^{\mu_1}}{\left(p_j\cdot p_l \right)\left(p_l\cdot (p_j+p_k) \right)}\bigg].
\eeqa
Squaring eq.(\ref{eq.factampgg}) and summing over the polarisations of the soft gluons we find
\beq\label{eq.wellknownfactgg}
|\cm_m(\ldots,i,j,k,l,\ldots)|^2 \stackrel{^{p_j,p_k \rightarrow 0}}{\longrightarrow} \soft{i}{j}{k}{l}(m_i,m_l)|\cm_{m-2}(\ldots,i,l,\ldots)|^2,
\eeq
with the massive double soft eikonal factor
\beqa
&&\hspace{-0.3in}\soft{i}{j}{k}{l}(m_i,m_l)=\frac{2s_{il}^2}{s_{ij}s_{kl}(s_{ij}+s_{ik})(s_{jl}+s_{kl})}+\frac{2s_{il}}{s_{jk}}\bigg[\frac{1}{s_{ij}s_{kl}}+\frac{1}{s_{ij}(s_{jl}+s_{kl})}\nonumber\\
&&+\frac{1}{s_{kl}(s_{ij}+s_{ik})}-\frac{4}{(s_{ij}+s_{ik})(s_{jl}+s_{kl})}\bigg]+\frac{2(1-\e)}{s_{jk}^2}\bigg(1-\frac{s_{ij}}{s_{ij}+s_{ik}}-\frac{s_{kl}}{s_{jl}+s_{kl}} \bigg)^2\nonumber\\
&&-\frac{4m_i^2}{s_{ij}s_{jk}(s_{ij}+s_{ik})}\bigg(\frac{s_{ik}}{s_{ij}+s_{ik}}+\frac{s_{il}s_{jk}}{s_{ij}s_{kl}}+\frac{s_{jl}}{s_{kl}}-\frac{s_{jl}}{s_{jl}+s_{kl}}\bigg)+\frac{4m_i^4}{s_{ij}^2(s_{ij}+s_{ik})^2}\nonumber\\
&&-\frac{4m_l^2}{s_{jk}s_{kl}(s_{jl}+s_{kl})}\bigg(\frac{s_{jl}}{s_{jl}+s_{kl}}+\frac{s_{il}s_{jk}}{s_{ij}s_{kl}}+\frac{s_{ik}}{s_{ij}}-\frac{s_{ik}}{s_{ij}+s_{ik}}\bigg)+\frac{4m_l^4}{s_{kl}^2(s_{jl}+s_{kl})^2}\nonumber\\
&&+\frac{4m_i^2m_l^2}{s_{ij}^2s_{kl}^2}.
\eeqa
This result converges to the massless colour-ordered double soft factor of \cite{Campbell:1997hg} in the limit where $m_i,m_l\rightarrow 0$.


\subsubsection{Soft-collinear factor in the colour-connected configuration}
Soft-collinear singularities occur in those regions of phase space in which a gluon becomes soft and two other massless partons become simultaneously collinear. The factorisation of colour-ordered matrix elements in these limits is different depending on the colour connection of the unresolved particles. When the soft gluon $j$ is colour connected to the collinear particles $k$ and $l$, the soft-collinear factorisation is given by
\beq
|\cm_m(\ldots,i,j,k,l,\ldots)|^2 \stackrel{^{p_k||p_l,p_j\rightarrow 0}}{\longrightarrow}\frac{1}{s_{kl}}P_{kl\rightarrow m}(z)Sc_{i,jkl}(m_i)|\cm_{m-2}(\ldots,i,m,\ldots)|^2,
\eeq
where $P_{kl\rightarrow m}(z)$ is one of the single collinear splitting functions in eqs.(\ref{eq.splitting1}-\ref{eq.splitting3}). If either parton $k$ or $l$ are in the initial state, $P_{kl\rightarrow m}(z)$ will be an initial-final Altarelli-Parisi splitting function. 

In the final-final case, the soft-collinear factor $Sc_{i,jkl}(m_i)$ reads
\beq
Sc_{i,jkl}(m_i)=\frac{2(s_{ik}+s_{il})}{s_{ij}(s_{jk}+s_{jl})}-\frac{2m_i^2}{s_{ij}^2},
\eeq
whereas in initial-final configurations they are
\beq
Sc_{i,jk\hat{l}}(m_i)=\ssoft{i}{j}{l}(m_i,0)\hspace{1in}Sc_{i j\hat{k}l}(m_i)=\ssoft{i}{j}{k}(m_i,0),
\eeq
with $\ssoft{i}{j}{k}(m_i,0)$ being the single massive soft factor given in eq.(\ref{eq.eikonalmassive}).


\subsubsection{Triple collinear factor}
In those regions of phase space where three colour-connected massless partons $(i,j,l)$ become collinear, a generic colour-ordered amplitude squared denoted by $|\cm_n(\ldots,i,j,k,\ldots)|^2$ factorises as
\beq
|\cm_{m}(\ldots,i,j,k,\ldots)|^2 \stackrel{^{p_i||p_j||p_k\rightarrow 0}}\longrightarrow P_{ijk \rightarrow l}|\cm_{m-2}(\ldots,l,\ldots)|^2,
\eeq
where the three colour-connected final state particles $(i,j,k)$ cluster to form a single parent particle $l$. The limit is approached in phase space when
\beq
p_i\rightarrow z_1 p_l\hspace{0.75in}p_j\rightarrow z_2 p_l\hspace{0.75in}p_k\rightarrow z_3 p_l
\eeq
with
\beq
 z_1+z_2+z_3=1 \hspace{1.25in}\text{and  }p_l^2\rightarrow 0.
\eeq
The triple collinear splitting functions generally depend on the momentum fractions $z_1$, $z_2$ and $z_3$, as well as on the invariants $s_{ij}$, $s_{jk}$, $s_{ik}$. The explicit functional form of $ P_{ijk \rightarrow l}$ varies according to the flavours of the three collinear particles as well as on their colour connection. There are two triple collinear splitting functions involving a fermion and two gluons, each of which applies to different colour orderings. In this paper we will need the one corresponding to a colour-ordering of the form $\ldots;;\q{i},\gl{j},\gl{k},\ldots$, in which case the splitting function is
\beqa
\lefteqn{P_{q_ig_jg_k \rightarrow q_l}(z_1,z_2,z_3,s_{ij},s_{ik},s_{jk}) =}\nonumber\\
&&\phantom{+} \frac{1}{s_{ij}s_{jk}} \bigg[ (1-\e) \left(\frac{1+z_1^2}{z_3}+\frac{1+(1-z_3)^2}{(1-z_1)} \right)+2\e \left( \frac{z_1}{z_3}+\frac{1-z_3}{1-z_1} \right) \bigg] \nonumber \\
&&+\frac{1}{s_{ij}s_{ijk}} \bigg[ (1-\e) \left( \frac{(1-z_3)^3+z_1(1-z_2)-2z_3}{z_3(1-z_1)} \right)- \e \left( \frac{2(1-z_3)(z_3-z_1)}{z_3(1-z_1)} -z_2 \right) -\e^2 z_2 \bigg] \nonumber\\
&&+\frac{1}{s_{jk}s_{ijk}} \bigg[ (1-\e) \left( \frac{(1-z_3)^2 (2-z_3)+z_2^3+2z_2z_1-2-z_3}{z_3(1-z_1)} \right)+2\e \frac{(z_2z_1-z_3-2z_3z_1)}{z_3(1-z_1)}\bigg] \nonumber \\
&&+(1-\e) \bigg[ \frac{2\left( z_2{s_{ijk}}-(1-z_1)s_{ij} \right)^2}{s_{jk}^2s_{ijk}^2(1-z_1)^2}+\frac{1}{s_{ijk}^2} \left( 4\frac{s_{ij}}{s_{jk}}+(1-\e) \frac{s_{jk}}{s_{ij}} + (3-\e) \right) \bigg].\nonumber\\
\eeqa

This triple collinear splitting function corresponds to a configuration in which the three collinear particles are in the final state. However, in the double real corrections for top pair production, only collinear limits of an initial state parton and two final state particles are relevant, given the fact that the tree-level matrix elements contain only two massless final state particles. The initial-final triple collinear splitting functions can always be obtained from their final-final counterparts. For example, the splitting function for the clustering $(\hat{i},j,k)\rightarrow \hat{l}$ can be related to the final-final case $(i,j,k)\rightarrow l$ as \cite{deFlorian:2001zd}
\beqa
\lefteqn{P_{\hat{i}jk \rightarrow \hat{l}}(z_1,z_2,z_3,s_{ij},s_{ik},s_{jk})=}\nonumber\\
&&(-1)^{\Delta}P_{ijk \rightarrow l}\left(\frac{1}{1-z_2-z_3},-\frac{z_2}{1-z_2-z_3},-\frac{z_3}{1-z_2-z_3},-s_{ij},-s_{ik},s_{jk}\right),
\eeqa
where $\Delta=0$ if the number of incoming fermions is the same before and after the crossing, and $\Delta=1$ otherwise.


\subsection{Infrared limits of $A_4^0(\Q{1},\gl{3},\gl{4},\qi{2})$}
The four-parton tree-level initial-final massive flavour-violating A-type antenna function denoted by $A_4^0(\Q{1},\gl{3},\gl{4},\qi{2})$ has the following single and double unresolved limits
\beqa
&&A_4^0(\Q{1},\gl{3},\gl{4},\qi{2})\stackrel{^{\gl{3},\gl{4}\rightarrow0}}{\longrightarrow}\soft{1}{3}{4}{2}(m_Q,0)\phantom{\frac{a}{b}}\\
&&A_4^0(\Q{1},\gl{3},\gl{4},\qi{2})\stackrel{^{\qi{2}||\gl{4},\, \gl{3}\rightarrow0}}{\longrightarrow}\frac{1}{s_{24}}\,P_{\hat{q}g\rightarrow \hat{q}}(z)\,Sc_{1;34\hat{2}}(m_Q)\\
&&A_4^0(\Q{1},\gl{3},\gl{4},\qi{2})\stackrel{^{\qi{2}||\gl{3}||\gl{4}}}{\longrightarrow}P_{\hat{q}gg\rightarrow \hat{q}}(z_1,z_2,z_3,s_{23},s_{24},s_{34})\phantom{\frac{a}{b}}\\
&&A_4^0(\Q{1},\gl{3},\gl{4},\qi{2})\stackrel{^{\gl{3}\rightarrow 0}}{\longrightarrow}\ssoft{1}{3}{4}(m_Q,0)A_3^0(\Q{1},\gl{4},\qi{2})\phantom{\frac{a}{b}}\\
&&A_4^0(\Q{1},\gl{3},\gl{4},\qi{2})\stackrel{^{\gl{4}\rightarrow 0}}{\longrightarrow}\ssoft{3}{4}{2}(0,0)A_3^0(\Q{1},\gl{3},\qi{2})\phantom{\frac{a}{b}}\\
&&A_4^0(\Q{1},\gl{3},\gl{4},\qi{2})\stackrel{^{\qi{2}||\gl{4}}}{\longrightarrow}\frac{1}{s_{24}}\,P_{\hat{q}g\rightarrow \hat{q}}(z)A_3^0(\Q{1},\gl{3},(\widehat{24})_q)\\
&&A_4^0(\Q{1},\gl{3},\gl{4},\qi{2})\stackrel{^{\gl{3}||\gl{4}}}{\longrightarrow}\frac{1}{s_{34}}\,P_{gg\rightarrow g}(z)A_3^0(\Q{1},\gl{(\wt{34})},\qi{2})+(\text{ang.}),
\eeqa 
In this last equation (ang.), stands for angular dependent terms. Those terms arise when a gluon splitting is involved in a collinear limit. In this case, the unresolved single collinear factor is not a spin-averaged Altarelli-Parisi splitting function as given in appendix \ref{sec.unresolvedfactors} but it also involves spin-dependent terms \cite{Abelof:2011ap}.


\section{The massive  initial-final antenna $A_3^1(\Q{1},\gl{3},\qi{2})$}\label{sec:A31loop}
The construction of a subtraction term for the real-virtual corrections to $t\bar{t}$ production in the $q\bar{q}$ channel in the leading-colour approximation requires a new initial-final  one-loop massive antenna function which we will present in this section. 

\subsection{ One-loop antenna functions}
Within the antenna formalism, the infrared limits of the real-virtual contributions are captured by  three-parton one-loop antennae \cite{GehrmannDeRidder:2005cm,GehrmannDeRidder:2011aa}. These are generally denoted as $X^1_3(i,j,k)$ and they depend on the antenna momenta $p_i,p_j,p_k$ as well as on the masses of the hard radiators in the massive case. In general,  these one-loop antenna functions are constructed out of colour-ordered  three-parton and two-parton matrix elements as
\beq\label{eq:X1def}
X_{3}^1(i,j,k) = S_{ijk,IK}\, \frac{\big|{\cal M}^1_{3}(i,j,k)\big|_{\NLO}^2}{|{\cal M}^0_2(I,K)|^2} - X_3^0(i,j,k)\, \frac{|{\cal M}^1_{2}(I,K)|^2}{|{\cal M}^0_2(IK)|^2} \;,
\eeq
where the tree-level antenna function, denoted by $X_3^0(i,j,k)$, is given by
\beq
X_3^0(i,j,k) = S_{ijk,IK}\, \frac{|{\cal M}^0_{ijk}|^2}{|{\cal M}^0_{IK}|^2}. 
\eeq
$S_{ijk,IK}$ denotes the symmetry factor associated with the antenna, which accounts both for potential identical particle symmetries and for the presence of more than one antenna in the basic two-parton process. Initial-final and initial-initial antennae can be obtained from their final-final counterparts by the appropriate crossing of partons to the initial-state. This procedure is straightforward at tree-level but requires some care in the one-loop case, since  antennae contain polylogarithms or hypergeometric functions that must be analytically continued to the appropriate kinematical region \cite{Daleo:2009yj,Gehrmann:2011wi}.

In any of the three kinematical configurations, the  antenna functions can be conveniently decomposed according to their colour factors as follows:\footnote{In \cite{GehrmannDeRidder:2005cm}, the leading colour contribution was denoted by $X^{1}_3(i,j,k)$, the subleading colour part by $\tilde{X}^{1}_3(i,j,k)$, and the $N_F$ part was denoted by $\hat{X}^{1}_3(i,j,k)$. We shall not use this notation here but instead follow that of eq.(\ref{eq.antennadec}), which we find more transparent.}
\beq\label{eq.antennadec}
X_3^1(i,j,k)=N_c X_3^{1,lc}(i,j,k) + N_l X_3^{1,l}(i,j,k)+ N_h X_3^{1,h}(i,j,k)-\frac{1}{N_c}X_3^{1,slc}(i,j,k).
\eeq
In general the  sub-antennae have ultraviolet and infrared divergences of explicit and implicit nature. In order to remove the ultraviolet poles, we renormalise the  amplitudes in eq.(\ref{eq:X1def}) following the scheme of \cite{Bonciani:2008az,Bonciani:2009nb}, with the renormalisation constants given in eq.(\ref{eq:Zcst}). We find that the renormalisation prescription of the different  sub-antennae is
\beqa\label{eq.renant1}
&&\hspace{-0.25in}X_3^{1,lc}(i,j,k)=X_{3,b}^{1,lc}(i,j,k)-\mu^{-2\e}\:\cepb\:\frac{b_0}{\e}X_3^0(i,j,k)\nonumber\\
&&\hspace{0.5in}-m_Q^{-2\e}\left(4\pi\right)^\e\Gamma(1+\e)\bigg(\frac{3}{2\e}+\frac{2}{1-2\e}\bigg)X_{3,1M}^0(i,j,k)\\
&&\hspace{-0.25in}X_3^{1,N_l}(i,j,k)=X_{3,b}^{1,N_l}(i,j,k)-\mu^{-2\e}\:\cepb\:\frac{b_{0,F}}{\e}X_3^0(i,j,k)\\
&&\hspace{-0.25in}X_3^{1,N_h}(i,j,k)=X_{3,b}^{1,N_h}(i,j,k)-\mu^{-2\e}\:\cepb\:\frac{b_{0,F}}{\e}X_3^0(i,j,k)\\
&&\hspace{-0.25in}X_3^{1,slc}(i,j,k)=X_{3,b}^{1,slc}(i,j,k)+m_Q^{-2\e}\left(4\pi\right)^\e\Gamma(1+\e)\bigg(\frac{3}{2\e}+\frac{2}{1-2\e}\bigg)X_{3,1M}^0(i,j,k),
\eeqa
where $b_0=11/6$ and $b_{0,F}=-1/3$ are the colour-ordered components of the QCD  beta function. We have also defined 
\beq
X_{3,1M}^0(i,j,k)=S_{ijk,IK}\frac{\re\,\left(\cmb_{3,1M}^0(i,j,k)\:(\cmb_3^0(i,j,k))^{\dagger}\right)}{|\cmb_2^0(I,K)|^2},
\eeq
where $\cmb_{3,1M}^0(i,j,k)$  is the tree-level amplitude with a mass insertion in the massive propagators. Interestingly, the wave function renormalisation counter terms coming from $\cmb_3^1(i,j,k)$ cancel against those coming from $\cmb_2^1(I,K)$, in such a way that the antenna function itself does not require wave function renormalisation. 

The  antennae that we employ in the real-virtual subtraction terms are renormalised at $\mu^2=|s_{ijk}|$. To ensure that the  matrix elements in the real-virtual contributions and the  antennae are renormalised at the same scale, we must substitute
\beqa\label{eq.substitutions}
&& X_{ijk}^{1,lc}\rightarrow X_{ijk}^{1,lc} +  \frac{b_0}{\e} X_{ijk}^0 \left( (|s_{ijk}|)^{-\e}-(\mu^2)^{-\e}\right)\\
&&X_{ijk}^{1,N_l}\rightarrow X_{ijk}^{1,N_l} + \frac{b_{0,F}}{\e} X_{ijk}^0 \left( (|s_{ijk}|)^{-\e}-(\mu^2)^{-\e}\right)\\
&&X_{ijk}^{1,N_h}\rightarrow X_{ijk}^{1,N_h} + \frac{b_{0,F}}{\e} X_{ijk}^0 \left( (|s_{ijk}|)^{-\e}-(\mu^2)^{-\e}\right)\\
&& X_{ijk}^{1,slc}\rightarrow X_{ijk}^{1,slc}.\phantom{\frac{b_{0,F}}{\e}}
\eeqa

After UV renormalisation,  one-loop antennae still have explicit and implicit infrared divergences. The structure of the former can be entirely captured by colour-ordered infrared singularity operators; the latter occur when massless partons in the antenna become soft or collinear.


\subsection{Single unresolved factors at one-loop}
The factorisation properties of  colour-ordered amplitudes in their soft and collinear limits has been extensively studied in \cite{Weinzierl:2003ra,Bern:1994zx,Bern:1998sc,Kosower:1999xi,Kosower:1999rx,Bern:1999ry,Catani:2000pi,Kosower:2002su,Kosower:2003cz,Catani:2003vu,Bern:2004cz,Badger:2004uk,Bierenbaum:2011gg}. Like at tree-level, the interference of a  one-loop amplitude with its tree-level counterpart yields  soft eikonal factors and collinear splitting functions in its soft and collinear limits respectively. Those singular factors are also found in the unresolved limits of  antennae. 

In general, when a gluon becomes soft or a pair of massless partons become collinear, the interference of a  one-loop and a tree-level colour-ordered amplitude factorises as 
\beq
\big|\cmb_m^{1}\big|_{\NLO}^2\rightarrow \text{Sing}^{(0)}_1 \big|\cmb^1_{m-1}\big|_{\NLO}^2+\text{Sing}^{(1)}_1 \big|\cmb^0_{m-1}\big|^2,
\eeq
where $\text{Sing}^{(1)}_1$ is a  single unresolved factor and $\big|\cmb^1_{m-1}\big|_{\NLO}^2$ is the interference of a  reduced one-loop sub-amplitude and its tree-level counterpart. Following the decomposition of the  colour-ordered amplitudes into primitives, the  unresolved factors can be decomposed as
\beq
\text{Sing}^{(1)}_1=N_c \,\text{Sing}^{(1),[lc]}_1+N_l \, \text{Sing}^{(1),[l]}_1+N_h \, \text{Sing}^{(1),[h]}_1-\frac{1}{N_c} \, \text{Sing}^{(1),[slc]}_1.
\eeq

In the following we shall present the explicit form of the  singular factors that must be considered in the construction of subtraction terms for the leading-colour real-virtual corrections to top pair production in the $q\bar{q}$ channel.


\subsubsection{ Collinear splitting functions}
For the  partonic process that we are presently considering, i.e.~$q\bar{q}\rightarrow t\bar{t}g$, the  splitting function that occurs when the final-state gluon becomes collinear to either of the incoming fermions is $P^{1}_{qg\leftarrow Q}(z)$. In the leading-colour approximation, only the $N_c$ part of this splitting function is needed, and it is given by
\beqa\label{eq.splittingloop}
&&\hspace{-0.3in}P^{(1),[lc]}_{\hat{q}g\rightarrow \hat{q}}(z)=\bigg[-\frac{b_0}{\e}-\frac{c_{\Gamma}}{\cepb}\bigg(\frac{s_{qg}}{\mu^2}\bigg)^{-\e}\bigg( \frac{\Gamma(1-\e)}{\e^2}\bigg(\frac{z}{(1+\e)\Gamma(-\e)}\gaussf{1}{1+\e}{2+\e}{z}\nonumber\\ 
&&+(-z)^{-\e}\Gamma(1+\e)\bigg)-\frac{\e}{2}\bigg]P_{\hat{q}g\rightarrow \hat{q}}(z)-\frac{c_{\Gamma}}{\cepb}\bigg(\frac{s_{qg}}{\mu^2}\bigg)^{-\e}\bigg(\frac{2\e+z-z^2(1-\e^2)}{z(1-z)}\bigg).
\eeqa
In this equation, $z$ is the momentum fraction carried by the gluon and $P_{\hat{q}g\rightarrow \hat{q}}(z)$ is the tree-level splitting function whose expression is
\beq
P_{\hat{q}g\rightarrow \hat{q}}(z)=\frac{1+(1-z)^2-\e z^2}{z(1-z)}.
\eeq


\subsubsection{Massive soft factors}
As it occurs at tree-level, when a soft gluon is emitted between massive fermions in the colour chain, the soft  factor contains mass dependent terms. While at tree-level the massless soft factor can be obtained from the massive one by setting the massess of the hard radiators to zero, this is no longer the case at the  one-loop level: masses are present in the arguments of logarithms that diverge in the massless limit. We must therefore consider separately the soft factors with: (a) two massless hard radiators, (b) one massless and one massive hard radiator, (c) two massive hard radiators. When treating the real-virtual corrections to top pair hadro-production within the leading-colour approximation, only case (b) must be considered. Furthermore, only the $N_c$ part of the soft currents and eikonal factors are needed.

When a gluon $j$ becomes soft in a  primitive amplitude where it is colour-connected to the hard particles $i$ and $k$, the amplitude factorises as
\beqa
&&\hspace{-0.5in}\cmb^{1,[X]}_{m}(\ldots,i,j,k,\ldots)\stackrel{^{p_j\rightarrow 0}}{\longrightarrow}\,\e^{\mu}(p_j,\lambda)J_{\mu}(p_i,p_j,p_k)\cmb^{1,[X]}_{m-1}(\ldots,i,k,\ldots)\phantom{\Big(}\nonumber\\
&&\hspace{-0.62in}\phantom{\cmb^{1,[X]}_{m}(\ldots,i,j,k,\ldots)\stackrel{^{p_j\rightarrow 0}}{\longrightarrow}}+\e^{\mu}(p_j,\lambda)J_{\mu}^{(1),[X]}(p_i,p_j,p_k;m_i,m_k)\cm_{m-1}(\ldots,i,k,\ldots),
\eeqa
where $X=lc,l,h,slc$, and the tree-level current $J_{\mu}(p_i,p_j,p_k)$ is given by
\beq\label{eq.currentg}
J_{\mu}(p_i,p_j,p_k)=\frac{p_i^{\mu}}{\sqrt{2}p_i\cdot p_j}-\frac{p_k^{\mu}}{\sqrt{2}p_j\cdot p_k}.
\eeq
The primitive  currents $J_{\mu}^{(1),[X]}(p_i,p_j,p_k;m_i,m_k)$ take a different form depending on whether $m_i$ and/or $m_k$ vanish. These massive  soft currents were derived in \cite{Bierenbaum:2011gg} as tensors in colour space that describe the soft factorisation of full amplitudes rather than of colour-ordered sub-amplitudes. The renormalised colour-ordered currents can be obtained from their results. In the case of one vanishing mass, the leading-colour current that we are presently interested in, reads
\beqa
&&\hspace{-0.4in}J_{\mu}^{(1),[lc]}(p_i,p_j,p_k;m_i,0)=-\frac{1}{2}\bigg\{ \frac{b_0}{\e}+\left(\frac{\mu^2s_{ik}}{s_{ij}s_{jk}}\right)^{-\e}\bigg[ \frac{1}{2\e^2}+\frac{i\pi}{2\e}-\frac{5\pi^2}{12}\nonumber\\
&&\hspace{-0.1in}+\frac{m_i^2s_{jk}}{2(s_{ij}s_{ik}-m_i^2s_{jk})}\bigg(2i\pi\ln\left(\frac{m_i^2s_{jk}}{s_{ij}s_{ik}}\right)+\ln^2\left(\frac{m_i^2s_{jk}}{s_{ij}s_{ik}}\right) \bigg)+\order{\e}\bigg]\bigg\}J_{\mu}(p_i,p_j,p_k).\nonumber\\
\eeqa

From the  one-loop and tree-level soft currents, massive  soft eikonal factors are obtained as
\beq
S^{(1),[X]}_{ijk}(m_i,m_k)=-2\,g^{\mu\nu}\,\re\left(J_{\mu}^{(1),[X]}(p_i,p_j,p_k;m_i,m_k)J_{\nu}(p_i,p_j,p_k)\right).
\eeq
We find
\beqa\label{eq.softloop}
&&\hspace{-0.35in}S^{(1),[lc]}_{ijk}(m_i,0)=-\bigg\{ \frac{b_0}{\e}+\left(\frac{\mu^2s_{ik}}{s_{ij}s_{jk}}\right)^{-\e}\bigg[ \frac{1}{2\e^2}-\frac{5\pi^2}{12}+\frac{m_i^2s_{jk}}{2(s_{ij}s_{ik}-m_i^2s_{jk})}\ln^2\left(\frac{m_i^2s_{jk}}{s_{ij}s_{ik}}\right)\nonumber\\
&&\hspace{1in}+\order{\e}\bigg]\bigg\}\ssoft{i}{j}{k}(m_i,0)
\eeqa
with the massive tree-level eikonal factor $\ssoft{i}{j}{k}(m_i,0)$ given in appendix \ref{sec.softfactor}.


\subsection{Infrared properties of $A_3^{1,lc} (\Q{1},\gl{3},\qi{2})$}
As mentioned above, in the context of this paper, a new massive  antenna is needed to subtract the unresolved infrared limits of the real-virtual contributions related to the  partonic process $q \bar{q} \to t \bar{t} g$. It is a flavour-violating quark-antiquark antenna denoted by $A_3^1(\Q{1},\gl{3},\qi{2})$, which we compute directly in the initial-final kinematics following the definition of eq.(\ref{eq:X1def}). Working in the leading-colour approximation, only the leading-colour part of the antenna $A_3^{1,lc} (\Q{1},\gl{3},\qi{2})$ needs to be considered. The full expression of this sub-antenna is too lengthy to be presented in this paper, but its pole part can be compactly written in terms of colour-ordered $\ione{i}{j}$ operators.
This pole part will be explicitly needed in section \ref{sec.RVsub} and is given by
\beq\label{eq.polesA13iflc}
\poles \left( A_3^{1,lc}(\Q{1},\gl{3},\qi{2})\right)=2\left(\ione{Q}{g}(\e,s_{13})+\ione{q}{g}(\e,s_{23})-\ione{Q}{\bar{q}}(\e,s_{123})\right)A_3^0(\Q{1},\gl{3},\qi{2}).
\eeq

Also the unresolved limits of $A_3^{1,lc} (\Q{1},\gl{3},\qi{2})$ will be required in section \ref{sec.RVsub} in the context of the construction of our real-virtual subtraction terms. They read
\beqa
&&A_3^{1,lc}(1_Q,3_g,\qi{2})\mathop{\longrightarrow}^{p_3\rightarrow 0}S^{(1),[lc]}_{132}(m_Q^2,0)\\
&&A_3^{1,lc}(1_Q,3_g,\qi{2})\mathop{\longrightarrow}^{p_2||p_3}\frac{1}{s_{23}}P^{1,[lc]}_{\hat{q}g\rightarrow \hat{q}}(z),
\eeqa
with the  soft and collinear factors defined in eqs.(\ref{eq.softloop}) and (\ref{eq.splittingloop}) respectively.
The integrated form of $A_3^{1,lc} (\Q{1},\gl{3},\qi{2})$ is not known at present. It will be part of 
$ \ds_{\NNLO}^{\rU}$ given in eq. \ref{eq.Udef} which will be derived elsewhere.


\section{Double real contributions to $q\bar{q} \to t\bar{t}$ at leading colour}\label{sec.RR}
It is the purpose of this section to present the structure of the double real contributions associated to the tree-level process $ q \bar{q} \to t \bar{t} g g $ at leading colour, and to construct the corresponding subtraction terms.


\subsection{The double real contribution $\ds_{q\bar{q},\NNLO,N_c^2}^{\RR}$}
The colour-decomposition of the tree-level amplitude for the partonic process $ q \bar{q} \to t \bar{t} g g $ reads
\beqa
&&\hspace{-0.2in}M_6^0(q_1 \bar{q}_2 \rightarrow Q_3 \bar{Q}_4 g_5 g_6)= 2\:g_s^4\sum_{(i,j)\in P(5,6)}\bigg[ \nonumber\\
&&\hspace{1in}(T^{a_i}T^{a_j})_{i_3i_1}\delta_{i_2i_4}\cm_6(\Q{3},\gl{i},\gl{j},\qi{1};;\qbi{2},\Qb{4})\phantom{\frac{1}{N_c}} \nonumber\\
&&\hspace{0.9in}+ (T^{a_i})_{i_3i_1}(T^{a_j})_{i_2i_4}\cm_6(\Q{3},\gl{i},\qi{1};;\qbi{2},\gl{j},\Qb{4})\phantom{\frac{1}{N_c}}\nonumber\\
&&\hspace{0.9in}+ \delta_{i_3i_1}(T^{a_i}T^{a_j})_{i_2i_4}\cm_6(\Q{3},\qi{1};;\qbi{2},\gl{i},\gl{j},\Qb{4})\phantom{\frac{1}{N_c}}\nonumber\\
&&\hspace{0.9in} - \frac{1}{N_c}(T^{a_i}T^{a_j})_{i_3i_4}\delta_{i_2i_1}\cm_6(\Q{3},\gl{i},\gl{j},\Q{4};;\qbi{2},\qi{1})\phantom{\frac{1}{N_c}}\nonumber\\
&&\hspace{0.9in}- \frac{1}{N_c}(T^{a_i})_{i_3i_4}(T^{a_j})_{i_2i_1}\cm_6(\Q{3},\gl{i},\Q{4};;\qbi{2},\gl{j},\qi{1})\phantom{\frac{1}{N_c}}\nonumber\\
&&\hspace{0.9in}- \frac{1}{N_c}\delta_{i_3i_4}(T^{a_i}T^{a_j})_{i_2i_1}\cm_6(\Q{3},\Q{4};;\qbi{2},\gl{i},\gl{j},\qi{1})\bigg].
\eeqa
Squaring this expression, combining it with all appropriate prefactors, phase space and measurement function, and retaining only the terms multiplied by $N_c^2$, we have
\beqa\label{eq.doublereal}
&&\hspace{-0.1in}\ds_{q\bar{q},\NNLO,N_c^2}^{\RR}=\frac{1}{2}\,\norm_{\NNLO}^{q\bar{q},\RR}\,N_c^2\hspace{-0.075in}\sum_{(i,j)\in P(5,6)}\hspace{-0.075in}\dphi_4(p_3,p_4,p_5,p_6;p_1,p_2)\bigg[|\cm_6(\Q{3},\gl{i},\gl{j},\qi{1};;\qbi{2},\Qb{4})|^2\nonumber\\
&&\hspace{0.5in}+|\cm_6(\Q{3},\gl{i},\qi{1};;\qbi{2},\gl{j},\Qb{4})|^2+|\cm_6(\Q{3},\qi{1};;\qbi{2},\gl{i},\gl{j},\Qb{4})|^2\bigg]J_{2}^{(4)}(p_3,p_4,p_5,p_6),\nonumber\\
\eeqa
where the overall factor $1/2$ accounts for the identical gluons in the final state. The normalisation factor is
\beq
\norm_{\NNLO}^{q\bar{q},\RR}=\norm_{\LO}^{q\bar{q}}\left(\frac{\asmu}{2\pi}\right)^2\:\frac{\cepb}{\cep},
\eeq
and $\norm_{\LO}^{q\bar{q}}$ has been given in eq.(\ref{eq.normlo}).

This contribution is singular in several single and double unresolved limits namely 
\begin{itemize}
\item Single soft limits: $p_i\rightarrow 0$ with $i=5,6$
\item Single collinear limits: $p_5||p_6$, and $p_i||p_j$ with $i=5,6$, $j=1,2$
\item Double soft limit: $p_5,p_6\rightarrow 0$
\item Triple collinear limits: $p_i||p_5||p_6$ with $i=1,2$
\item Soft-collinear limits: $p_5\rightarrow 0,\, p_i||p_6$ and $p_6\rightarrow 0,\, p_i||p_6$ with $i=1,2$
\item Double collinear limits: $p_i||p_5,\, p_j||p_6$ with $i,j=1,2\,\,i\neq j$
\end{itemize}


\subsection{The double real subtraction term $\ds^{\rS}_{q\bar{q},\NNLO,N_c^2}$}
The general structure of the double real subtraction terms obtained within the framework of the antenna formalism has been presented in \cite{GehrmannDeRidder:2005cm,Glover:2010im} in the massless case and extended to the massive case in \cite{Abelof:2011ap,Abelof:2012rv}. Without entering into the details of this structure, let us recall that in general, double real antenna subtraction terms, which reproduce the behaviour of the double real contributions in all their single and double unresolved limits, contain five different configurations corresponding to:
\begin{itemize}
\item[(a)] One unresolved parton 
\item[(b)] Two colour-connected unresolved partons (colour-connected)
\item[(c)] Two unresolved partons that are not colour-connected but share a common radiator (almost colour-connected)
\item[(d)] Two unresolved partons that are well separated from each other in the colour chain (colour-unconnected)
\item[(e)] Compensation terms for the over subtraction of large angle soft emission.
\end{itemize}
The antenna content of the subtraction terms for each of these configurations is the same for the final-final, initial-final and initial-initial configurations and it is summarised in Table \ref{tab:S1breakdown}, which is taken from \cite{GehrmannDeRidder:2011aa}.\footnote{As discussed in \cite{Abelof:2011ap,Currie:2013dwa} for example, this content is strictly valid only for leading-colour like double real contributions which involve colour-ordered matrix elements squared. For subleading colour contributions involving interferences of colour-ordered matrix elements, more antenna functions are needed.}
\begin{table}[t!]
\begin{center}
{\small
\begin{tabular}{|c|ccccc|}
\hline
 & $a$ & $b_4$ & $b_{3\times 3},c$ & $d$ & $e$   \\\hline
$\ds_{\NNLO}^{\rS}$ 
& $X_3^0 |{\cal M}^0_{m+3}|^2$ 
& $X_4^0 |{\cal M}^0_{m+2}|^2$ 
& $X_3^0 X_3^0 |{\cal M}^0_{m+2}|^2$
& $X_3^0 X_3^0 |{\cal M}^0_{m+2}|^2$ 
& $S X_3^0 |{\cal M}^0_{m+2}|^2$   \\
$\int_1 \ds_{\NNLO}^{\rS,1}$ 
& ${\cal X}_3^0 |{\cal M}^0_{m+3}|^2$  
& -- 
&  ${\cal X}_3^0 X_3^0 |{\cal M}^0_{m+2}|^2$ 
& -- 
& ${\cal S} X_3^0 |{\cal M}^0_{m+2}|^2$ \\
$\int_2 \ds_{\NNLO}^{\rS,2}$ 
& -- 
&  ${\cal X}_4^0 |{\cal M}^0_{m+2}|^2$ 
& -- 
&${\cal X}_3^0 {\cal X}_3^0 |{\cal M}^0_{m+2}|^2$ 
&  --   \\ \hline
\end{tabular}
}
\end{center}
\caption{Type of contribution to the double real subtraction term ${\rm{d}}\hat\sigma_{\NNLO}^{\rS}$, together with the integrated form of each term.   The unintegrated antenna and soft functions are denoted as $X_3^0$, $X_4^0$ and $S$ while their integrated forms are ${\cal X}_3^0$, ${\cal X}_4^0$ and ${\cal S}$ respectively.  ${\cal M}^0_{n}$ denotes an $n$-particle tree-level colour-ordered amplitude.}
\label{tab:S1breakdown}
\end{table}

For the evaluation of the NNLO corrections to heavy quark pair production in the $q\bar{q}$ channel, the configurations $(c)$ and $(e)$, which always occur together, are not needed and will not be discussed here either. Only $(S,a)$, $(S,b)$ and $(S,d)$ subtraction terms are needed to approximate the double real contributions of eq.(\ref{eq.doublereal}), such that the total subtraction term is given by
\beq\label{eq.Sqqbar}
\ds^{\rS}_{q\bar{q},\NNLO,N_c^2}=\ds^{\rS,a}_{q\bar{q},\NNLO,N_c^2}+\ds^{\rS,b}_{q\bar{q},\NNLO,N_c^2}+\ds^{\rS,d}_{q\bar{q},\NNLO,N_c^2}.
\eeq

The $(S,a)$ type subtraction term, denoted by $\ds_{q\bar{q},\NNLO,N_c^2}^{\rS,a}$, subtracts the single unresolved limits and it is built with products of a tree-level three-parton antennae and five-parton reduced matrix elements. It is given by, 
\beqa\label{eq.subtermqqbRRa}
&&\hspace{-0.2in}\ds_{q\bar{q},\NNLO,N_c^2}^{\rS,a}=\frac{1}{2}\,\norm_{\NNLO}^{q\bar{q},\RR}\,N_c^2\sum_{(i,j)\in P(5,6)}\dphi_4(p_3,p_4,p_5,p_6;p_1,p_2)\nonumber\\
&&\hspace{0.25in}\times\bigg[A_3^0(\Q{3},\gl{i},\qi{1})|\cm_5(\Q{(\wt{3i})},\qi{\bar{1}};;\qbi{2},\gl{j},\Qb{4})|^2 J_2^{(3)}(\wt{p_{3i}},p_4,p_j)\nonumber\\
&&\hspace{0.325in}+A_3^0(\Qb{4},\gl{j},\qbi{2})|\cm_5(\Q{3},\gl{i},\qi{1};;\qbi{\bar{2}},\Qb{(\wt{4j})})|^2 J_2^{(3)}(p_3,\wt{p_{4j}},p_i)\phantom{\bigg[}\nonumber\\
&&\hspace{0.325in}+d_3^0(\Q{3},\gl{i},\gl{j})|\cm_5(\Q{(\wt{3i})},\gl{(\wt{ij})},\qi{1};;\qbi{2},\Qb{4})|^2 J_2^{(3)}(\wt{p_{3i}},p_4,\wt{p_{ij}})\phantom{\bigg[}\nonumber\\
&&\hspace{0.325in}+d_3^0(\Qb{4},\gl{j},\gl{i})|\cm_5(\Q{3},\qi{1};;\qbi{2},\gl{(\wt{ji})},\Qb{(\wt{4j})})|^2 J_2^{(3)}(p_3,\wt{p_{4j}},\wt{p_{ji}})\phantom{\bigg[}\nonumber\\
&&\hspace{0.325in}+d_3^0(\qi{1},\gl{j},\gl{i})|\cm_5(\Q{3},\gl{(\wt{ji})},\qi{\bar{1}};;\qbi{2},\Qb{4})|^2 J_2^{(3)}(p_3,p_4,\wt{p_{ji}})\phantom{\bigg[}\nonumber\\
&&\hspace{0.325in}+d_3^0(\qbi{2},\gl{i},\gl{j})|\cm_5(\Q{3},\qi{1};;\qbi{\bar{2}},\gl{(\wt{ij})},\Qb{4})|^2 J_2^{(3)}(p_3,p_4,\wt{p_{ij}})\bigg].
\eeqa
All three-parton antennae present in this subtraction term have been derived in unintegrated and in integrated form in \cite{Abelof:2011jv,Daleo:2006xa,GehrmannDeRidder:2009fz}. Furthermore, as can be seen from Table \ref{tab:S1breakdown}, the integrated form of $\ds_{q\bar{q},\NNLO,N_c^2}^{\rS,a}$ must be added back at the three-parton level, and it will therefore contribute to the real-virtual counter term $\ds_{q\bar{q},\NNLO,N_c^2}^{\rT}$, which will be presented in section \ref{sec.RVsub}.

The $(S,b)$ type subtraction term, denoted by $\ds_{q\bar{q},\NNLO,N_c^2}^{\rS,b}$, takes care of the double unresolved limits of the double real contributions in those subamplitudes in which both final state gluons are colour-connected. It is given by,
\beqa\label{eq.subtermqqbRRb}
&&\hspace{-0.3in}\ds_{q\bar{q},\NNLO,N_c^2}^{\rS,b}=\frac{1}{2}\,\norm_{\NNLO}^{q\bar{q},\RR}\,N_c^2\sum_{(i,j)\in P(5,6)}\dphi_4(p_3,p_4,p_5,p_6;p_1,p_2)\nonumber\\
&&\hspace{-0.145in}\times\bigg[\bigg( A_4^0(\Q{3},\gl{i},\gl{j},\qi{1})-d_3^0(\Q{3},\gl{i},\gl{j})A_3^0(\Q{(\wt{3i})},\gl{(\wt{ij})},\qi{1})\nonumber\\
&&\hspace{0.4in}-d_3^0(\qi{1},\gl{j},\gl{i})A_3^0(\Q{3},\gl{(\wt{ji})},\qi{\bar{1}})\bigg) |\cm_4(\Q{(\wt{3ij})},\Qb{4},\qbi{2},\qi{\bar{1}})|^2 J_2^{(2)}(\wt{p_{3ij}},p_4)\nonumber\\
&&\hspace{-0.075in}+\bigg( A_4^0(\Qb{4},\gl{j},\gl{i},\qbi{2})-d_3^0(\Qb{4},\gl{j},\gl{i})A_3^0(\Qb{(\wt{4j})},\gl{(\wt{ji})},\qbi{2})\nonumber\\
&&\hspace{0.4in}-d_3^0(\qbi{2},\gl{i},\gl{j})A_3^0(\Qb{4},\gl{(\wt{ij})},\qbi{\bar{2}})\bigg) |\cm_4(\Q{3},\Qb{(\wt{4ij})},\qbi{\bar{2}},\qi{1})|^2 J_2^{(2)}(p_3,\wt{p_{4ij}})\bigg].
\eeqa
Two different kinds of structures are involved in this subtraction term: $X_4^0\times |\cm_4|^2$ and $X_3^0\times X_3^0\times |\cm_4|^2$. The former subtracts the double unresolved limits while introducing spurious single unresolved singularities, whereas the latter removes these spurious limits ensuring that the four-parton antenna is only active in the double unresolved regions. The four-parton antenna $A_4^0(\Q{3},\gl{i},\gl{j},\qi{1})$ present in this equation appears in a subtraction term for the first time. It  was discussed in section \ref{sec.ant4} together with its infrared limits and its explicit form can be found in appendix \ref{sec.a04}.

As shown in Table \ref{tab:S1breakdown}, the brackets in $\ds_{q\bar{q},\NNLO,N_c^2}^{\rS,b}$ should be expanded in order to combine its integrated form with the three and two-parton contributions. The pieces involving products of three-parton antennae, which we denote as $\ds_{q\bar{q},\NNLO,N_c^2}^{\rS,b\,3\times 3}$ should be included in the three-parton contributions $\ds_{q\bar{q},\NNLO,N_c^2}^{\rT}$ with only the ``outer'' antenna integrated, while the terms involving a four-parton antenna, $\ds_{q\bar{q},\NNLO,N_c^2}^{\rS,b\,4}$, are naturally added in integrated form to the two-parton counter term $\ds_{q\bar{q},\NNLO,N_c^2}^{\rU}$. The integration of $\ds_{q\bar{q},\NNLO,N_c^2}^{\rS,b\,4}$ will require analogous methods as developed in \cite{Abelof:2012he}  and will be addressed  elsewhere.

Finally, the subtraction term of type $(S,d)$, denoted by $\ds_{q\bar{q},\NNLO,N_c^2}^{\rS,d}$, is built out of products of two three-parton antennae and four-parton reduced matrix elements squared. Its role in the partonic process that we are presently considering is to ensure the correct subtraction of the initial-final double collinear limits of the double real contributions given in eq.(\ref{eq.doublereal}). It is given by
\beqa\label{eq.subtermqqbRRd}
&&\hspace{-0.2in}\ds_{q\bar{q},\NNLO,N_c^2}^{\rS,d}=-\frac{1}{2}\,\norm_{\NNLO}^{q\bar{q},\RR}\,N_c^2\sum_{(i,j)\in P(5,6)}\dphi_4(p_3,p_4,p_5,p_6;p_1,p_2)\nonumber\\
&&\hspace{0.125in}\times A_3^0(\Q{3},\gl{i},\qi{1})A_3^0(\Qb{4},\gl{j},\qbi{2})|\cm_4(\Q{(\wt{3i})},\Qb{(\wt{4j})},\qbi{\bar{2}},\qi{\bar{1}})|^2 J_2^{(2)}(\wt{p_{3i}},\wt{p_{4j}}).
\eeqa
This subtraction term will be added back to the two-parton counter term $\ds_{q\bar{q},\NNLO,N_c^2}^{\rU}$ with both three-parton antennae integrated over their corresponding antenna phase space. We shall not discuss this integration in this paper.

In section \ref{sec.tests} we will present a series of numerical tests that show that the subtraction term $\ds_{q\bar{q},\NNLO,N_c^2}^{\rS}$ of eq.(\ref{eq.Sqqbar}) correctly approximates the double real contributions $\ds_{q\bar{q},\NNLO,N_c^2}^{\RR} $ in all its single and double unresolved limits.


\section{General structure of the real-virtual contributions to $q\bar{q} \to t\bar{t}$ at leading-colour}\label{sec.structureRV}
The real-virtual contributions to top-antitop production in the
quark-antiquark channel are obtained using the interference of the one-loop
and tree-level amplitudes for the partonic process $ q\bar{q} \to t \bar{t}
g$.  The colour decomposition of the  matrix element reads,
\beqa\label{eq.colourdecqqbRV}
&&\hspace{-0.1in}M_5^1(q_1\bar{q}_2\rightarrow Q_3\bar{Q}_4g_5)= \nonumber\\
&&\hspace{-0.05in}\sqrt{2}\,g_s^6\,\cep\,\bigg\{\bigg[ (T^{a_5})_{i_3i_1}\delta_{i_2i_4}\cmb_5^1(\Q{3},\gl{5},\qi{1};;\qbi{2},\Qb{4})+(T^{a_5})_{i_2i_4}\delta_{i_3i_1}\cmb_5^1(\Q{3},\qi{1};;\qbi{2},\gl{5},\Qb{4})\bigg]\nonumber\\
&&\hspace{0.45in}-\frac{1}{N_c}\bigg[(T^{a_5})_{i_3i_4}\delta_{i_2i_1}\cmb_5^1(\Q{3},\gl{5},\Qb{4};;\qbi{2},\qi{1})+(T^{a_5})_{i_2i_1}\delta_{i_3i_4}\cmb_5^1(\Q{3},\Qb{4};;\qbi{2},\gl{5},\qi{1})\bigg]\bigg\},\nonumber\\
\eeqa
where each of the sub-amplitudes has the following decomposition into primitives
\beq
\cmb_5^1(...)=N_c\cmb_5^{[lc]}(...)+N_l\cmb_5^{[l]}(...)+ N_h\cmb_5^{[h]}(...)-\frac{1}{N_c}\cmb_5^{[slc]}(\dots).
\eeq
Interfering the  matrix element in eq.(\ref{eq.colourdecqqbRV}) with the tree-level amplitude in eq.(\ref{eq.colourdecqqbR}), combining the result with the phase space and the jet function, and retaining only the terms proportional to $N_c^2$, we obtain 
\beqa\label{eq.realvirtual}
&&\hspace{-0.3in}\ds_{q\bar{q},\NNLO,N_c^2}^{\RV}=\norm_{\NNLO}^{\RV,\,q\bar{q}}\,N_c^2\int\frac{{\rm d}x_1}{x_1}\frac{{\rm d}x_2}{x_2}\,\dphi_3(p_3,p_4,p_5;x_1 p_1,x_2 p_2)\delta(1-x_1)\delta(1-x_2)\nonumber\\
&&\times\Big( \big|\cmb_5^{[lc]}(\Q{3},\gl{5},\qi{1};;\Qb{4},\qbi{2})\big|_{\NLO}^2+\big|\cmb_5^{[lc]}(\Q{3},\qi{1};;\qbi{2},\gl{5},\Qb{4})\big|_{\NLO}^2\bigg)J_2^{(3)}(p_3,p_4,p_5),
\nonumber\\
\eeqa
where the trivial dependence on $x_1$ and $x_2$ is introduced for later convenience. The overall factor $\norm_{\NNLO}^{\RV,\,q\bar{q}}$ is 
\beq
\norm_{NNLO}^{q\bar{q},RV}=\norm_{LO}^{q\bar{q}}\frac{\cepb^2}{\cep}=\norm_{NNLO}^{q\bar{q},RR}\:\cep.
\eeq
The  leading-colour primitive amplitudes in eq.(\ref{eq.realvirtual}) contain ultraviolet poles that must be removed by renormalisation. Following the scheme of \cite{Bonciani:2008az,Bonciani:2009nb}, which was described in section \ref{sec.virtualnlo}, we renormalise the primitive amplitudes as
\beqa
\label{eq:nloUVcts}
&&\hspace{-0.2in}\cmb_{5,ren}^{[lc]}(\ldots)=\cmb_{5,b}^{[lc]}(\ldots)-\frac{3}{2}\cepb\frac{b_0}{\e}\cm_5(\ldots)\nonumber\\
&&\hspace{0.2in}-\frac{1}{2}(4\pi)^\e\Gamma(1+\e)\bigg(\frac{m_Q}{\mu} \bigg)^{-2\e}\bigg(\frac{3}{2\e}+\frac{2}{1-2\e}\bigg)\Big(\cm_5(\ldots)-\cm_{5,1M}(\ldots)\Big),
\eeqa
where $b_0=11/6$ and $\cm_{5,1M}(\ldots)$ denotes the tree-level amplitude with a mass insertion in the heavy fermion propagators.

After UV renormalisation, the real-virtual contributions $\ds_{q\bar{q},\NNLO,N_c^2}^{\RV}$ still contain infrared divergences of implicit and explicit types. The explicit ones originate from the loop integration of the partial amplitudes $\big|\cmb_5^{[lc]}(...)\big|_{\NLO}^2$ and can be written as the following combination of colour-ordered infrared singularity operators
\beqa\label{eq.polesvirtuallc}
&&\hspace{-0.1in}{\cal P}oles\left( \ds_{q\bar{q},\NNLO,N_c^2}^{\RV} \right)=\nonumber\\
&&\norm_{\NNLO}^{\RV,\,q\bar{q}}\,N_c^2\int\frac{{\rm d}x_1}{x_1}\frac{{\rm d}x_2}{x_2}\,\dphi_3(p_3,p_4,p_5;x_1 p_1,x_2 p_2)\delta(1-x_1)\delta(1-x_2)\nonumber\\
&&\hspace{0.1in}\times \bigg[\bigg( 2\ione{Q}{g}(\e,s_{35}) + 2\ione{q}{g}(\e,s_{15}) + 2\ione{Q}{\bar{q}}(\e,s_{24}) \bigg)|\cm_5(\Q{3},\gl{5},\qi{1};;\qbi{2},\Qb{4})|^2\nonumber\\
&&\hspace{0.175in}+\bigg( 2\ione{Q}{g}(\e,s_{45}) + 2\ione{q}{g}(\e,s_{25}) + 2\ione{Q}{\bar{q}}(\e,s_{13}) \bigg)|\cm_5(\Q{3},\qi{1};;\qbi{2},\gl{5},\Qb{4})|^2\bigg] J_2^{(3)}(p_3,p_4,p_5).\nonumber\\
\eeqa
As we shall see in section \ref{sec.RVsub}, these poles will be canceled by the singly integrated double real subtraction terms and mass factorisation counter terms.

The implicit infrared poles, on the other hand, originate from the configurations where the final state gluon becomes soft or collinear to either of the incoming particles. Those will be dealt  with the genuine real-virtual subtraction term $\ds^{VS}_{q\bar{q},\NNLO,N_c^2}$ which will also be constructed in section \ref{sec.RVsub}.


\section{Real-virtual contributions to top-antitop production in the quark-antiquark channel with \OpenLoops}\label{sec.openloops}
For the calculation of the  matrix elements that enter the real-virtual contributions in eq.(\ref{eq.realvirtual}) we employ \OpenLoops~\cite{Cascioli:2011va}, a fully automated generator of one-loop
 corrections to Standard Model processes. As discussed in the following, \OpenLoops builds 
Feynman diagrams with a recursive algorithm that allows for a 
fast and numerically stable evaluation of loop amplitudes.
The reduction of  amplitudes to scalar integrals 
can be achieved by interfacing \OpenLoops 
to tensor-integral~\cite{Denner:2002ii,Denner:2005nn} or OPP reduction 
libraries~\cite{Ossola:2006us,Ossola:2007ax,Mastrolia:2010nb}.

In the context of NNLO calculations,
the integration of (subtracted)  contributions over soft and collinear 
regions poses non trivial technical challenges as compared to 
conventional NLO applications. In particular,
the loss of precision resulting from the cancellation
between amplitudes and subtraction terms
in the soft and collinear regions 
needs to be compensated by sufficiently high numerical accuracy.
However, this is quite challenging since infrared singularities
tend to amplify numerical instabilities 
that arise from spurious singularities (like inverse Gram determinants)
in the  reduction algorithms.
It is thus quite interesting to investigate to which extend automated 
 generators can guarantee an adequate level of numerical stability
for NNLO calculations. 
In this respect \OpenLoops has already been shown to be successfully applicable
to the calculation of the NNLO corrections to $pp\to Z\gamma$ \cite{Grazzini:2013bna}. 
In this case,  using the $q_T$-subtraction technique~\cite{Catani:2007vq},
it was found that the tensor-reduction library \Collier~\cite{Denner:2014gla}, 
which implements the methods of~\cite{Denner:2002ii,Denner:2005nn,Denner:2010tr},
is sufficiently stable to perform the entire calculation in double precision.
Very recently, \OpenLoops was also applied to 
$t\bar t$ production in association with up to two jets at NLO~\cite{Hoeche:2014qda},
which is closely related to the present NNLO calculation.

In this work, \OpenLoops is used to evaluate the  amplitudes for 
$q\bar q\to t\bar t g$. The interference with the related Born amplitudes,
the sums over external colours and helicity states, as well as the ultraviolet 
renormalisation (\ref{eq:nloUVcts}) are performed in a fully automated way.
The UV-finite but still IR-divergent  result is returned in the form
of a  Laurent series,
\begin{equation}
\left|\cal{M}\right|^2_{\NLO}=\frac{(4\pi)^\epsilon}{\Gamma(1-\epsilon)}\sum_{k=-2}^0
{\cal A}_k \epsilon^k,
\end{equation}
which  must be combined with the corresponding subtraction terms. 
For consistency with the helicity amplitudes implemented in 
\OpenLoops, the tree matrix elements in eq.(\ref{eq.polesvirtuallc})
need to be evaluated in $D=4$ dimensions. 

Tree amplitudes ($\calM^0$) and loop amplitudes ($\calM^1$) are 
expressed as sums over corresponding Feynman diagrams, 
\begin{equation}
\calM^k 
= 
\sum_{d} \calC^{(d)}\calA_k^{(d)},
\end{equation}
where the colour factors 
$\calC^{(d)}$ associated with individual diagrams are factorised, and 
the corresponding colour-stripped amplitudes are denoted as
$\calA_k^{(d)}$.
All colour structures are reduced to a standard basis 
$\{\calC_i\}$, and the colour information  needed 
to build colour-summed squared matrix elements is encoded in the
colour-interference matrix,
\begin{equation}
\label{eq:colint}
\calK_{ij}=
\sum_{\col}\calC_i^* \calC_j.
\end{equation}
These colour bookkeeping operations are done only once, using a generic and 
automated algebraic algorithm, during the generation of the numerical code for a 
particular process. This approach provides high flexibility in the colour 
treatment, and the leading-colour approximation used in this paper could be 
easily implemented via a $1/N_c$ expansion of the colour-interference matrix 
(\ref{eq:colint}).
Additionally, in order to obtain the leading colour contribution of the 
counter-term amplitude, the substitutions $C_F\to N_c$, $C_A\to 0$, $T_F\to 0$ are 
applied to colour factors which are attributed to renormalisation constants.

The calculation of colour-stripped loop amplitudes within \OpenLoops is based on the 
representation
\begin{eqnarray}
\label{eq:npointloop}
\calA_1^{(d)} 
=
\int
\frac{\rd^Dq\; \calN^{(d)} (q)}{D_0D_1\dots D_{n-1}},
\end{eqnarray}
where the denominators $D_i=(q+p_i)^2-m_i^2+\ri\varepsilon$
depend on the loop momentum $q$, external momenta $p_i$, and internal
masses $m_i$. The numerator $\calN^{(d)} (q)$ corresponds to a particular
diagram or to a set of diagrams with the same loop topology.
It is expressed as a polynomial of degree $R\le n$ 
in the loop momentum,
\begin{eqnarray}
\label{eq:Npoly}
\calN^{(d)}(q)&=& \sum_{r=0}^R \calN^{(d)}_{\mu_1\dots \mu_r}\; q^{\mu_1}\dots q^{\mu_r}.
\end{eqnarray}
In contrast to traditional approaches, where the above expressions are 
constructed via explicit insertion of the Feynman rules, 
the \OpenLoops method consists of a numerical recursion 
that builds the polynomial coefficients $\calN^{(d)}_{\mu_1\dots \mu_r}$
in a iterative way starting from related coefficients for lower-point topologies, 
i.e.~topologies with a lower number of loop propagators.
The recursion is formulated in $D=4$ dimensions, and rational terms 
resulting from $\mathcal{O}(D-4)$ contributions to the 
numerator are easily obtained in a process-independent way 
via so-called $R_2$ counter terms~\cite{Draggiotis:2009yb}.

For the reduction of  amplitudes to scalar integrals, the 
\OpenLoops representation (\ref{eq:npointloop})--(\ref{eq:Npoly}) allows one to 
use the tensor-integral or OPP reduction techniques. In the former case,
the reduction is performed at the level of process-independent tensor integrals,
\begin{eqnarray}
\label{eq:TI}
%
T_{n,r}^{\mu_1\dots \mu_r}=
\int
\frac{\rd^Dq\; q^{\mu_1}\dots q^{\mu_r}}{D_0D_1\dots D_{n-1}},
\end{eqnarray}
which are then combined with the corresponding coefficients. 
In this approach, the \Collier library implements 
systematic expansions in Gram determinants 
and other kinematic quantities~\cite{Denner:2005nn}, which 
avoid numerical instabilities due to 
spurious singularities.
In the OPP reduction framework, the reduction is performed at the level 
of the full integrand in eq.(\ref{eq:npointloop}). This requires 
multiple evaluations of the numerator function, and using
the representation (\ref{eq:Npoly}) in combination with the
\OpenLoops coefficients, $\calN^{(d)}_{\mu_1\dots \mu_r}$,
renders OPP reduction similarly fast as tensor reduction~\cite{Cascioli:2011va}.

In Section~\ref{sec.tests} we will 
investigate  the numerical stability of the  amplitudes 
in the soft and collinear regions using \OpenLoops in combination with 
the OPP reduction library \Cuttools~\cite{Ossola:2007ax}. In this context we will
exploit the quadruple precision mode of \Cuttools both as a rescue system for 
matrix elements that are not sufficiently stable in double precision, and 
for precision tests of the real-virtual cancellations in the deep infrared regime.


\section{Real-virtual subtraction terms}\label{sec.RVsub}
The purpose of the real-virtual counter term $\ds^{\rT}_{q\bar{q},\NNLO,N_c^2}$ is to cancel the explicit $\e$-poles of the real-virtual contributions $\ds^{\RV}_{q\bar{q},\NNLO,N_c^2}$ and to simultaneously subtract their infrared limits in such a way that the difference $\ds^{\RV}_{q\bar{q},\NNLO,N_c^2}-\ds^{\rT}_{q\bar{q},\NNLO,N_c^2}$ can be safely integrated numerically in four dimensions. The generic antenna content of this counter term has been derived for the massless case in \cite{GehrmannDeRidder:2011aa,GehrmannDeRidder:2012dg}, and it remains unchanged in the massive case. We will here follow the formalism developed in these references, to which the reader is referred for details. 

In general, real-virtual antenna counter terms contain singly integrated double real subtraction terms, NNLO mass factorisation counter terms and genuine real-virtual subtraction terms. For the leading-colour contributions to top pair production in the $q\bar{q}$ channel the counter term has the following structure
\beqa\label{eq.Tqqbar}
\lefteqn{\ds^{\rT}_{q\bar{q},\NNLO,N_c^2}=-\left( \int_1 \ds^{\rS,a}_{q\bar{q},\NNLO,N_c^2} + \ds^{\MF,1a}_{q\bar{q},\NNLO,N_c^2}\right)}\nonumber\\ 
&&+ \bigg[ \ds^{\VS,a}_{q\bar{q},\NNLO,N_c^2}+\ds^{\VS,b}_{q\bar{q},\NNLO,N_c^2}   +\ds^{\VS,d}_{q\bar{q},\NNLO,N_c^2}- \int_1 \ds^{\rS,b\:3\times 3}_{q\bar{q},\NNLO,N_c^2} -\ds^{\MF,1b}_{q\bar{q},\NNLO,N_c^2} \bigg].\nonumber\\
\eeqa
In the most general case, real-virtual subtraction terms contain yet another component, labelled $(\VS,c)$ \cite{GehrmannDeRidder:2011aa,GehrmannDeRidder:2012dg}, whose absence in this particular case is related to the absence of the subtraction terms labelled $(S,c)$ and $(S,e)$ at the double real level. Furthermore, in eq.(\ref{eq.Tqqbar}) we have splitted the mass factorisation counter term  $\ds^{\MF,1}_{q\bar{q},\NNLO,N_c^2}$ into two terms $\ds^{\MF,1a}_{q\bar{q},\NNLO,N_c^2}$ and $\ds^{\MF,1b}_{q\bar{q},\NNLO,N_c^2}$. In the following, we shall present all the pieces of $\ds^{\rT}_{q\bar{q},\NNLO,N_c^2}$, starting with the explicit expressions of these two mass factorisation counter terms.


\subsection{The mass factorisation counter term $\ds_{\NNLO}^{\MF,1}$}
For a given partonic process initiated by partons labelled $i$ and $j$ the mass factorisation counter term $\ds_{ij,\NNLO}^{\MF,1}$  is related to the NLO real
emission partonic cross sections $\ds_{kl,\NLO}^{\rR}$ and its corresponding antenna subtraction term $\ds_{kl,\NLO}^{\rS}$. It is given by
\beq\label{eq:MFNNLOone}
\ds_{ij,\NNLO}^{\MF,1}(p_1,p_2)= -\cepb
\sum_{k,l}\int\frac{{\rm d}x_1}{x_1} \frac{{\rm d}x_2}{x_2}\, \GG^{(1)}_{ij;kl}(x_1,x_2)\,\bigg[\ds_{kl,\NLO}^{\rR}-\ds_{kl,\NLO}^{\rS}\bigg] (x_1p_1,x_2p_2), 
\eeq  
with the  kernel $\GG^{(1)}_{ij;kl}(x_1,x_2)$ defined in eq.(\ref{eq.oneloopkernelgeneral}). It is useful to further decompose this mass factorisation counter term as follows:
\beq
\ds_{ij,\NNLO}^{\MF,1}=\ds_{ij,\NNLO}^{\MF,1a}+\ds_{ij,\NNLO}^{\MF,1b}
\eeq
with 
\beqa
&&\hspace{-0.2in}\ds_{ij,\NNLO}^{\MF,1a}(p_1,p_2)=-\cepb
\;\sum_{k,l}\int\frac{{\rm d}x_1}{x_1}\frac{{\rm d}x_2}{x_2}{\bf \Gamma}^{(1)}_{ij,kl}(x_1,x_2) \ds_{kl,\NLO}^{\rR}(x_1p_1,x_2p_2)\label{eq.mfnnlo1a},\\ \nonumber\\
&&\hspace{-0.2in}\ds_{ij,\NNLO}^{\MF,1b}(p_1,p_2)=+\cepb
\;\sum_{k,l}\int\frac{{\rm d}x_1}{x_1}\frac{{\rm d}x_2}{x_2}{\bf \Gamma}^{(1)}_{ij,kl}(x_1,x_2) \ds_{kl,\NLO}^{\rS}(x_1p_1,x_2p_2),\label{eq.mfnnlo1b}.
\eeqa

In the context of this paper,  the mass factorisation counter term denoted by  $\ds^{\MF,1a}_{q\bar{q},\NNLO,N_c^2}$ is constructed as in eq.(\ref{eq.mfnnlo1a}) with $\ds^{\rR}_{q\bar{q},\NLO}$ given in eq.(\ref{eq.qqbR}). Retaining only the terms with an overall $N_c^2$ we have
\beqa\label{eq.polesmf1alc}
&&\hspace{-0.15in} \ds^{\MF,1a}_{q\bar{q},\NNLO,N_c^2}=-\norm_{\NNLO}^{\RV,\,q\bar{q}}\,N_c^2\int \frac{{\rm d}x_1}{x_1} \frac{{\rm d}x_2}{x_2}\,\dphi_3(p_3,p_4,p_5;x_1p_1,x_2p_2)\nonumber\\
&&\times\bigg\{\frac{1}{2}\bigg(\Gamma^{(1)}_{qq}(x_1)\delta(1-x_2)+\Gamma^{(1)}_{qq}(x_2)\delta(1-x_1)\bigg)|\cm_5(\Q{3},\gl{5},\qi{\bar{1}};;\qbi{\bar{2}},\Qb{4})|^2\nonumber\\
&&\hspace{0.11in}+\frac{1}{2}\bigg(\Gamma^{(1)}_{qq}(x_1)\delta(1-x_2)+\Gamma^{(1)}_{qq}(x_2)\delta(1-x_1)\bigg)|\cm_5(\Q{3},\qi{\bar{1}};;\qbi{\bar{2}},\gl{5},\Qb{4})|^2\bigg\}J_2^{(3)}(p_3,p_4,p_5).\nonumber\\
\eeqa
We note that  this contribution contains five-parton matrix elements. It can therefore  develop spurious single unresolved limits which have to be compensated for by other terms in $\ds^{\rT}_{q\bar{q},\NNLO,N_c^2}$ as we shall see below.
 
Furthermore, the mass factorisation counter term $\ds^{\MF,1b}_{q\bar{q},\NNLO,N_c^2}$ is constructed as in eq.(\ref{eq.mfnnlo1b}) with the NLO subtraction term $\ds^{\rS}_{q\bar{q},\NLO}$ given in eq.(\ref{eq.qqbS}). Retaining only the terms with an overall $N_c^2$ colour we have
\beqa\label{eq.mfqqb1b}
&&\hspace{-0.1in}\ds^{\MF,1b}_{q\bar{q},\NNLO,N_c^2}=\frac{1}{2}\norm_{\NNLO}^{q\bar{q},\RV}N_c^2\int \frac{{\rm d}x_1}{x_1}\frac{{\rm d}x_2}{x_2}\dphi_3(p_3,p_4,p_5;x_1p_1,x_2p_2)\nonumber\\
&&\hspace{0.4in}\times\bigg(\Gamma^{(1)}_{qq}(x_1)\delta(1-x_2)+\Gamma^{(1)}_{qq}(x_2)\delta(1-x_1) \bigg)\nonumber\\
&&\hspace{0.9in}\times \bigg[A_3^0(\Q{3},\gl{5},\qi{\bar{1}})|\cm_4(\Q{(\wt{35})},\Qb{4},\qbi{\bar{2}},\qi{\bar{\bar{1}}})|^2 J_2^{(2)}(\wt{p_{35}},p_4)\nonumber\\
&&\hspace{0.975in}+A_3^0(\Qb{4},\gl{5},\qi{\bar{2}})|\cm_4(\Q{3},\Qb{(\wt{45})},\qbi{\bar{\bar{2}}},\qi{\bar{1}})|^2 J_2^{(2)}(p_3,\wt{p_{45}})\bigg]. \\ \nonumber
\eeqa


\subsection{Cancellation of explicit infrared poles in $\ds_{\NNLO,q\bar{q},N_c^2}^{\RV}$}
We continue with the construction of $ \ds^{\rT}_{q\bar{q},\NNLO,N_c^2}$ by showing that the explicit infrared poles present in the real-virtual contributions and given in eq.(\ref{eq.polesvirtuallc}) are cancelled as
\beq\label{eq.poles1} 
\poles \left( \ds_{\NNLO,q\bar{q},N_c^2}^{\RV}+ \int_1\ds^{\rS,a}_{q\bar{q},\NNLO,N_c^2} +\ds^{\MF,1a}_{q\bar{q},\NNLO,N_c^2} \right)=0.
\eeq

The integrated subtraction term $\int_1 \ds^{\rS,a}_{q\bar{q},\NNLO,N_c^2}$ can be obtained from eq.(\ref{eq.subtermqqbRRa}) by integrating each of the three-parton antenna functions over the appropriate antenna phase spaces. It reads
\beqa\label{eq.subintalc}
&&\hspace{-0.5in}\int_1 \ds^{\rS,a}_{q\bar{q},\NNLO,N_c^2}=\norm_{\NNLO}^{\RV,\,q\bar{q}}\,N_c^2 \int\frac{{\rm d}x_1}{x_1}\frac{{\rm d}x_2}{x_2}\,\dphi_3(p_3,p_4,p_5;x_1p_1,x_2p_2)\nonumber\\
&&\hspace{-0.3in}\times \bigg[ \bigg(\frac{1}{2}{\cal D}^0_{Qgg}(\e,s_{35},x_1,x_2)+ \frac{1}{2}{\cal D}^0_{q,gg}(\e,s_{\bar{1}5},x_1,x_2)\nonumber\\
&&\hspace{0.75in} +{\cal A}^0_{q,Qg}(\e,s_{\bar{2}4},x_2,x_1)\bigg) |\cm_5(\Q{3},\gl{5},\qi{\bar{1}};;\qbi{\bar{2}},\Qb{4})|^2\nonumber\\
&&\hspace{-0.225in} +\bigg(\frac{1}{2}{\cal D}^0_{Qgg}(\e,s_{45},x_1,x_2) + \frac{1}{2}{\cal D}^0_{q,gg}(\e,s_{\bar{2}5},x_2,x_1)\nonumber\\
&&\hspace{0.75in} +{\cal A}^0_{q,Qg}(\e,s_{\bar{1}3},x_1,x_2)\bigg) |\cm_5(\Q{3},\qi{\bar{1}};;\qbi{\bar{2}},\gl{5},\Qb{4})|^2\bigg]J_2^{(3)}(p_3,p_4,p_5).
\eeqa
The integrated antennae in the equation above have been derived in \cite{Abelof:2011jv}. Only their pole parts will be needed in the context of this paper. The poles of the flavour violating antenna ${\cal A}^0_{q,Qg}$ were given in eq.(\ref{eq.A03flint}), and those of the D-type antennae are given by
\beqa
&&\hspace{-0.3in}\poles\left({\cal D}^0_{Qgg}(\e,s_{ij},x_1,x_2)\right)=-4\ione{Q}{g}(\e,s_{ij})\delta(1-x_1)\delta(1-x_2)\label{eq.polesd1}\\
&&\hspace{-0.3in}\poles\left({\cal D}^0_{q,gg}(\e,s_{ij},x_1,x_2)\right)=-4\ione{q}{g}(\e,s_{ij})\delta(1-x_1)\delta(1-x_2)+\Gamma^{(1)}_{qq}(x_1)\delta(1-x_2).\label{eq.polesd2}
\eeqa
The pole part of the singly integrated real subtraction term denoted as $(S,a)$ is therefore given by
\beqa\label{eq.polesintsubtermlc}
&&\hspace{-0.1in}{\cal P}oles\left( \int_1 \ds^{\rS,a}_{q\bar{q},\NNLO,N_c^2}\right) =\norm_{\NNLO}^{\RV,\,q\bar{q}}\,N_c^2\int \frac{{\rm d}x_1}{x_1}\frac{{\rm d}x_2}{x_2}\,\dphi_3(p_3,p_4,p_5;x_1p_1,x_2p_2)\nonumber\\
&&\hspace{-0.1in}\times \bigg\{\bigg[ - \delta(1-x_1)\delta(1-x_2)\bigg( 2\ione{Q}{g}(\e,s_{35})+ 2\ione{q}{g}(\e,s_{15})+2\ione{Q}{\bar{q}}(\e,s_{24})\bigg)\nonumber\\
&&\hspace{0.175in}+\frac{1}{2}\bigg(\Gamma^{(1)}_{qq}(x_1)\delta(1-x_2)+\Gamma^{(1)}_{qq}(x_2)\delta(1-x_1)\bigg)\bigg]|\cm_5(\Q{3},\gl{5},\qi{\bar{1}};;\qbi{\bar{2}},\Qb{4})|^2\nonumber\\
&&\hspace{0.025in}+\bigg[ - \delta(1-x_1)\delta(1-x_2)\bigg( 2\ione{Q}{g}(\e,s_{45})+ 2\ione{q}{g}(\e,s_{25})+ 2\ione{Q}{\bar{q}}(\e,s_{13})\bigg)\nonumber\\
&&\hspace{0.175in}+\frac{1}{2}\bigg(\Gamma^{(1)}_{qq}(x_1)\delta(1-x_2)+\Gamma^{(1)}_{qq}(x_2)\delta(1-x_1)\bigg)\bigg]|\cm_5(\Q{3},\qi{\bar{1}};;\qbi{\bar{2}},\gl{5},\Qb{4})|^2\bigg\} J_2^{(3)}(p_3,p_4,p_5).\nonumber\\
\eeqa
The measurement function $J_2^{(3)}$ in eq.(\ref{eq.polesintsubtermlc}) allows the final state gluon in the reduced five-particle matrix elements squared to become unresolved. The corresponding singular limits of this subtraction term are spurious, since they do not correspond to any physical limits of the real-virtual contribution $\ds^{\RV}_{q\bar{q},\NNLO,N_c^2}$. Those must therefore be cancelled by other terms in $\ds^{\rT}_{q\bar{q},\NNLO,N_c^2}$. We shall shortly see below that this is indeed the case. 

Combining eqs.(\ref{eq.polesvirtuallc}), (\ref{eq.polesintsubtermlc}) and (\ref{eq.polesmf1alc}) it can easily seen that eq.(\ref{eq.poles1}) holds.


\subsection{Construction of $\ds^{\VS}_{q\bar{q},\NNLO,N_c^2}$}
The real-virtual subtraction term $\ds^{\VS}_{q\bar{q},\NNLO,N_c^2}$ has three components:
\beq
\ds_{q\bar{q},\NNLO,N_c^2}^{\VS}=\ds_{q\bar{q},\NNLO,N_c^2}^{\VS,a}+\ds_{q\bar{q},\NNLO,N_c^2}^{\VS,b}+\ds_{q\bar{q},\NNLO,N_c^2}^{\VS,d}.
\eeq
The $(\VS,a)$ piece subtracts the single unresolved limits of the real-virtual contributions, the $(\VS,d)$ subtraction term corrects for the different renormalisation scales in the  matrix elements and in the  antennae, while the $(\VS,b)$ part has the twofold purpose of removing the spurious unresolved limits of $\int_1 \ds^{\rS,a}_{q\bar{q},\NNLO,N_c^2}$ and achieving at the same time the following explicit pole cancellation:
\beq\label{eq.conditiondsvsb}
{\cal P}oles\left( \ds_{q\bar{q},\NNLO,N_c^2}^{\VS,a}+\ds_{q\bar{q},\NNLO,N_c^2}^{\VS,b}+\int_1 \ds^{\rS,b\:3\times 3}_{q\bar{q},\NNLO,N_c^2}-\ds^{\MF,1b}_{q\bar{q},\NNLO,N_c^2}\right)=0.
\eeq
We shall present these three subtraction contributions separately below.


\subsubsection{Construction of $\ds^{\VS,a}_{q\bar{q},\NNLO,N_c^2}$}
Following the general framework described in \cite{GehrmannDeRidder:2011aa}, in order to subtract the single unresolved limits of the real-virtual contributions given in eq.(\ref{eq.realvirtual}) we construct our subtraction terms of the type $(\VS,a)$ with  one-loop antennae multiplied by reduced tree-level matrix elements and  one-loop matrix elements multiplied by tree-level antennae. They read,   
\beqa\label{eq.subvsa}
&&\hspace{-0.35in}\ds_{q\bar{q},\NNLO,N_c^2}^{\VS,a}=\hspace{-0.02in}\norm_{\NNLO}^{\RV,\,q\bar{q}}\,N_c^2\int\frac{{\rm d}x_1}{x_1}\frac{{\rm d}x_2}{x_2}\,\dphi_3(p_3,p_4,p_5; x_1p_1,x_2p_2)\delta(1-x_1)\delta(1-x_2)\nonumber\\
&&\hspace{1.1in}\times\bigg\{A_3^0(\Q{3},\gl{5},\qi{\bar{1}})\big|\cmb_{4,1}^{[lc]}(\Q{(\wt{35})},\Qb{4},\qbi{\bar{2}},\qi{\bar{\bar{1}}})\big|_{\NLO}^2
J_2^{(2)}(p_{\wt{35}},p_4)\nonumber\\
&&\hspace{1.2in}+ A_3^{1,lc}(\Q{3},\gl{5},\qi{\bar{1}})|\cm_4(\Q{(\wt{35})},\Qb{4},\qbi{\bar{2}},\qi{\bar{\bar{1}}})|^2 J_2^{(2)}(p_{\wt{35}},p_4)\phantom{\bigg[}\nonumber\\
&&\hspace{1.2in} + A_3^0(\Qb{4},\gl{5},\qbi{\bar{2}})\big|\cmb_{4,1}^{[lc]}(\Q{3},\Qb{(\wt{45})},\qbi{\bar{\bar{2}}},\qi{\bar{1}})\big|_{\NLO}^2 J_2^{(2)}(p_3,p_{\wt{45}})\phantom{\bigg[}\nonumber\\
&&\hspace{1.2in} + A_3^{1,lc}(\Qb{4},\gl{5},\qbi{\bar{2}})|\cm_4(\Q{3},\Qb{(\wt{45})},\qbi{\bar{\bar{2}}},\qi{\bar{1}})|^2 J_2^{(2)}(p_3,p_{\wt{45}})\bigg\}.
\eeqa
The three-parton  antenna $A_3^{1,lc}$ appears here in a subtraction term for the first time. This antenna has been presented together with its singular limits in section \ref{sec:A31loop}. Its integration over the antenna phase space will as for $\ds^{S,b,4}$ require application  of the methods presented in
\cite{Abelof:2012he}.


\subsubsection{Construction of $\ds^{\VS,b}_{q\bar{q},\NNLO,N_c^2}$}
In order to construct our $(\VS,b)$-type subtraction terms in such a way that the pole cancellation of eq.(\ref{eq.conditiondsvsb}) holds, we have to examine the pole parts of $\ds_{\bar{q},\NNLO,N_c^2}^{\VS,a}$, $\int_1\ds^{\rS,b\:3\times 3}_{q\bar{q},\NNLO,N_c^2}$ and $\ds^{\MF,1b}_{q\bar{q},\NNLO,N_c^2}$ with the latter expression given before in eq.(\ref{eq.mfqqb1b}). The poles of \linebreak $\ds_{q\bar{q},\NNLO,N_c^2}^{\VS,a}$ are simply obtained using the expressions of the pole part of the  four-parton matrix element and of the  antenna in eq.(\ref{eq.subvsa}). The explicit infrared poles of the one loop antenna have been given in eq.(\ref{eq.polesA13iflc}), and those of the matrix elements are given by: 
\beqa
&&\hspace{-0.1in}{\cal P}oles\left( \big|\cmb_{4,1}^{[lc]}(\Q{3},\Qb{4},\qbi{2},\qi{1})\big|_{\NLO}^2\right)=\nonumber\\
&&\hspace{0.75in} 2\bigg( \ione{Q}{\bar{q}}(\e,s_{13},m_Q^2)+ \ione{Q}{\bar{q}}(\e,s_{24},m_Q^2)\bigg)|\cm_4(\Q{3},\Qb{4},\qbi{2},\qi{1})|^2.
\eeqa
Relabelling the final state momenta, we find
\beqa\label{eq.polesdsvsa}
&&\hspace{-0.1in}\poles \left( \ds_{q\bar{q},\NNLO,N_c^2}^{\VS,a} \right)=\norm_{\NNLO}^{q\bar{q},\RV}N_c^2\int\frac{{\rm d}x_1}{x_1}\frac{{\rm d}x_2}{x_2} \dphi_3(p_3,p_4,p_5;p_1,p_2)\nonumber\\
&&\hspace{0.3in}\times \bigg\{ 2\bigg[ \ione{Q}{\bar{q}}(\e,s_{24})+\ione{Q}{g}(\e,s_{35})+\ione{q}{g}(\e,s_{15})\bigg]\nonumber\\
&&\hspace{1.3in}\times A_3^0(\Q{3},\gl{5},\qi{1})|\cm_4(\Q{(\wt{35})},\Qb{4},\qbi{2},\qi{\bar{1}})|^2 J_2^{(2)}(\wt{p_{35}},p_4)\nonumber\\
&&\hspace{0.4in}+2\bigg[ \ione{Q}{\bar{q}}(\e,s_{13})+\ione{Q}{g}(\e,s_{45})+\ione{q}{g}(\e,s_{25})\bigg]\nonumber\\
&&\hspace{1.3in}\times A_3^0(\Qb{4},\gl{5},\qbi{2})|\cm_4(\Q{3},\Qb{(\wt{45})},\qbi{\bar{2}},\qi{1})|^2 J_2^{(2)}(p_3,\wt{p_{45}})\bigg\}.
\eeqa

The singly integrated subtraction term, $\int_1 \ds^{\rS,b\:3\times 3}_{q\bar{q},\NNLO,N_c^2}$, on the other hand, is obtained by integrating the ``outer'' antennae in eq.(\ref{eq.subtermqqbRRb}) over the corresponding three-parton antenna phase space. We find
\beqa
&&\hspace{-0.75in}\int_1 \ds^{\rS,b\:3\times 3}_{q\bar{q},\NNLO,N_c^2}=-\norm_{\NNLO}^{\RV,\,q\bar{q}}\,N_c^2\int \frac{{\rm d}x_1}{x_1}\frac{{\rm d}x_2}{x_2}\,\dphi_3(p_3,p_4,p_5;x_1p_1,x_2p_2)\nonumber\\
&&\hspace{-0.3in}\times\bigg\{ \bigg( \frac{1}{2}{\cal D}^0_{Qgg}(\e,s_{35},x_1,x_2)+  \frac{1}{2}{\cal D}^0_{q,gg}(\e,s_{\bar{1}5},x_1,x_2)\bigg)\nonumber\\
&&\hspace{0.75in}\times A_3^0(\Q{3},\gl{5},\qi{\bar{1}})|\cm_4(\Q{(\wt{35})},\Qb{4},\qbi{\bar{2}},\qi{\bar{\bar{1}}})|^2 J_2^{(2)}(p_{\wt{35}},p_4)\nonumber\\
&&\hspace{-0.2in}+\bigg( \frac{1}{2}{\cal D}^0_{Qgg}(\e,s_{45},x_1,x_2)+  \frac{1}{2}{\cal D}^0_{q,gg}(\e,s_{\bar{2}5},x_2,x_2)\bigg)\nonumber\\
&&\hspace{0.75in}\times A_3^0(\Qb{4},\gl{5},\qbi{\bar{2}})|\cm_4(\Q{3},\Qb{(\wt{45})},\qbi{\bar{\bar{2}}},\qi{\bar{1}})|^2 J_2^{(2)}(p_3,p_{\wt{45}})\bigg\},
\eeqa
and using eqs.(\ref{eq.A03flint}), (\ref{eq.polesd1}) and (\ref{eq.polesd2}) we get
\beqa\label{eq.polesdsb33}
&&\hspace{-0.1in}\poles \left( \int_1 \ds^{\rS,b\:3\times 3}_{q\bar{q},\NNLO,N_c^2} \right)=\norm_{\NNLO}^{q\bar{q},\RV} \,N_c^2\int \frac{{\rm d}x_1}{x_1}\frac{{\rm d}x_2}{x_2}\dphi_3(p_3,p_4,p_5;x_1p_1,x_2p_2) \nonumber\\
&&\times \bigg\{ \bigg[ \delta(1-x_1)\delta(1-x_2)\bigg( 2\ione{Q}{g}(\e,s_{35})+ 2\ione{q}{g}(\e,s_{15})\bigg)-\frac{1}{2}\Gamma^{(1)}_{qq}(x_1)\delta(1-x_2)\bigg]\nonumber\\
&&\hspace{1.8in}\times A_3^0(\Q{3},\gl{5},\qi{\bar{1}})|\cm_4(\Q{(\wt{35})},\Qb{4},\qbi{\bar{2}},\qi{\bar{\bar{1}}})|^2 J_2^{(2)}(\wt{p_{35}},p_4)\phantom{\bigg]}\nonumber\\
&&\hspace{0.105in}+\bigg[  \delta(1-x_1)\delta(1-x_2)\bigg( 2 \ione{Q}{g}(\e,s_{45})+ 2\ione{q}{g}(\e,s_{25})\bigg)-\frac{1}{2}\Gamma^{(1)}_{qq}(x_2)\delta(1-x_1)\bigg]\nonumber\\
&&\hspace{1.8in}\times A_3^0(\Qb{4},\gl{5},\qbi{\bar{2}})|\cm_4(\Q{3},\Qb{(\wt{45})},\qbi{\bar{\bar{2}}},\qi{\bar{1}})|^2 J_2^{(2)}(p_3,\wt{p_{45}}) \bigg\}.\nonumber\\
\eeqa

Combining equations (\ref{eq.polesdsvsa}), (\ref{eq.polesdsb33}) and (\ref{eq.mfqqb1b}) we find that
\beqa\label{eq.condit2}
&&\hspace{-0.1in}\poles\left( \ds_{q\bar{q},\NNLO,N_c^2}^{\VS,a}-\int_1 \ds^{\rS,b\:3\times 3}_{q\bar{q},\NNLO,N_c^2}-\ds^{\MF,1b}_{q\bar{q},\NNLO,N_c^2}\right)=\nonumber\\
&&\norm_{\NNLO}^{q\bar{q},\RV}\,N_c^2\int \frac{{\rm d}x_1}{x_1}\frac{{\rm d}x_2}{x_2}\dphi_3(p_3,p_4,p_5;x_1p_1,x_2p_2)\nonumber\\
&&\times\bigg\{2\ione{Q}{\bar{q}}(\e,s_{24},m_Q^2)\delta(1-x_1)\delta(1-x_2)A_3^0(\Q{3},\gl{5},\qi{\bar{1}})|\cm_4(\Q{(\wt{35})},\Qb{4},\qbi{\bar{2}},\qi{\bar{\bar{1}}})|^2 J_2^{(2)}(\wt{p_{35}},p_4)\nonumber\\
&&\hspace{0.1in}+2\ione{Q}{\bar{q}}(\e,s_{13},m_Q^2)\delta(1-x_1)\delta(1-x_2)A_3^0(\Qb{4},\gl{5},\qbi{\bar{2}})|\cm_4(\Q{3},\Qb{(\wt{45})},\qbi{\bar{\bar{2}}},\qi{\bar{1}})|^2 J_2^{(2)}(p_3,\wt{p_{45}})\phantom{\bigg[}\nonumber\\
&&\hspace{0.1in}-\frac{1}{2}\Gamma^{(1)}_{qq}(x_1)\delta(1-x_2)A_3^0(\Qb{4},\gl{5},\qi{\bar{2}})|\cm_4(\Q{3},\Qb{(\wt{45})},\qbi{\bar{\bar{2}}},\qi{\bar{1}})|^2 J_2^{(2)}(p_3,\wt{p_{45}})\phantom{\bigg[}\nonumber\\
&&\hspace{0.1in}-\frac{1}{2}\Gamma^{(1)}_{qq}(x_2)\delta(1-x_1)A_3^0(\Q{3},\gl{5},\qi{\bar{1}})|\cm_4(\Q{(\wt{35})},\Qb{4},\qbi{\bar{2}},\qi{\bar{\bar{1}}})|^2 J_2^{(2)}(\wt{p_{35}},p_4)\bigg\}.
\eeqa
In order for eq.(\ref{eq.conditiondsvsb}) to be satisfied, $\ds_{q\bar{q},\NNLO,N_c^2}^{\VS,b}$ must be constructed in such a way that its pole part is opposite to the equation above. We must therefore identify the integrated antenna functions that yield the $\ione{i}{j}$ operators and splitting kernels in eq.(\ref{eq.condit2}). In this case, the integrated antennae that should be employed are initial-final flavour-violating A-type antennae, and it can be seen that eq.(\ref{eq.conditiondsvsb}) is satisfied if we write 
\beqa\label{eq.subvsb}
&&\hspace{-0.1in}\ds_{q\bar{q},\NNLO,N_c^2}^{\VS,b}=\norm_{\NNLO}^{q\bar{q},\RV}\,N_c^2\int \frac{{\rm d}x_1}{x_1}\frac{{\rm d}x_2}{x_2}\dphi_3(p_3,p_4,p_5;x_1p_1,x_2 p_2)\nonumber\\
&&\hspace{0.2in}\times\bigg\{ {\cal A}^0_{q,Qg}(\e,s_{\bar{1}3},x_1,x_2)A_3^0(\Qb{4},\gl{5},\qbi{\bar{2}})|\cm_4(\Q{3},\Qb{(\wt{45})},\qbi{\bar{\bar{2}}},\qi{\bar{1}})|^2 J_2^{(2)}(p_3,\wt{p_{45}})\nonumber\\
&&\hspace{0.29in}+{\cal A}^0_{q,Qg}(\e,s_{\bar{2}4},x_2,x_1)A_3^0(\Q{3},\gl{5},\qi{\bar{1}})|\cm_4(\Q{(\wt{35})},\Qb{4},\qbi{\bar{2}},\qi{\bar{\bar{1}}})|^2 J_2^{(2)}(\wt{p_{35}},p_4)\bigg\}.\nonumber\\
\eeqa
Furthermore the subtraction term $\ds_{q\bar{q},\NNLO,N_c^2}^{\VS,b}$ together with \linebreak $\int_1\ds_{q\bar{q},\NNLO,N_c^2}^{\rS,b\,3\times3}$ and $\ds^{\MF,1b}_{q\bar{q},\NNLO,N_c^2}$ reproduces the spurious single unresolved behaviour of $\int_1\ds_{q\bar{q},\NNLO,N_c^2}^{\rS,a}$ and $\ds^{\MF,1a}_{q\bar{q},\NNLO,N_c^2}$ which is the second requirement the subtraction term $(\VS,b)$ has to fulfill.


\subsubsection{Construction of $\ds^{\VS,d}_{q\bar{q},\NNLO,N_c^2}$}
Finally, the ultraviolet type subtraction term denoted by $\ds_{q\bar{q},\NNLO,N_c^2}^{\VS,d}$ is proportional to the leading colour part of $\beta_{0}$, $b_0=11/6$. It reads
\beqa\label{eq.subvsd}
&&\hspace{-0.5in}\ds_{q\bar{q},\NNLO,N_c^2}^{\VS,d}=\norm_{\NNLO}^{\RV,\,q\bar{q}}\,N_c^2\int \frac{{\rm d}x_1}{x_1}\frac{{\rm d}x_2}{x_2}\,\dphi_3(p_3,p_4,p_5;x_1p_1,x_2 p_2)\delta(1-x_1)\delta(1-x_2)\nonumber\\
&&\times\bigg\{b_0\log \left( \frac{\mu^2}{|s_{135}|}\right)A_3^0(\Q{3},\gl{5},\qi{\bar{1}})|\cm_4(\Q{(\wt{35})},\Qb{4},\qbi{\bar{2}},\qi{\bar{\bar{1}}})|^2 J_2^{(2)}(p_{\wt{35}},p_4)\nonumber\\
&&\hspace{0.125in}+b_0\log \left( \frac{\mu^2}{|s_{245}|}\right)A_3^0(\Qb{4},\gl{5},\qbi{\bar{2}})|\cm_4(\Q{3},\Qb{(\wt{45})},\qbi{\bar{\bar{2}}},\qi{\bar{1}})|^2 J_2^{(2)}(p_3,p_{\wt{45}})\bigg\}.\\ \nonumber
\eeqa


\subsubsection{The complete real-virtual subtraction term  $\ds^{\rT}_{q\bar{q},\NNLO,N_c^2}$ }
Putting everything together, the three-parton level contribution $\ds^{\rT}_{q\bar{q},\NNLO,N_c^2}$ to be combined with the real-virtual contributions $\ds^{\RV}_{q\bar{q},\NNLO,N_c^2}$ can be conveniently written in the following way
\beqa\label{eq.dstcomplete}
&&\hspace{-0.14in}\ds^{\rT}_{q\bar{q},\NNLO,N_c^2}=\norm_{\NNLO}^{q\bar{q},\RV} N_{c}^2 \int \frac{{\rm d}x_1}{x_1}\frac{{\rm d}x_2}{x_2}\,\dphi_3(p_3,p_4,p_5;x_1p_1,x_2 p_2)\nonumber\\
&&\hspace{-0.115in}\times\bigg\{-\bigg(\frac{1}{2}{\cal D}_{Qgg}^0(\e,s_{35},x_1,x_2)+\frac{1}{2}{\cal D}_{q,gg}^0(\e,s_{\bar{1}5},x_1,x_2)+{\cal A}_{q,Qg}^0(\e,s_{\bar{2}4},x_2,x_1)\nonumber\\
&&\hspace{0.24in}-\frac{1}{2}\Gamma_{qq}^{(1)}(x_1)\delta(1-x_2)-\frac{1}{2}\Gamma_{qq}^{(1)}(x_2)\delta(1-x_1)\bigg)|\cm_5(\Q{3},\gl{5},\qi{\bar{1}};;\qbi{\bar{2}},\Qb{4})|^2 J_2^{(3)}(p_3,p_4,p_5)\nonumber\\
&&\hspace{0.09in}-\bigg(\frac{1}{2}{\cal D}_{Qgg}^0(\e,s_{45},x_1,x_2)+\frac{1}{2}{\cal D}_{q,gg}^0(\e,s_{\bar{2}5},x_2,x_1)+{\cal A}_{q,Qg}^0(\e,s_{\bar{1}3},x_1,x_2)\nonumber\\
&&\hspace{0.24in}-\frac{1}{2}\Gamma_{qq}^{(1)}(x_1)\delta(1-x_2)-\frac{1}{2}\Gamma_{qq}^{(1)}(x_2)\delta(1-x_1)\bigg)|\cm_5(\Q{3},\qi{\bar{1}};;\qbi{\bar{2}},\gl{5},\Qb{4})|^2 J_2^{(3)}(p_3,p_4,p_5)\nonumber\\
&&\hspace{0.09in}+\bigg[ A_3^{1,lc}(\Q{3},\gl{5},\qi{\bar{1}})\delta(1-x_1)\delta(1-x_2) + \bigg(\frac{1}{2}{\cal D}^0_{Qgg}(\e,s_{35},x_1,x_2)+\frac{1}{2}{\cal D}^0_{q,gg}(\e,s_{\bar{1}5},x_1,x_2)\nonumber\\
&&\hspace{0.35in}-{\cal A}_{q,Qg}^0(\e,s_{\bar{1}35},x_1,x_2)\bigg)A_3^0(\Q{3},\gl{5},\qi{\bar{1}})\bigg]|\cm_4(\Q{(\wt{35})},\Qb{4},\qbi{\bar{2}},\qi{\bar{\bar{1}}})|^2 J_2^{(2)}(p_{\wt{35}},p_4)\phantom{\bigg(}\nonumber\\
&&\hspace{0.09in}+\bigg[ A_3^{1,lc}(\Qb{4},\gl{5},\qbi{\bar{2}})\delta(1-x_1)\delta(1-x_2) + \bigg(\frac{1}{2}{\cal D}^0_{Qgg}(\e,s_{45},x_1,x_2)+\frac{1}{2}{\cal D}^0_{q,gg}(\e,s_{\bar{2}5},x_2,x_1)\nonumber\\
&&\hspace{0.35in}-{\cal A}_{q,Qg}^0(\e,s_{\bar{2}45},x_2,x_1)\bigg)A_3^0(\Qb{4},\gl{5},\qbi{\bar{2}})\bigg]|\cm_4(\Q{3},\Qb{(\wt{45})},\qbi{\bar{\bar{2}}},\qi{\bar{1}})|^2 J_2^{(2)}(p_3,p_{\wt{45}})\phantom{\bigg(}\nonumber\\
&&\hspace{0.09in}+A_3^0(\Q{3},\gl{5},\qi{\bar{1}})\bigg[ \big|\cmb_{4,1}^{[lc]}(\Q{(\wt{35})},\Qb{4},\qbi{\bar{2}},\qi{\bar{\bar{1}}})\big|_{\NLO}^2\delta(1-x_1)\delta(1-x_2)\nonumber\\
&&\hspace{0.35in}+\bigg( {\cal A}_{q,Qg}^0(\e,s_{\bar{1}35},x_1,x_2)+ {\cal A}_{q,Qg}^0(\e,s_{\bar{2}4},x_2,x_1)\nonumber\\
&&\hspace{0.35in}-\frac{1}{2}\Gamma_{qq}^{(1)}(x_1)\delta(1-x_2)-\frac{1}{2}\Gamma_{qq}^{(1)}(x_2)\delta(1-x_1)\bigg)|\cm_4(\Q{(\wt{35})},\Qb{4},\qbi{\bar{2}},\qi{\bar{\bar{1}}})|^2\bigg]J_{2}^{(2)}(p_{\wt{35}},p_4)\nonumber\\
&&\hspace{0.09in}+A_3^0(\Qb{4},\gl{5},\qbi{\bar{2}})\bigg[ \big|\cmb_{4,1}^{[lc]}(\Q{3},\Qb{(\wt{45})},\qbi{\bar{\bar{2}}},\qi{\bar{1}})\big|_{\NLO}^2\delta(1-x_1)\delta(1-x_2)\nonumber\\
&&\hspace{0.35in}+\bigg( {\cal A}_{q,Qg}^0(\e,s_{\bar{2}45},x_2,x_1)+ {\cal A}_{q,Qg}^0(\e,s_{\bar{1}3},x_1,x_2)\nonumber\\
&&\hspace{0.35in}-\frac{1}{2}\Gamma_{qq}^{(1)}(x_1)\delta(1-x_2)-\frac{1}{2}\Gamma_{qq}^{(1)}(x_2)\delta(1-x_1)\bigg)|\cm_4(\Q{3},\Qb{(\wt{45})},\qbi{\bar{\bar{2}}},\qi{\bar{1}})|^2\bigg]J_{2}^{(2)}(p_3,p_{\wt{45}})\nonumber\\
&&\hspace{0.09in}+b_0\log \left( \frac{\mu^2}{|s_{\bar{1}35}|}\right)\hspace{-0.02in}A_3^0(\Q{3},\gl{5},\qi{\bar{1}})\delta(1-x_1)\delta(1-x_2)|\cm_4(\Q{(\wt{35})},\Qb{4},\qbi{\bar{2}},\qi{\bar{\bar{1}}})|^2 J_2^{(2)}(p_{\wt{35}},p_4)\nonumber\\
&&\hspace{0.09in}+b_0\log \left( \frac{\mu^2}{|s_{\bar{2}45}|}\right)\hspace{-0.02in}A_3^0(\Qb{4},\gl{5},\qbi{\bar{2}})\delta(1-x_1)\delta(1-x_2)|\cm_4(\Q{3},\Qb{(\wt{45})},\qbi{\bar{\bar{2}}},\qi{\bar{1}})|^2 J_2^{(2)}(p_3,p_{\wt{45}})\bigg\}.\nonumber\\
\eeqa
The pole part of the terms which contain tree-level five-parton matrix elements squared exactly cancel the explicit $\e$-poles of the real-virtual contributions $\ds^{\RV}_{q\bar{q},\NNLO,N_c^2}$. On the other hand, the content of the square brackets $[\ldots]$ is free of poles in $\e$. 

From all terms in $\ds^{\rT}_{q\bar{q},\NNLO,N_c^2}$, only those corresponding to real-virtual subtraction terms 
\beq
\ds_{q\bar{q},\NNLO,N_c^2}^{\VS}=\ds_{q\bar{q},\NNLO,N_c^2}^{\VS,a}+\ds_{q\bar{q},\NNLO,N_c^2}^{\VS,b}+ +\ds_{q\bar{q},\NNLO,N_c^2}^{\VS,d}
\eeq
must be integrated and added back at the two-parton level with the double virtual contributions. The individual contributions in this sum were given in eqs.(\ref{eq.subvsa}), (\ref{eq.subvsb}) and (\ref{eq.subvsd}).

We have shown in this section that, by construction, the counter term $\ds^{\rT}_{q\bar{q},\NNLO,N_c^2}$ exactly cancels the explicit infrared poles of the real-virtual contributions. For the three-parton final state $\ds^{\RV}_{q\bar{q},\NNLO,N_c^2}-\ds^{\rT}_{q\bar{q},\NNLO,N_c^2}$ to be numerically integrable in four dimensions, it remains to be shown that the real-virtual counter term $ds^{\rT}_{q\bar{q},\NNLO,N_c^2}$ constitutes a good approximation of the real-virtual contributions in the soft and collinear limits. We shall address this issue in the next section with a series of numerical tests employing the amplitudes obtained with \OpenLoops as described in section \ref{sec.openloops}.


\section{Numerical tests of soft and collinear cancellations}\label{sec.tests}

The double real and real-virtual contributions to heavy quark pair
production in the $q\bar{q}$ channel presented in sections \ref{sec.RR} and
\ref{sec.openloops} have been implemented in a {\tt Fortran} code together
with the corresponding subtraction terms $\ds^{\rS}_{q\bar{q},\NNLO,N_c^2}$
and $\ds^{\rT}_{q\bar{q},\NNLO,N_c^2}$.  In this section we investigate how
well these subtraction terms fulfil their purpose of approximating
$\ds^{\RR}_{q\bar{q},\NNLO,N_c^2}$ and $\ds^{\RV}_{q\bar{q},\NNLO,N_c^2}$ in
all unresolved limits.  In the case of the real-virtual contributions, the
quality of the cancellations in the infrared regions provides also important
insights into the numerical stability of the  amplitudes.

For each singular region we used a series of phase-space samples generated
with {\tt RAMBO}~\cite{Kleiss:1985gy} by requiring an increasingly small
distance, parametrised in terms of appropriate parameters $x_k$,
from the relevant singularity.  In the next two sections, we
will quantify the level of the real-real and real-virtual cancellations as

\beq
\label{eq:RRcanc}
\delta_{\RR}=
\left|\frac{\ds^{\RR}_{q\bar{q},\NNLO,N_c^2}}{\ds^{\rS}_{q\bar{q},\NNLO,N_c^2}}-1\right|,
\eeq
and
\beq 
\label{eq:RVcanc}
\delta_{\RV}=\left|\frac{
\ds^{\RV}_{q\bar{q},\NNLO,N_c^2}}{
\ds^{\rT}_{q\bar{q},\NNLO,N_c^2}}-1\right|,
\eeq 
respectively. To demonstrate the consistency and stability of the subtractions we will show that the $\delta_{\RR}$ and $\delta_{\RV}$ distributions converge to zero in all relevant $x_k\to 0$ limits. On the right-hand-side of (\ref{eq:RVcanc}) the consistent subtraction of explicit infrared singularities in the numerator and denominator is implicitly understood. Each of the employed samples consists of about $10^4$ points with $\sqrt{\hat{s}}=1$\,TeV 
\footnote{For simplicity, $\hat{s}$ will be denoted by $s$ in this section.} and $m_Q=174.3\,\gev$.


\subsection{Tests of the double real contributions}
We start by discussing infrared cancellations for the double real contribution $q\bar{q}\to Q\bar{Q}gg$ in leading colour approximation. To this end we generated $2\to 4$ phase space points near all possible single and double unresolved limits. The $2\to 4$ tree-level matrix elements in (\ref{eq:RRcanc}) have been computed with an in-house {\tt Mathematica} program based on {\tt Qgraf}~\cite{Nogueira:1991ex} and numerically checked against {\tt MadGraph}~\cite{Alwall:2011uj} for a few phase space points.


\subsubsection{Double soft limits}
\begin{figure}[t]
\begin{center}  
\subfigure[]{
\resizebox{0.35\linewidth}{!}{
\label{fig.dssketch}
\includegraphics{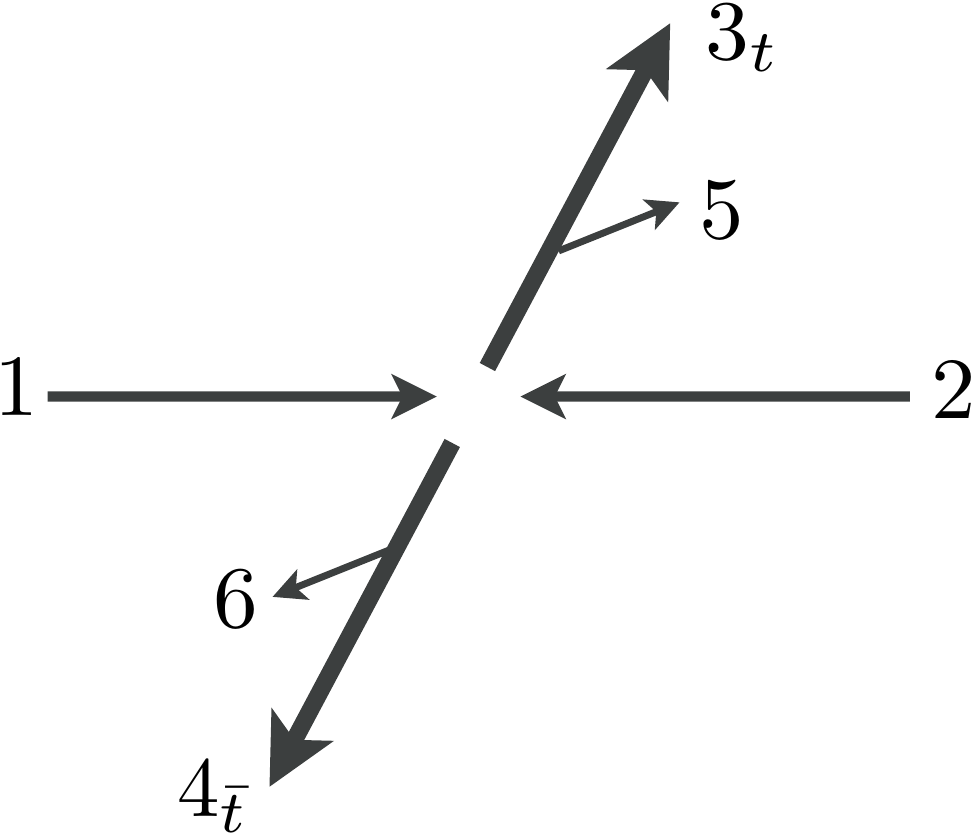}}}
\hspace{0.1in}
\subfigure[]{
\resizebox{0.6\linewidth}{!}{
\label{fig.DS}
\includegraphics{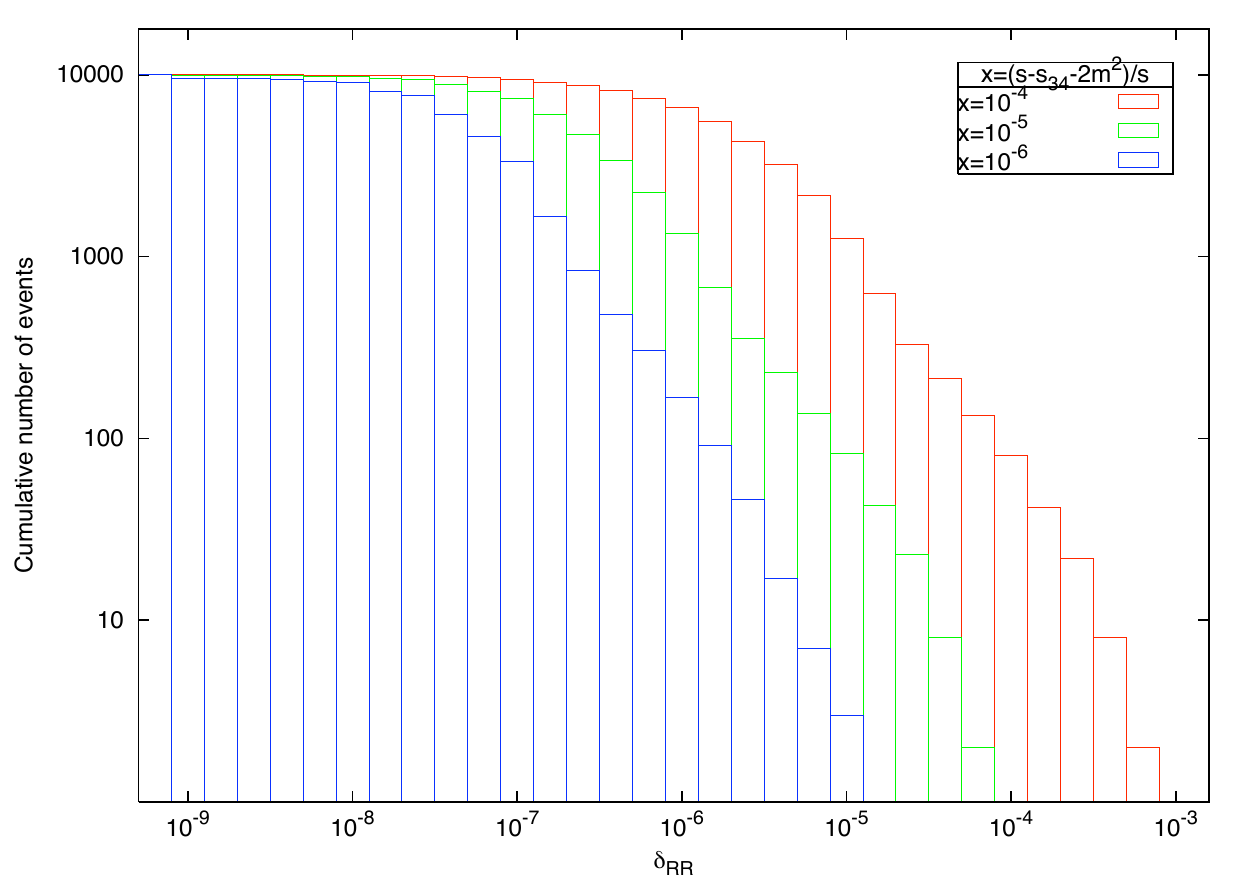}}}
\caption{(a) Sketch of a double soft event. (b) Cumulative distributions of $\delta_{\rm RR}$ for $10^4$ double soft phase space points with three different values of $x$}
\end{center}
\end{figure}

As shown in Fig.\ref{fig.dssketch}, a double soft phase space point is characterised by the heavy quark pair taking nearly the full energy of the event, and therefore a suitable variable  to control the proximity of the events to the singular limit is $x=(s-s_{34}-2m_Q^2)/s$. 
In Fig.\ref{fig.DS} we show cumulative distributions of $\delta_{\rm RR}$
for three different values of $x$.  
Each bin contains the total number of points for which 
the relative difference between matrix element and subtraction term
is larger than $\delta_{\rm RR}$.
The good convergence of the subtraction
terms to the double real contributions as the singularity is approached can
be seen in the fact that the events accumulate more rapidly near
$\delta_{\rm RR}=0$ as the control variable $x$ is taken to be smaller.


\subsubsection{Triple collinear limits}
\begin{figure}[t]
\begin{center}  
\subfigure[]{
\label{fig.tcsketch}
\resizebox{0.35\linewidth}{!}{
\includegraphics{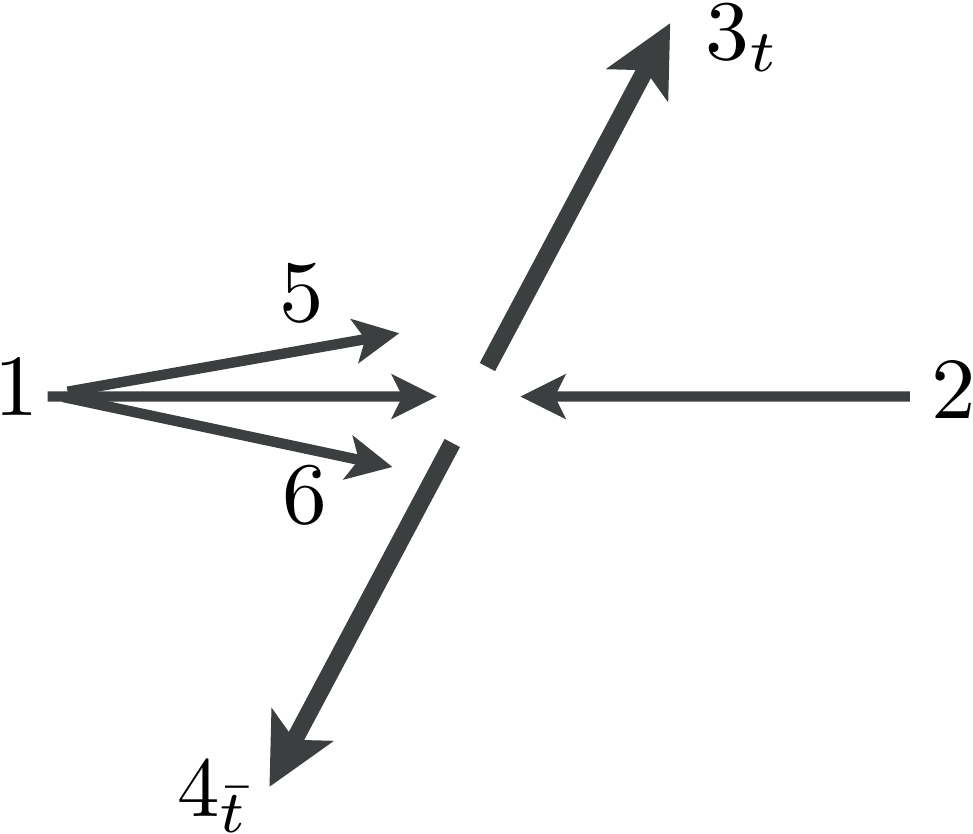}}}
\subfigure[]{
\label{fig.TC}
\resizebox{0.6\linewidth}{!}{
\includegraphics{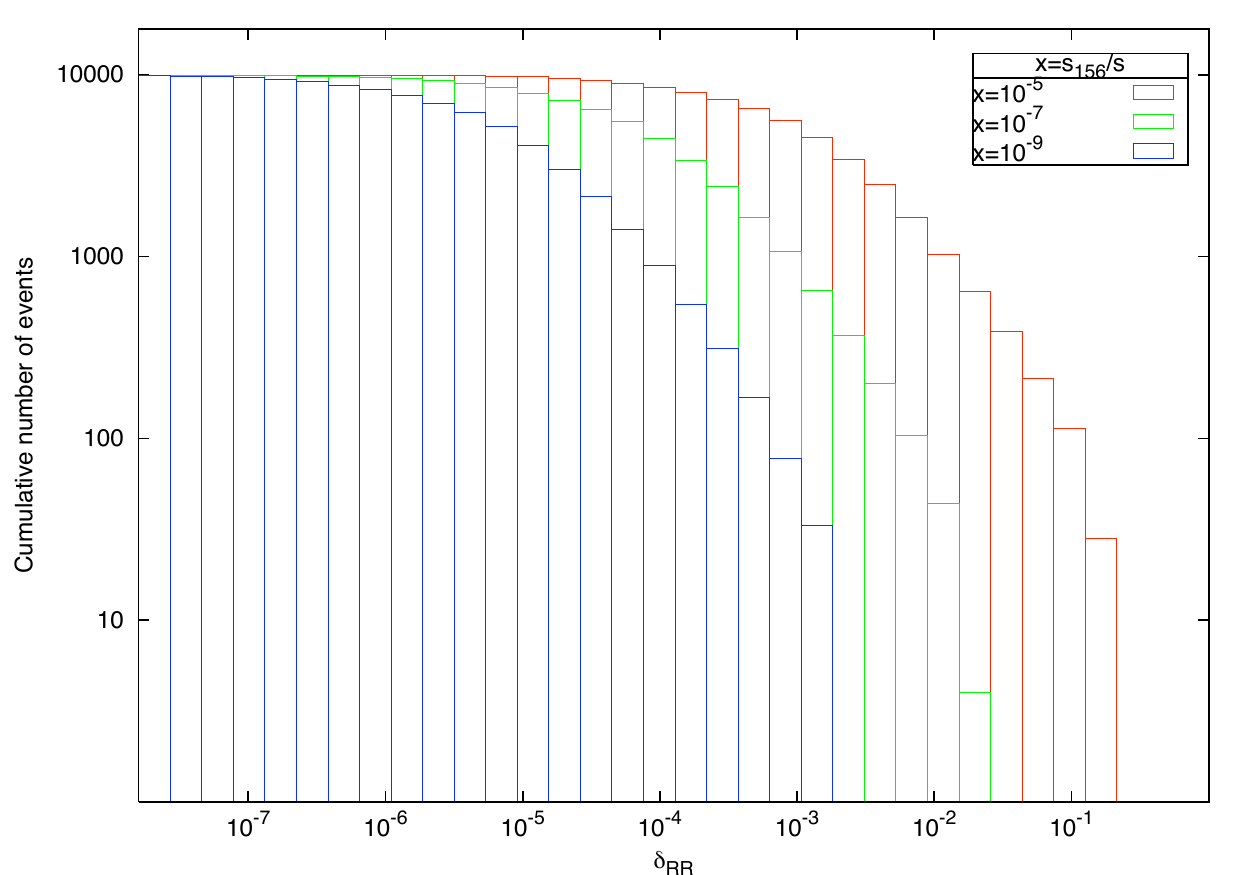}}}
\caption{(a) Sketch of a triple collinear event. (b) Cumulative distributions of $\delta_{\rm RR}$ for $10^4$ triple collinear phase space points with three different values of $x$}
\end{center}
\end{figure}

Since we do not subtract collinear limits involving the massive fermions because they are regulated by the large value of $m_Q$, the only types of triple collinear limits that we must consider are of initial-final nature as depicted in Fig.\ref{fig.tcsketch}. The control variable in this case is defined as $x=s_{i56}/s$, where $i=1,2$. In Fig.\ref{fig.TC} we show how, as we take smaller values of $x$, i.e.~as we get closer in phase space to the singularity of the real radiation matrix element, there is a more rapid accumulation of events around $\delta_{\rm RR}=0$, signalling again that the approximation is correct. These results correspond to the triple collinear limit $p_1||p_5||p_6$. Similar results are obtained for $p_2||p_5||p_6$.


\subsubsection{Soft-collinear limits}
\begin{figure}[t]
\begin{center}  
\subfigure[]{
\resizebox{0.35\linewidth}{!}{
\label{fig.softcollsketch}
\includegraphics{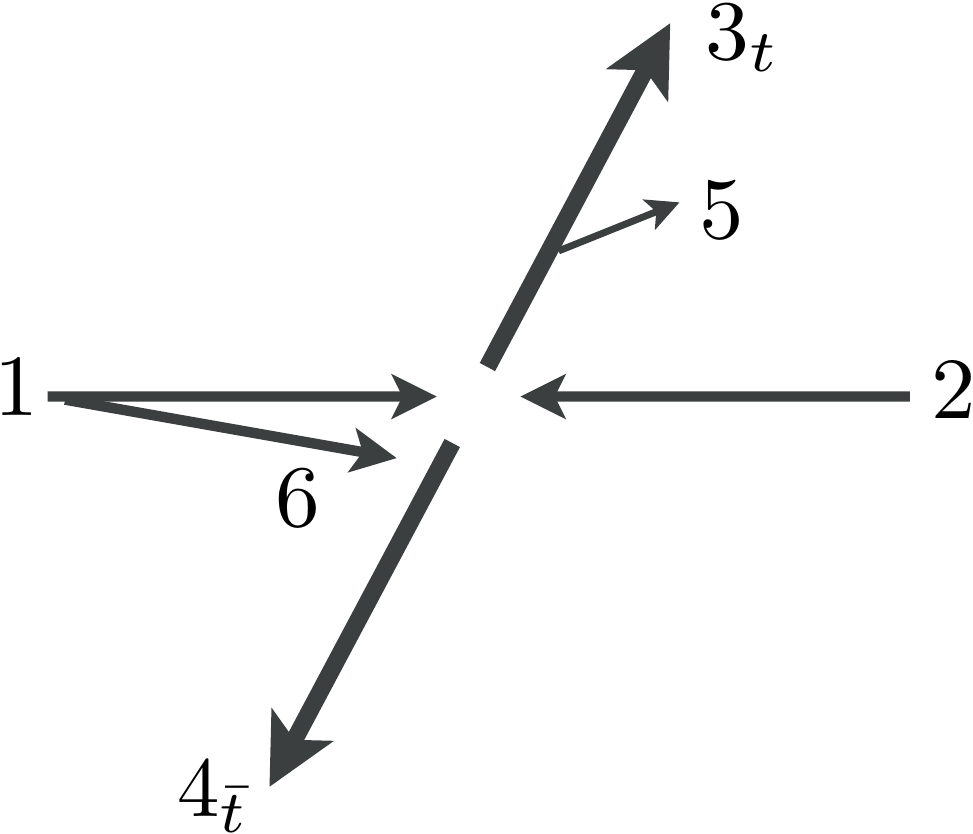}}}
\subfigure[$q\bar{q}\rightarrow t\bar{t}gg$]{
\resizebox{0.6\linewidth}{!}{
\label{fig.SOFTCOLL}
\includegraphics{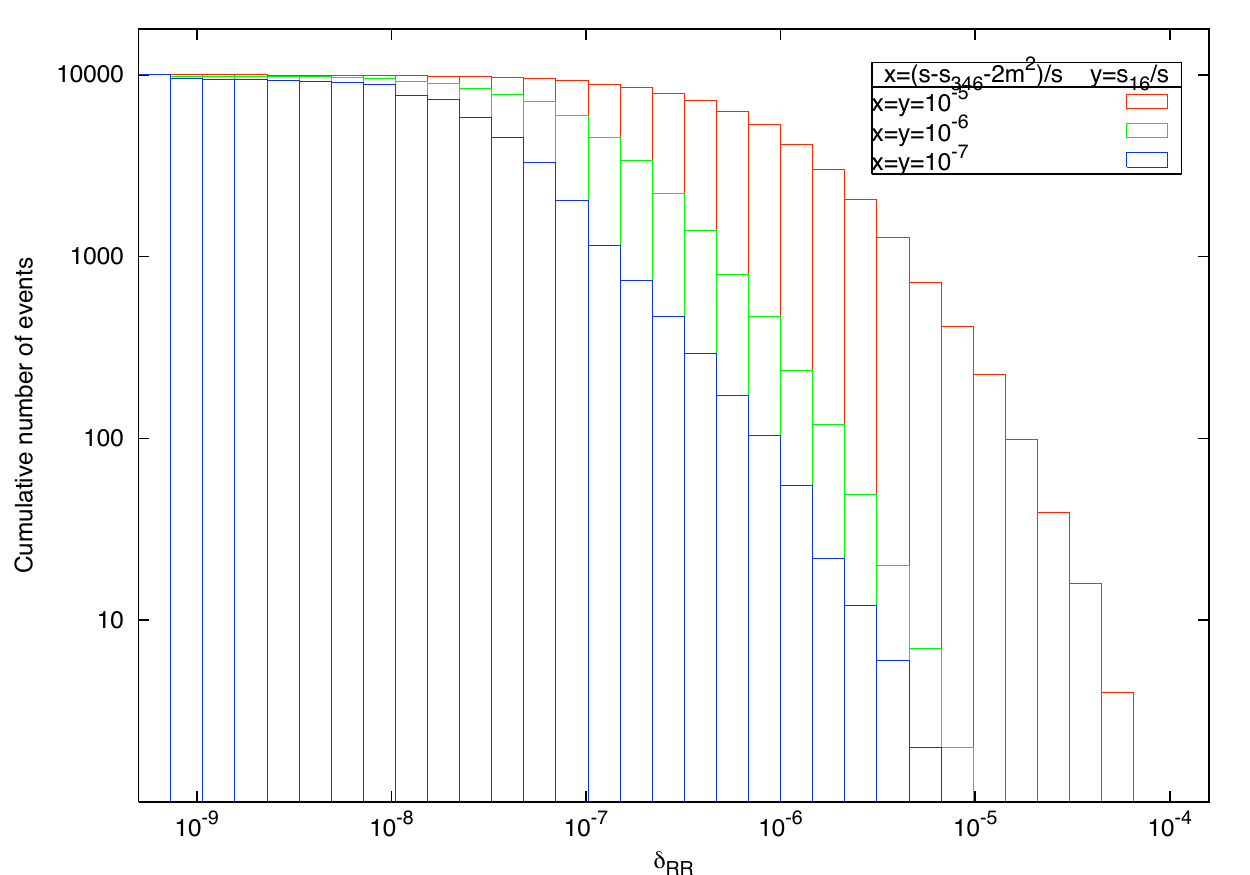}}}
\caption{(a) Sketch of a soft-collinear event. (b) Cumulative distributions of $\delta_{\rm RR}$ for $10^4$ soft-collinear phase space points with three different values of $x$ and $y$}
\end{center}
\end{figure}

As shown in Fig.\ref{fig.softcollsketch}, soft-collinear limits occur when one of the final state gluons becomes soft and the remaining one becomes collinear to an initial state leg. To probe the soft-collinear regions of phase space we generate events with a soft gluon and rotate the final state to make the hard gluon collinear to one of the initial state legs.
We employ two control variables $x$ and $y$. If we consider the limit where
gluon $(5)$ is soft and $(6)$ becomes collinear to the incoming leg $(1)$,
$x$ is defined as 
$x=(s-s_{346}-2m_Q^2)/s$ and $y$ is given by $y=s_{16}/s$. 
As can be seen if Fig.\ref{fig.SOFTCOLL} the convergence of the subtraction
term to the partonic double real contribution is once more achieved.


\subsubsection{Double collinear limits}
\begin{figure}[t]
\begin{center}  
\subfigure[]{
\label{fig.dcsketch}
\resizebox{0.35\linewidth}{!}{
\includegraphics{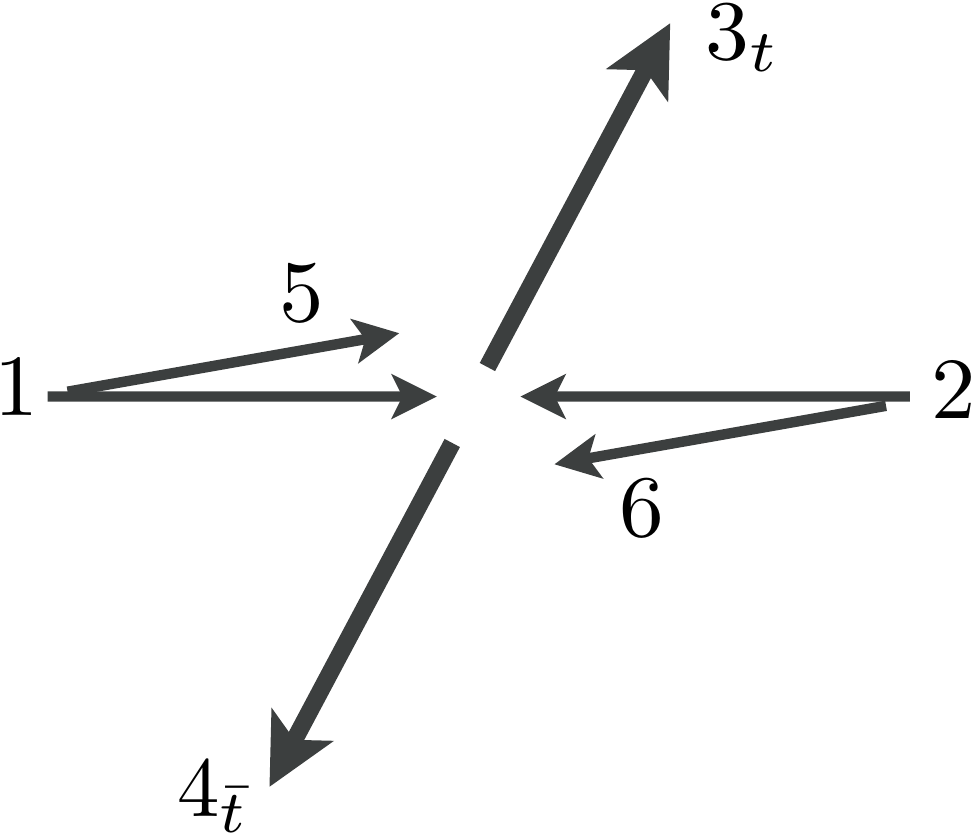}}}
\subfigure[]{
\label{fig.DC}
\resizebox{0.6\linewidth}{!}{
\includegraphics{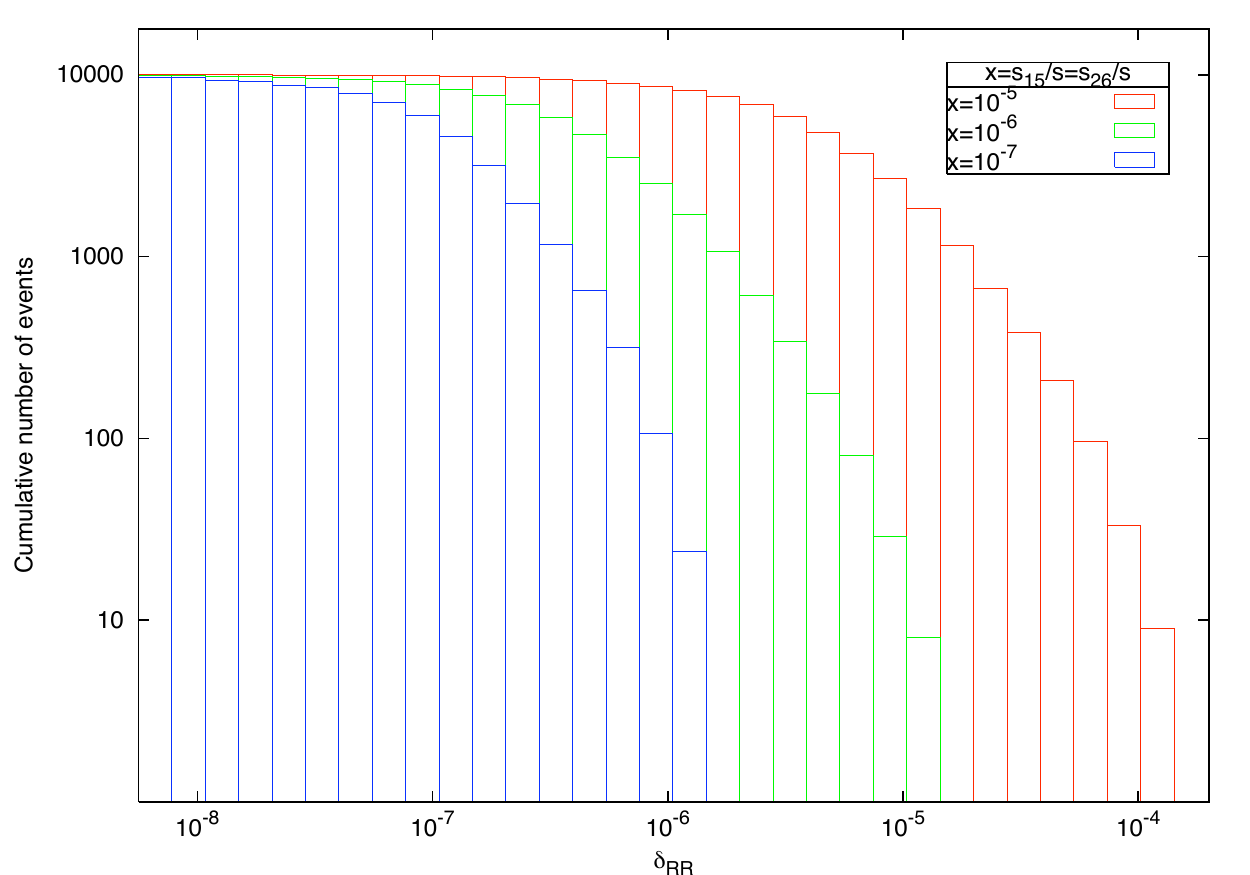}}}
\caption{(a) Sketch of a double collinear event. (b) Cumulative distributions of $\delta_{\rm RR}$ for $10^4$ double collinear phase space points with three different values of $x$}
\end{center}
\end{figure}

Due to the fact that the quasi-collinear limits involving the heavy (anti) quark do not require subtraction, the only double collinear limits in which the double real contributions can diverge are the two simultaneous single collinear limits depicted in Fig.\ref{fig.dcsketch}. To control the proximity of the phase space points to the double collinear singularity $p_1||p_5$, $p_2||p_6$ we employ the variable 
$x=s_{15}/s=s_{26}/s$.
As can be seen from Fig.\ref{fig.DC}, our numerical results show that behaviour of the double real corrections in their double collinear limits, is correctly described by our subtraction terms. Similar results are obtained for the double collinear limit $p_1||p_6$, $p_2||p_5$.


\subsubsection{Single soft limits}
Single soft limits are characterised by having the three hard final state particles taking nearly the full center-of-mass energy of the event leaving one of the final state gluons with an almost vanishing energy. Consequently, if the soft-gluon momentum is $p_5$, we define the control variable as $x=(s_{346}-s-2m_Q^2)/s$.
\begin{figure}[t]
\begin{center}  
\subfigure[]{
\label{fig.sketchss}
\resizebox{0.35\linewidth}{!}{
\includegraphics{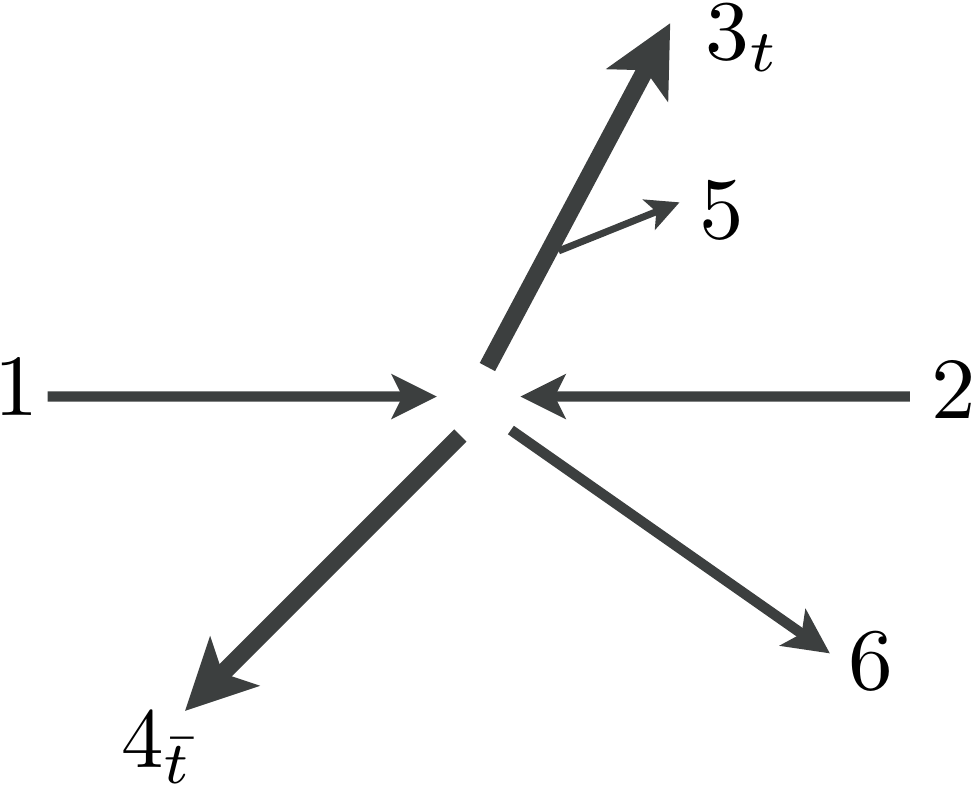}}}
\subfigure[]{
\label{fig.Soft}
\resizebox{0.6\linewidth}{!}{
\includegraphics{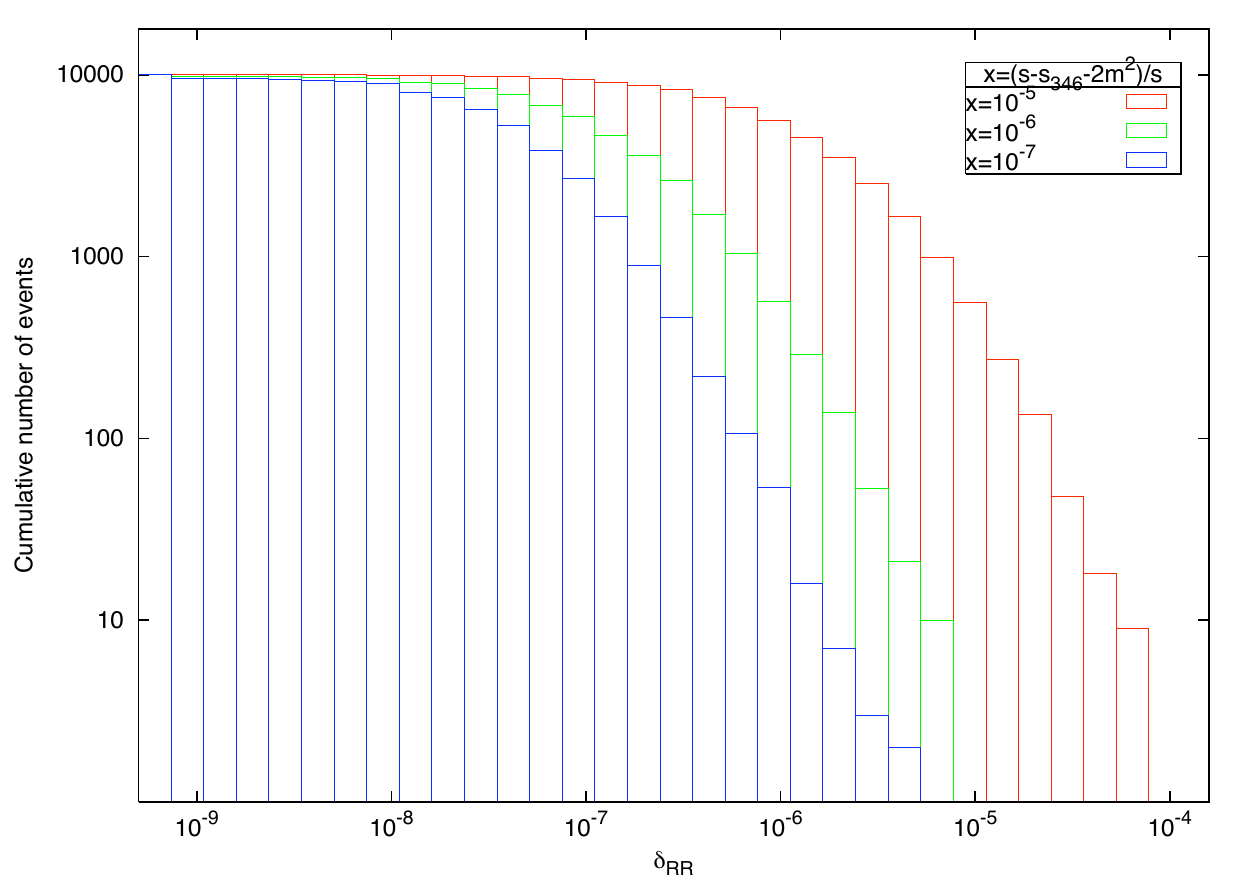}}}
\caption{(a) Sketch of a single soft event. (b) Cumulative distributions of $\delta_{\rm RR}$ for $10^4$ single soft phase space points with three different values of $x$}
\end{center}
\end{figure}
In Fig.\ref{fig.Soft} we show how as the singularity is approached by making $x$ closer to zero, events accumulate more rapidly near $\delta_{\rm RR}=0$.  Analogous results are obtained when the 
soft-gluon momentum is $p_6$.


\subsubsection{Final-final single collinear limit}
\begin{figure}[t]
\begin{center}  
\subfigure[]{
\label{fig.sketchsc}
\resizebox{0.35\linewidth}{!}{
\includegraphics{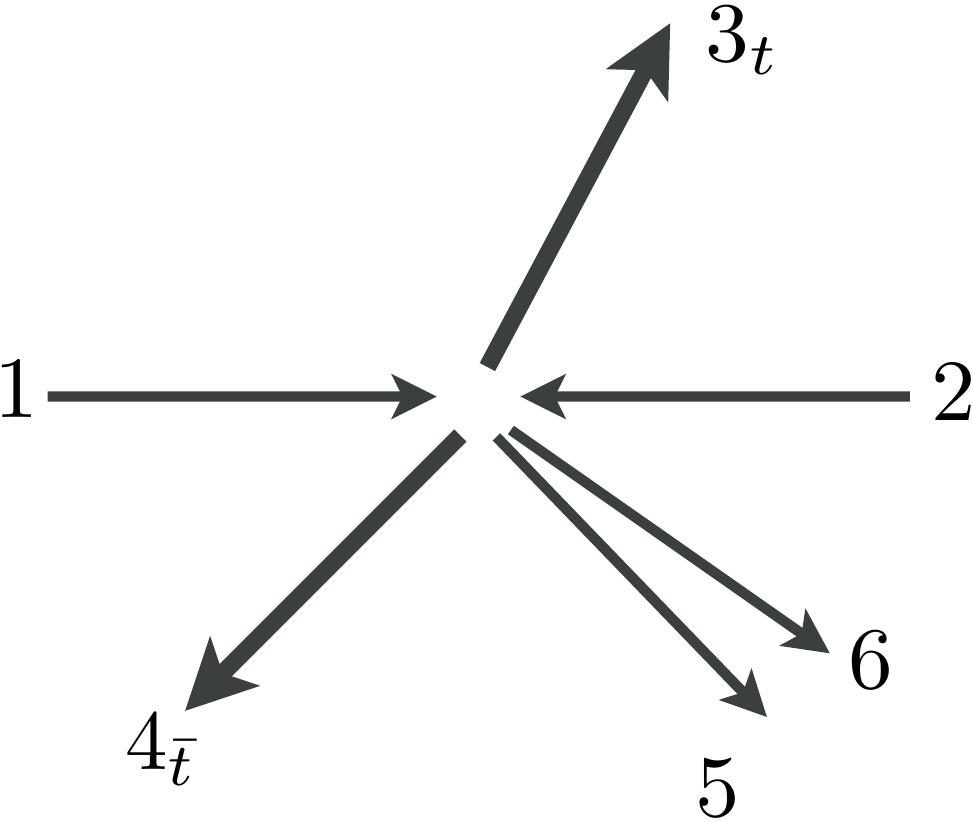}}}
\subfigure[]{
\label{fig.SCff}
\resizebox{0.6\linewidth}{!}{
\includegraphics{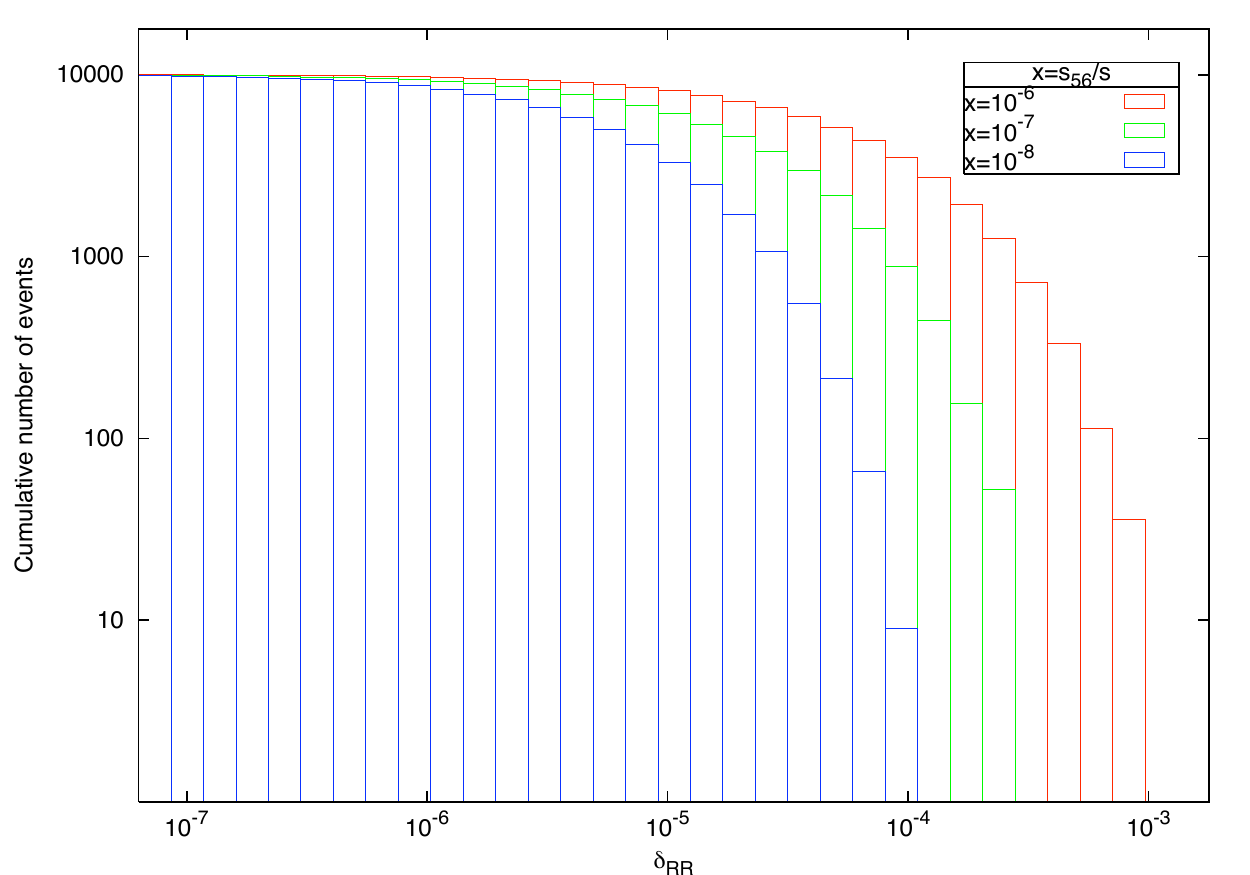}}}
\caption{(a) Sketch of a final-final single collinear limit. (b) Cumulative distributions of $\delta_{\rm RR}$ for $10^4$ final-final single collinear phase space points with three different values of $x$}
\end{center}
\end{figure}

As depicted in Fig.\ref{fig.sketchsc}, final-final collinear limits occur when the final state gluons with momentum $p_5$ and $p_6$ become collinear. This divergence is approached as the ratio $x=s_{56}/s$ gets closer to zero.

As discussed previously in
\cite{Abelof:2011ap,Abelof:2012rv,GehrmannDeRidder:2007jk,Glover:2010im},
because of the presence of angular correlations between the splitting
functions and the reduced matrix elements, in single collinear limits
corresponding to the gluon splittings $g\to gg$ and $g\to q\bar{q}$, antenna
subtraction terms do not reproduce the behaviour of the real radiation
matrix elements in an exact point-by-point manner but in a two-to-two point
manner.  This is due to the fact that the angular correlations which spoil
the convergence are averaged out when a single collinear phase space point
is combined with another single collinear point which differs from the
original by a $\pi/2$ rotation of the collinear pair around the collinear
axis.  A thorough discussion of this issue can be found in
\cite{Glover:2010im}.  In the histogram of Fig.\ref{fig.SCff} the
aforementioned angular averaging has been performed.


\subsubsection{Initial-final single collinear limits}
\begin{figure}[t]
\begin{center}  
\subfigure[]{
\label{fig.sketchscif}
\resizebox{0.35\linewidth}{!}{
\includegraphics{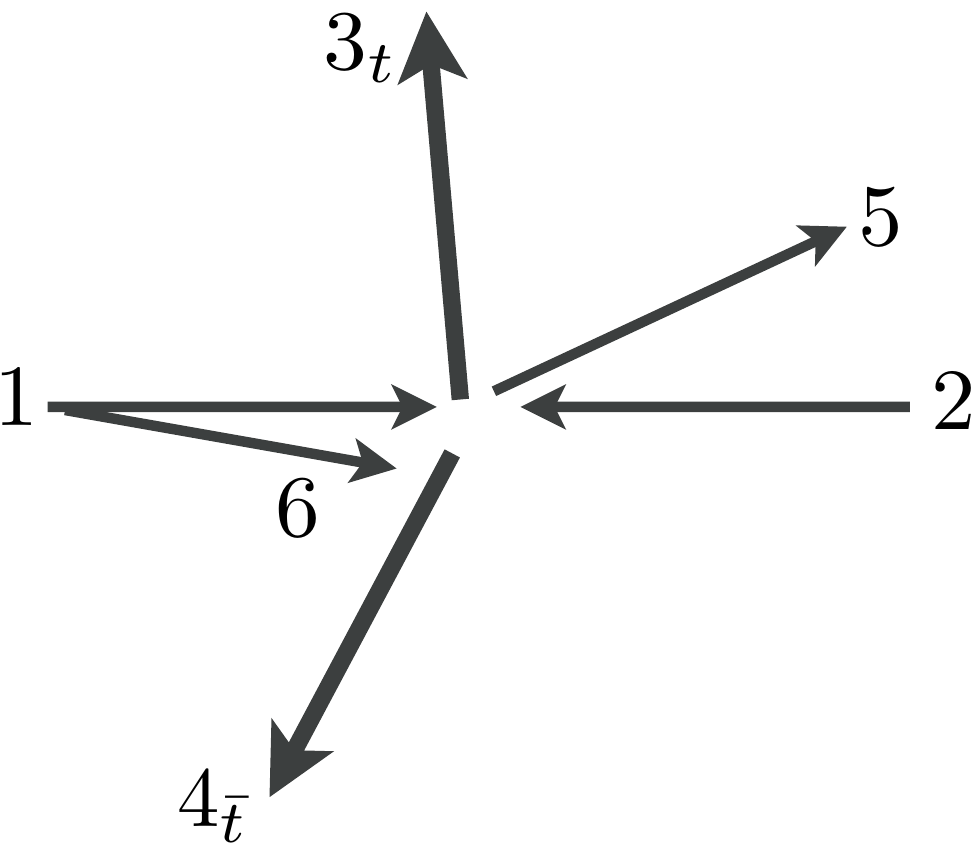}}}
\subfigure[]{
\label{fig.SCif}
\resizebox{0.6\linewidth}{!}{
\includegraphics{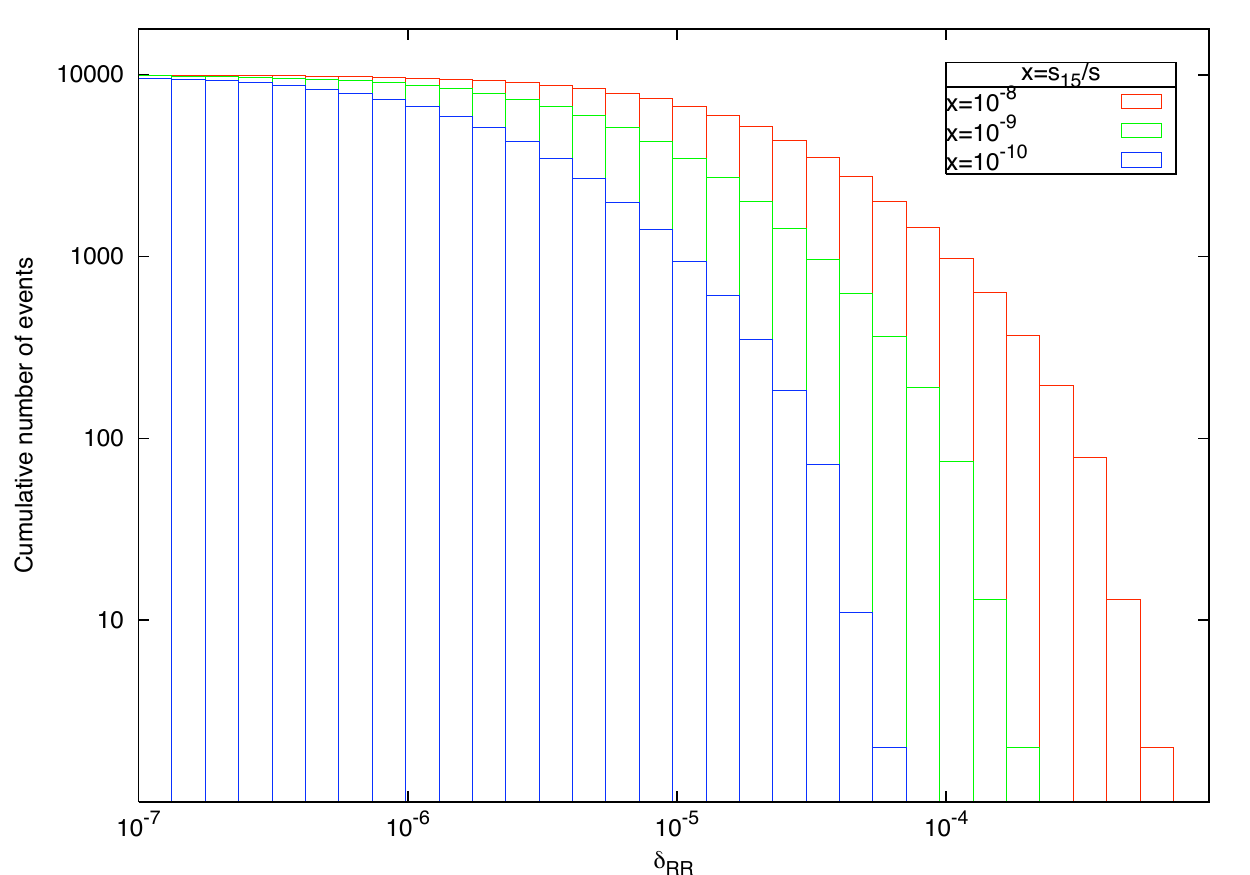}}}
\caption{(a) Sketch of an initial-final single collinear limit. (b) Cumulative distributions of $\delta_{\rm RR}$ for $10^4$ initial-final single collinear phase space points with three different values of $x$}
\end{center}
\end{figure}

The topology of the single initial-final collinear events is illustrated in Fig.\ref{fig.sketchscif}, and the corresponding control variable is defined analogously to the final-final case. There are four different collinear limits in the partonic process $q\bar{q}\rightarrow Q\bar{Q}gg$, namely $p_i||p_j$ with $i=1,2$ and $j=5,6$.
Fig.~\ref{fig.SCif} contains our results for the limit $p_1||p_6$, which
clearly show that the subtraction terms correctly approximate the double
real radiation contributions in this limit.  
The singularity in Fig.~\ref{fig.SCif} is parametrised in terms of $x=s_{16}/s$, and
similar histograms are obtained
for the other three limits of this kind.


\subsection{Tests of the real-virtual contributions}

In this section we study the cancellation between the real-virtual matrix
elements and the corresponding subtraction terms.  Due to the lower
multiplicity of the $2\to 3$ final state and the fact that the heavy quark
mass regulates all final-final single collinear limits, the singular
structure of the real-virtual contributions is simpler than that of the
double real pieces.  Indeed, only the soft limit $p_5\to 0$ and the
initial-final collinear limits $p_i||p_5$ ($i=1,2$) must be considered.

The real-virtual cancellations provide a strong check both of the correctness
of the subtraction terms presented in section \ref{sec.RVsub} and of the
numerical stability of the \OpenLoops  amplitudes discussed in
section \ref{sec.openloops}.  In the vicinity of the soft and collinear
singularities matrix elements and subtraction terms are strongly enhanced,
and the cancellation can amount to several digits.  While this requires
augmented numerical accuracy in the unsubtracted amplitudes, numerical
instabilities related to Gram determinants can be strongly amplified in the
vicinity of the singularities.  It is thus crucial to prevent that the
infrared cancellations are spoiled by numerical instabilities of the 
amplitudes.
To this end, \OpenLoops implements an instability trigger, which monitors
the numerical accuracy of the results by means of a scaling test.  The
 amplitudes are evaluated a second time by rescaling all
dimensionful input parameters by a constant factor $\xi$, and the output is
rescaled back by a factor $\xi^{-d}$ depending on its mass dimension $d$.  The
agreement with the original matrix element serves as an accuracy estimate,
and phase-space points that are not sufficiently stable are automatically
reevaluated with a rescue system. 
Results presented in the following 
have been obtained with \Cuttools as a reduction back end of 
\OpenLoops, using the quadruple-precision mode of \Cuttools as a rescue system 
for unstable points.
Matrix elements are first evaluated in double precision and 
are reevaluated in 
quadruple precision if their estimated double-precision accuracy 
is less than 3 correct digits or smaller than the observed cancellation
$\delta_{\RV}$ with the subtraction term. The stability of the 
quadruple-precision output is assessed with an additional scaling test.
Due to the fact that the scaling test tends to overestimate 
the accuracy, following a universal distribution, one must demand for an 
accuracy which is higher than the cancellation by about the width of this 
distribution. For calibration we determine the width from double precision 
scalings, using a quadruple precision result as reference point, finding a width 
of around one decimal digit. If needed, the accuracy estimate can be improved 
using multiple scalings.

\begin{figure}[t]
  \begin{center}
    \subfigure[]{
      \label{fig.sketchsoftrv}
      \resizebox{0.35\linewidth}{!}{\includegraphics{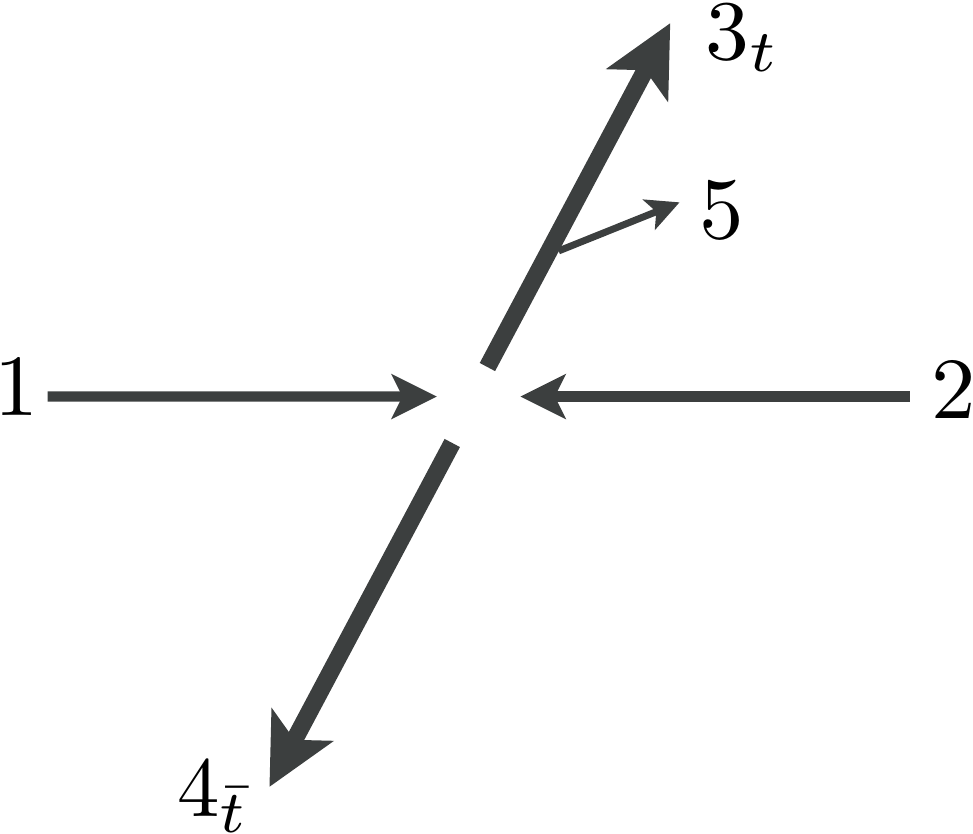}}
    }
    \subfigure[]{
      \label{fig.SoftRV}
      \resizebox{0.6\linewidth}{!}{\includegraphics{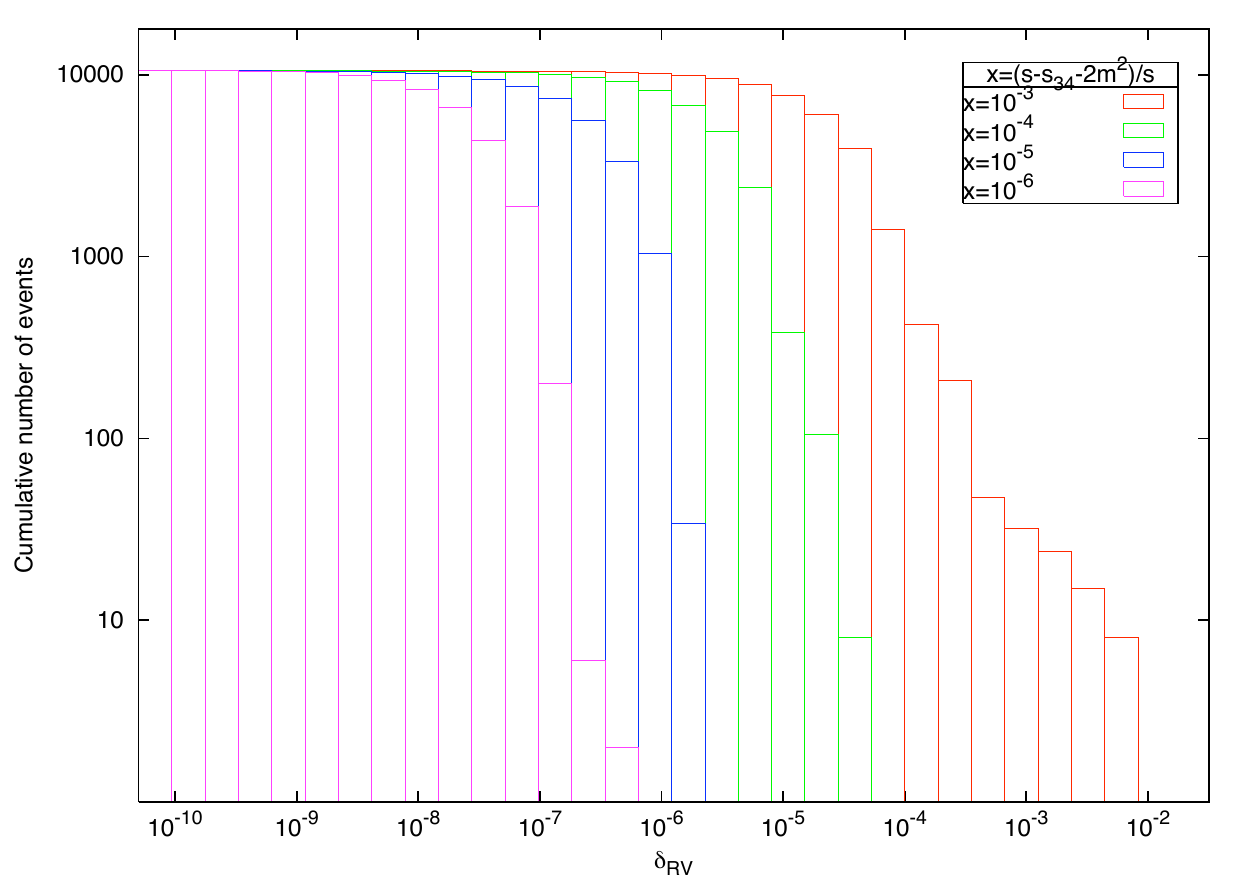}}
    }
    \caption{(a) Sketch of soft event limit. (b) Distribution of $R$ for $10^4$
             soft phase space points with three different values of $x$}
  \end{center}
\end{figure}

\begin{figure}[t]
  \begin{center}
    \subfigure[]{
      \label{fig.sketchcollrv}
      \resizebox{0.35\linewidth}{!}{\includegraphics{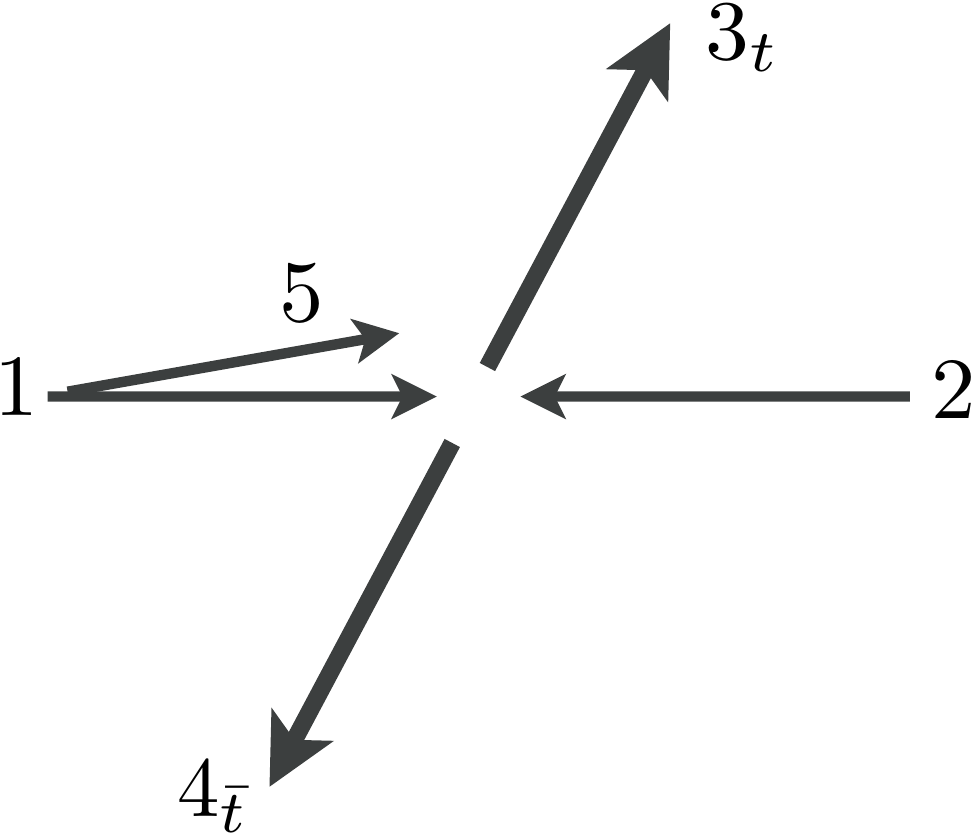}}
    }
    \subfigure[]{
      \label{fig.CollRV}
      \resizebox{0.6\linewidth}{!}{\includegraphics{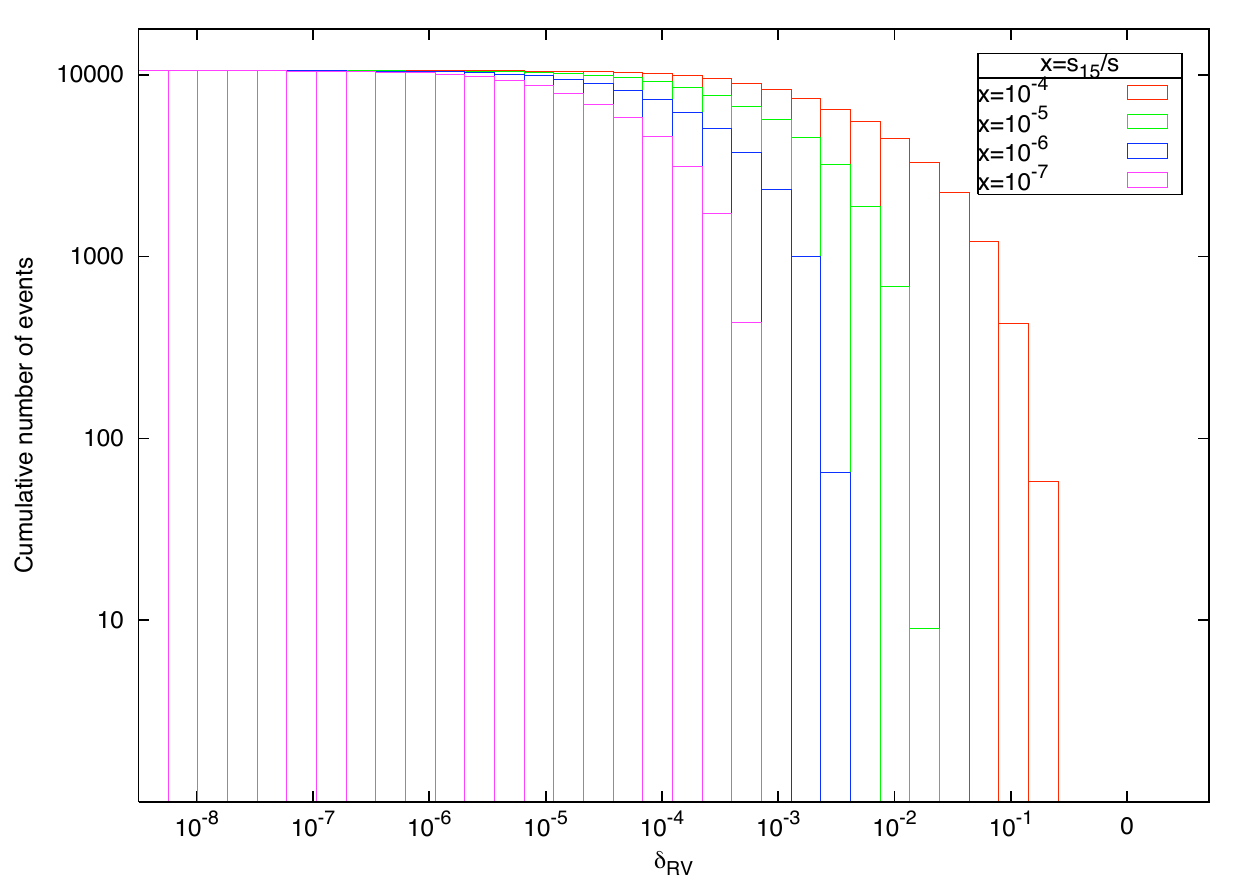}}
    }
    \caption{(a) Sketch of collinear event limit. (b) Distribution of $R$ for
             $10^4$ collinear phase space points with three different values of $x$}
  \end{center}
\end{figure}

Figure \ref{fig.SoftRV} shows the degree of cancellation $\delta_{\RV}$ in
the soft region for samples of $10^4$ phase space points for several values
of the control variable $x=(s-s_{34}-2m_Q^2)/s$, which describes the
softness of the phase space points.  As the singularity is approached with
smaller values of $x$, the subtraction term
$\ds^{\rT}_{q\bar{q},\NNLO,N_c^2}$ converges  to the real-virtual corrections
$\ds^{\RV}_{q\bar{q},\NNLO,N_c^2}$ as expected.  Similarly, Figure \ref{fig.CollRV}
demonstrates the consistency of the cancellation in the collinear region,
parametrised by the control variable $x=s_{15}/s$.

For what concerns the numerical stability of the  matrix elements,
in the collinear region it turns out that for the depicted values of the
control variable, double precision provides sufficient stability (in the sense of
the criterion described above) for the vast majority of the phase space
points.  This also holds in the soft regions with $x=10^{-3}$ and
$x=10^{-4}$.  Starting at the soft sample with $x=10^{-5}$, a sizable
fraction of the matrix elements must be evaluated in quadruple precision. 
However, it should be pointed out that this deep infrared region
($x=10^{-5}$  corresponds to a gluon energy around 5 MeV) is not 
relevant for physical applications based on antenna subtraction.  In fact, as
will be shown in section \ref{sec.integration}, double precision 
results are sufficiently stable to obtain integrated
cross sections with permil level accuracy.

Detailed findings on the numerical stability 
and the reliability of the trigger system are
summarised in Table \ref{tab.stab.trigger}.
The trigger system to detect unstable points from scalings can lead to false 
positive results, meaning that points will be evaluated in quadruple precision 
although they were actually stable enough. This is a side effect of avoiding 
false negative results, meaning points which are regarded as stable although 
they are not. 
Note that in the $x=10^{-6}$ soft sample even quadruple precision is no more 
enough to observe full cancellation for all points, and $\mathcal{O}(5\%)$ of the 
points are tagged as unstable. This shows in the tail of the corresponding 
distribution in Figure \ref{fig.SoftRV}, where the two bins around $x=10^{-6.5}$ 
are populated only by unstable points.

\begin{table}
  \begin{center}
    \begin{tabular}{|l|rrr|}\hline
      sample                & unstable & triggered & false negative \\ \hline
      soft      $x=10^{-3}$ & 0.0004 & 0.0009 & 0      \\
      soft      $x=10^{-4}$ & 0.008  & 0.06   & 0.0001 \\
      soft      $x=10^{-5}$ & 0.31   & 0.70   & 0.002  \\
      soft      $x=10^{-6}$ & 0.96   & 1      & 0.001  \\
      collinear $x=10^{-4}$ & 0      & 0      & 0      \\
      collinear $x=10^{-5}$ & 0      & 0      & 0      \\
      collinear $x=10^{-6}$ & 0.0001 & 0.0008 & 0      \\
      collinear $x=10^{-7}$ & 0.009  & 0.12   & 0.0002 \\ \hline
    \end{tabular}
    \caption{For the samples of phase space points of Figs.~\ref{fig.SoftRV} and 
             \ref{fig.CollRV} the fraction of points is shown which are are unstable in 
             double precision (``unstable''), meaning that the accuracy is not high enough to 
             observe full cancellation between matrix element and subtraction term. 
             ``triggered'' is the fraction of points which is detected as unstable by the 
             trigger system described in the text, and subsequently evaluated in quadruple 
             precision, and ``false negative'' is the fraction of points which are unstable, 
             but not triggered.}
    \label{tab.stab.trigger}
  \end{center}
\end{table}


\section{Stability of the integration over the three-particle phase space}
\label{sec.integration}

As a further and more realistic test of the stability of the 
real-virtual matrix elements and of the related subtraction terms
we have integrated the 
difference ${\rm d}\sigma^{\RV}_{q\bar{q},\NNLO,N_c^2}-{\rm 
d}\sigma^{\rT}_{q\bar{q},\NNLO,N_c^2}$ inclusively over the three-particle phase 
space employing a parton level event generator. In this integration, we impose a 
technical cut on the gluon $p_T$ using the control variable $y_{\rm 
cut}=p_T^{g}/\sqrt{\hat{s}}$, in such a way that no events are generated too 
close to the soft and collinear singularities. Naturally, since the entire phase 
space ought to be covered in the integration, $y_{\rm cut}$ must be taken small. 
While the unsubtracted ${\rm d}\sigma^{\RV}$ contribution
would lead to a logarithmic divergence
in the limit $y_{\rm cut}\to 0$, the subtraction term
guarantees a smooth convergence 
at small $y_{\rm cut}$. In practice the integral should reach a plateau 
for a sufficiently small value of the cut, $y_{\rm cut}^{\rm max}$, 
i.e. for any $y_{\rm 
cut}<y_{\rm cut}^{\rm max}$ the integral of ${\rm 
d}\sigma^{\RV}-{\rm 
d}\sigma^{\rT}$ should remain stable within Monte Carlo integration errors.
\begin{figure}[t]
  \begin{center}
    \label{fig.ycuttest}
    \includegraphics{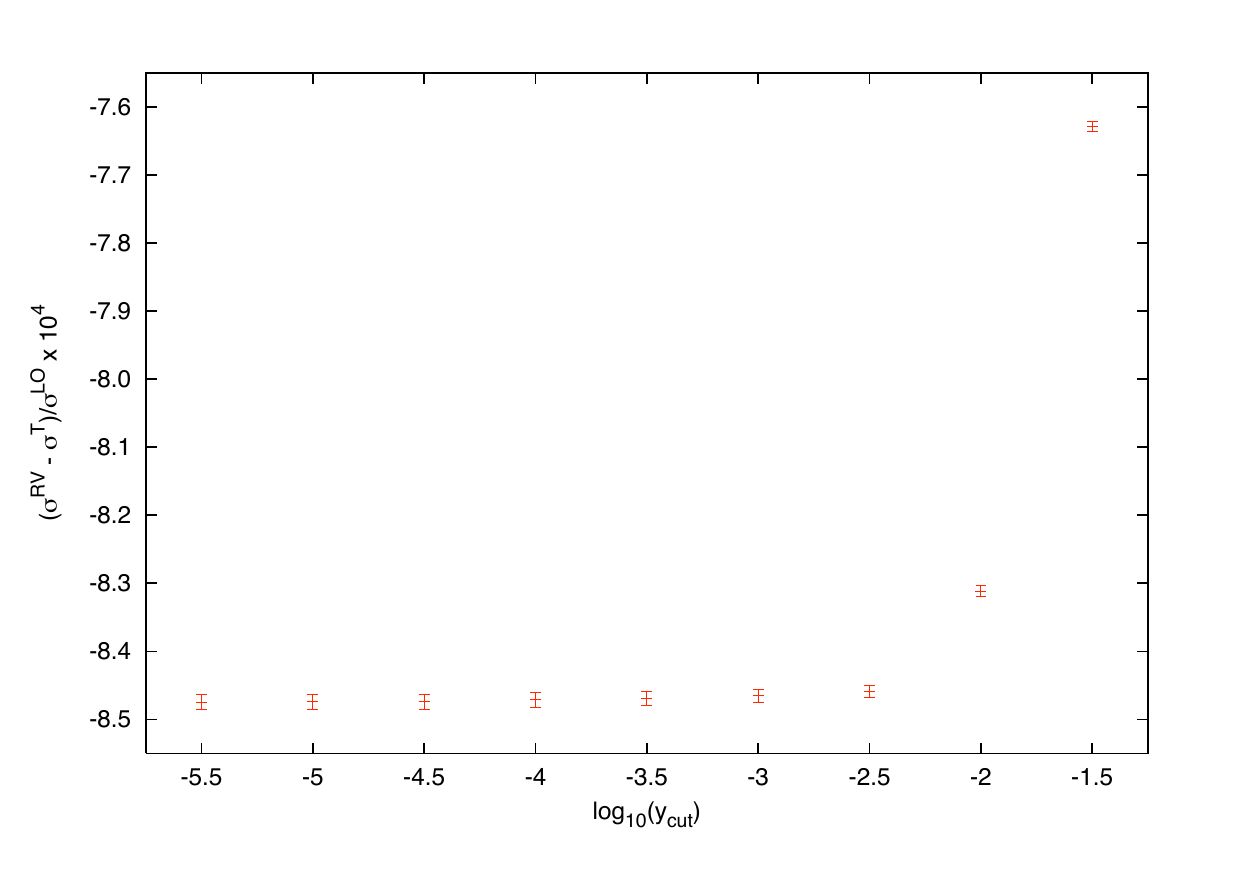}
    \caption{Inclusive phase space integral of
             ${\rm d}\sigma^{\RV}_{q\bar{q},\NNLO,N_c^2}-{\rm d}\sigma^{\rT}_{q\bar{q},\NNLO,N_c^2}$
             normalised to $\sigma_{q\bar{q},LO}$ for different values of $y_{\rm cut}$.
             The error bars correspond to the Monte Carlo uncertainty.}
  \end{center}
\end{figure}
This is clearly confirmed in Figure~\ref{fig.ycuttest}, where we plot the ratio 
\beq
  \frac{\sigma^{\RV}_{q\bar{q},\NNLO,N_c^2}-\sigma^{\rT}_{q\bar{q},\NNLO,N_c^2}}{\sigma_{q\bar{q},LO}}
\eeq
for $pp\to t\bar{t}$ as a function of $y_{\rm cut}$. For both the NNLO 
real-virtual subtracted contributions and the LO normalisation we used 
$\sqrt{s}=7\,{\rm TeV}$, $m_{t}=174.3\,{\rm GeV}$ and set the renormalisation 
and factorisation scales to $\mu_R=\mu_F=m_t$. We employed the MSTW2008nnlo90cl 
and MSTW2008lo90cl PDF sets for the NNLO and the LO contributions respectively. 
The high stability of the integration results for values of $y_{\rm cut}$ below $y_{\rm 
cut}^{max}\sim10^{-3}$ provides solid evidence of the correctness of the real-virtual subtraction terms 
of eq.(\ref{eq.dstcomplete}). 
Moreover, using \OpenLoops in combination with \Cuttools, 
it turns out that the stability plateau is reached before 
encountering significant instabilities in double precision.
For $y_{\rm cut}=10^{-3}(10^{-4})$ we find that only $1$ out of $10^5(10^4)$ events requires a quadruple 
precision reevaluation. This allows for a highly efficient evaluation of the
real-virtual contributions based on double precision 
for the vast majority of the phase space points.


\section{Summary and outlook}\label{sec.conclusions}

In this paper, we presented the double real and real-virtual NNLO
contributions to hadronic $t \bar{t}$ production in the quark-antiquark
annihilation channel.  The computation is performed in leading colour
approximation using the antenna subtraction method, which was extended to
deal with the presence of a massive fermion pair in the final state.  
The real-real subtraction terms, presented in section \ref{sec.RR}, approximate 
the corresponding $2\to 4$ tree matrix elements in all single and double 
unresolved limits, while
the real-virtual subtraction terms, presented in section \ref{sec.RVsub}, 
remove the explicit infrared poles present in the $2\to 3$ 
one-loop matrix elements, as well as the implicit
singularities that occur in the soft and collinear limits.  
The relevant new tree-level four-parton and
three-parton massive initial-final antennae functions, together with their
unresolved counterparts, have been derived in sections \ref{sec.ant4} and 
\ref{sec:A31loop}.

The correctness of the subtraction and its numerical stability have been
demonstrated with detailed cancellation checks in section \ref{sec.tests}.  To this end,
the convergence of the subtracted real-real and real-virtual contributions
was studied by means of event samples generated in several phase space
slices with increasingly small distance from all relevant single and
double-unresolved limits.

To compute the one-loop $q\bar q\to t\bar t g$ real-virtual contributions we
used \OpenLoops in combination with the \Cuttools implementation of OPP
reduction.  This provides interesting insights into the potential
benefits of new automated one-loop generators in the framework of NNLO
calculations.  While the high CPU speed of \OpenLoops represents an obvious
attractive feature, numerical instabilities could represent a
very serious issue for NNLO applications.  In fact,
while the strong cancellations between one-loop amplitudes
and related subtraction terms call for augmented numerical accuracy 
in the soft and collinear regions,
the typical Gram-determinant instabilities of one-loop amplitudes 
tend to be strongly enhanced in the infrared
regions.  It is thus important to make sure that the infrared 
subtractions are not spoiled by numerical instabilities of the one-loop
matrix elements.  To this end, using scaling tests as well as the 
quadruple precision mode of the \Cuttools library, 
we performed detailed studies of the interplay between 
one-loop instabilities and infrared cancellations.
On the one hand, it turns out that quadruple precision is
essential (and at some point even insufficient) 
to avoid excessive numerical instabilities in the deep infrared regime.
On the other hand, we found that such instabilities 
arise only at very small gluon energies and are 
essentially irrelevant for an  NNLO calculation
based on antenna subtraction.
In particular, using a realistic infrared cut-off, 
one-loop amplitudes in double precision turn out to be
sufficiently stable  for the vast majority (more than 99.99\%) 
of the phase space points.
The fact that quadruple precision can be avoided almost completely
implies a drastic efficiency improvement 
for the integration of the real-virtual NNLO contributions.

In order to complete the NNLO corrections to top-antitop production in the quark-antiquark channel at leading colour, the 2-parton contributions $\ds_{\NNLO}^{\VV}$ and its corresponding counterterm  
$\ds_{\NNLO}^{\rU}$ need to be added to the 3 and 4-parton contributions ( $\ds_{\NNLO}^{\RV} $ and $\ds_{\NNLO}^{RR}$) and their corresponding subtraction term $\ds_{\NNLO}^{\rT}$ and $\ds_{\NNLO}^{\rS}$ derived in this paper. The 2-loop contributions participating in  $\ds_{\NNLO}^{\VV}$ are known. 
However $\ds_{\NNLO}^{\rU}$ is  presently unknown. In it, essential unknown ingredients are the integrated 4-parton tree level  antenna  $A_4^0(\Q{1},\gl{3},\gl{4},\qi{2})$ and the integrated one-loop
antenna  $A_3^{1,lc} (\Q{1},\gl{3},\qi{2})$ which have been presented in unintegrated form together with their infrared limits for the first time in this paper. 

The results presented in this paper constitute a major step towards 
a complete NNLO calculation, based on antenna subtraction,
of top-pair production in the quark-antiquark channel.
Our final goal is then the construction of an NNLO parton-level 
event generator for the two, three and four partonic contributions, 
which  will be applicable to any fully differential observable at 
hadron colliders.

\acknowledgments
We are grateful to S.~Dittmaier and J. Pires for many useful discussions. 
This research was supported by the Swiss National Science Foundation
(SNF) under contracts PP00P2-139192, PP00P2-128552, and PBEZP2-145917 
and in part by the European
Commission through the 'LHCPhenoNet' (PITN-GA-2010-264564),
'HiggsTools' (PITN--GA--2012--316704) Initial Training Networks and the ERC Advanced Grant 
'MC@NNLO' (340983),
which are hereby acknowledged.

\appendix


\section{Appendix A: Single unresolved tree-level universal factors}\label{sec.unresolvedfactors}
In single unresolved limits, tree-level colour-ordered matrix elements squared and antenna functions yield universal single unresolved factors. These well-known universal factors associated to collinear limits are Altarelli-Parisi splitting functions \cite{Altarelli:1977zs}, and those occurring in soft limits are soft eikonal factors. Those needed in the context of this paper are given below.


\subsection{The collinear splitting functions}
In this paper we have considered the collinear radiation emitted from a massive fermion to be regulated by the mass of this fermion. 
Consequently, we shall here restrict ourselves to present the usual massless Altarelli-Parisi splitting functions arising in collinear configurations  involving only massless partons. 

When a pair of massless final state particles $i$ and $j$ with momentum $p_i$ and $p_j$ become collinear and cluster into a parent parton of momentum $p_k=p_i+p_j$ the kinematics of the collinear configuration can be described as 
\beq\label{eq.zff}
p_i\rightarrow z\,p_k\hspace{1in} p_j\rightarrow (1-z)p_k,
\eeq
where $z$ is the momentum fraction of one of the partons in the collinear pair. The specific form of the splitting function depends on the species of partons $i$ and $j$. There are three different splitting functions, corresponding to the three possible final-final parton-parton splittings. In conventional dimensional regularisation, they are given by
\beqa
&&\hspace{-0.3in} P_{qg\rightarrow q}(z)=\frac{1+(1-z)^2-\e z^2}{z}\phantom{\bigg[}\label{eq.splitting1}\\
&&\hspace{-0.3in} P_{q\bar{q}\rightarrow g}(z)=\frac{z^2+(1-z)^2-\e}{1-\e}\phantom{\bigg[}\label{eq.splitting2}\\
&&\hspace{-0.3in} P_{gg\rightarrow g}(z)=2\left[\frac{z}{1-z}+\frac{1-z}{z}+z(1-z)\right].\phantom{\bigg[}\label{eq.splitting3}
\eeqa

When one of the collinear particles is in the initial state, the kinematics of the collinear limit can be described as
\beq\label{eq.zif}
p_j\rightarrow z\,p_i\hspace{1in}p_k\rightarrow (1-z)p_i,
\eeq
and the four splitting functions corresponding to the four different parton-parton splittings read
\beqa
&&\hspace{-0.3in} P_{q\hat{q}\rightarrow g}(z)=\frac{1+z^2-\e(1-z)^2}{(1-\e)(1-z)^2}=\frac{1}{1-z}\frac{1}{1-\e}P_{qg\rightarrow q}(1-z)\phantom{\bigg[}\label{eq.splitting4}\\
&&\hspace{-0.3in} P_{\hat{q}g\rightarrow \hat{q}}(z)=\frac{1+(1-z)^2-\e z^2}{z(1-z)}=\frac{1}{1-z}P_{qg\rightarrow q}(z)\phantom{\bigg[}\label{eq.splitting5}\\
&&\hspace{-0.3in} P_{q\hat{g}\rightarrow \hat{\bar{q}}}(z)=\frac{z^2+(1-z)^2-\e}{1-z}=\frac{1-\e}{1-z}P_{q\bar{q}\rightarrow g}(z)\phantom{\bigg[}\label{eq.splitting6}\\
&&\hspace{-0.3in} P_{g\hat{g}\rightarrow \hat{g}}(z)=\frac{2(1-z+z^2)^2}{z(1-z)^2}=\frac{1}{1-z}P_{gg\rightarrow g}(z).\phantom{\bigg[}\label{eq.splitting7}
\eeqa
The additional factors $(1-\e)$ and $1/(1-\e)$ account for the different number of polarizations of quark and gluons in the cases in which the particle entering the hard processes changes its type. The antiquark-gluon splitting functions are identical to the quark-gluon ones due to the invariance of the splitting under charge conjugation. In this paper, only the latter splitting functions arising in initial-final collinear configurations are employed.


\subsection{The massive soft eikonal factor}\label{sec.softfactor}
When a gluon with momentum $p_j$ becomes soft in a colour-ordered tree-level amplitude where it is colour connected to partons $i$ and $k$ with masses $m_i$ and $m_k$ respectively, the associated soft factor is given by \cite{Abelof:2011jv,GehrmannDeRidder:2009fz}
\beq\label{eq.eikonalmassive}
S_{ijk}(m_i,m_k)=\frac{2s_{ik}}{s_{ij}s_{jk}}-\frac{2m_i^2}{s_{ij}^2}-\frac{2m_k^2}{s_{jk}^2}.
\eeq
When $m_i=m_k=0$ this factor reduces to the usual massless soft eikonal factor.


\section{Appendix B: Colour-ordered infrared singularity operators}\label{sec.iones}
The explicit pole structure of colour-ordered  matrix elements can be written in terms of colour-ordered infrared singularity operators $\ione{i}{j}$. Within the antenna subtraction method, the pole part of  antennae as well as that of integrated tree-level three-parton antennae can be also captured by these operators.

If only massless particles are involved, the following set 
of
operators is sufficient (in addition to the  splitting kernels $\Gamma^{(1)}_{ij}(x)$) to express the pole structure of a QCD  amplitude as well as that of a one-particle inclusive integral of a tree-level amplitude \cite{Daleo:2006xa,GehrmannDeRidder:2005cm}:
\beqa
&&\ione{q}{\bar{q}}(\e,s_{q\bar{q}})=-\frac{e^{\e\gamma_E}}{2\Gamma(1-\e)}\left(\frac{|s_{q\bar{q}}|}{\mu^2}\right)^{-\e}\left[  \frac{1}{\e^2}+\frac{3}{2\e} \right]\\
&&\ione{q}{g}(\e,s_{qg})=-\frac{e^{\e\gamma_E}}{2\Gamma(1-\e)}\left(\frac{|s_{qg}|}{\mu^2}\right)^{-\e}\left[  \frac{1}{\e^2}+\frac{5}{3\e} \right]\\
&&\ione{g}{g}(\e,s_{gg})=-\frac{e^{\e\gamma_E}}{2\Gamma(1-\e)}\left(\frac{|s_{gg}|}{\mu^2}\right)^{-\e}\left[  \frac{1}{\e^2}+\frac{11}{6\e} \right]\\
&&\ione{q}{g,F}(\e,s_{qg})=\frac{e^{\e\gamma_E}}{2\Gamma(1-\e)}\left(\frac{|s_{qg}|}{\mu^2}\right)^{-\e}\frac{1}{6\e}\label{eq.ioneqgf}\\ 
&&\ione{g}{g,F}(\e,s_{gg})=\frac{e^{\e\gamma_E}}{2\Gamma(1-\e)}\left(\frac{|s_{gg}|}{\mu^2}\right)^{-\e}\frac{1}{3\e}.
\eeqa
When massive fermions denoted by  $Q$ of mass $m_{Q}$  are involved, the following operators must also be considered \cite{Abelof:2011jv}
\beqa
&&\ione{Q}{\bar{Q}}(\e,s_{Q\bar{Q}})=-\frac{e^{\e\gamma_E}}{2\Gamma(1-\e)}\left(\frac{|s_{Q\bar{Q}}|}{\mu^2}\right)^{-\e}\left[ \frac{1}{\e}\left(1-\frac{1+r_0}{2\sqrt{r_0}}\ln\left(\frac{1+\sqrt{r_0}}{1-\sqrt{r_0}} \right)\right)\right]\\
&&\ione{Q}{\bar{q}}(\e,s_{Q\bar{q}})=-\frac{e^{\e\gamma_E}}{2\Gamma(1-\e)}\left(\frac{|s_{Q\bar{q}}|}{\mu^2}\right)^{-\e}\left[ \frac{1}{2\e^2}+\frac{5}{4\e}+\frac{1}{2\e}\ln\left( \frac{m_Q^2}{|s_{Q\bar{q}}|}\right)\right]\\
&&\ione{Q}{g}(\e,s_{Qg})=-\frac{e^{\e\gamma_E}}{2\Gamma(1-\e)}\left(\frac{|s_{Qg}|}{\mu^2}\right)^{-\e}\left[ \frac{1}{2\e^2}+\frac{17}{12\e}+\frac{1}{2\e}\ln\left( \frac{m_Q^2}{|s_{Q\bar{g}}|}\right)\right]\\
&&\ione{Q}{g,F}(\e,s_{Qg})=\frac{e^{\e\gamma_E}}{2\Gamma(1-\e)}\left(\frac{|s_{Qg}|}{\mu^2}\right)^{-\e} \frac{1}{6\e},\phantom{ \left( \frac{m_Q^2}{|s_{Q\bar{g}}|}\right)}\label{eq.ioneQgf}
\eeqa
with
\beq
r_0=1-\frac{4m_Q^2}{s_{Q\bar{Q}}+2m_Q^2}.
\eeq


\section{Appendix C: The complete expression of $A_4^0(\Q{1},\gl{3},\gl{4},\qi{2})$}\label{sec.a04}
The full expression of the initial-final flavour-violating antenna $A_4^0(\Q{1},\gl{3},\gl{4},\qi{2})$ discussed in section \ref{sec.ant4} is given by
\beqa
&&\hspace{-0.2in}A_4^0(\Q{1},\gl{3},\gl{4},\qi{2})=\frac{1}{(Q^2+m_Q^2)}\bigg[-\frac{2}{s_{24} s_{34}}\left(2 s_{12}+s_{13}-s_{134}-s_{234}\right)\nonumber\\
&&+\frac{1}{s_{13} s_{134}}\left(-9 s_{12}+5 s_{234}+3 s_{24}-8 s_{34}\right)+\frac{1}{s_{134}^2}\left(3 s_{12}+4 s_{13}-3 s_{234}+3 s_{34}\right)\nonumber\\
&&+\frac{1}{s_{13} s_{234}}\left(-8 s_{12}+5 s_{134}+5s_{24}-7 s_{34}\right)+\frac{1}{s_{234}^2}\left(3s_{12}-3 s_{134}-4 s_{24}+3 s_{34}\right)\nonumber\\
&&+\frac{1}{s_{13} s_{24}}\left(-2 s_{12}+s_{134}+s_{234}-4 s_{34}\right)+\frac{1}{s_{24}s_{234}^2}\left(-s_{34}^2-s_{12} s_{34}+s_{134} s_{34}\right)\nonumber\\
&&+\frac{1}{s_{24}s_{134}}\left(8 s_{12}+5 s_{13}-5 s_{234}+7 s_{34}\right)+\frac{1}{s_{13}s_{134}^2}\left(s_{34}^2+s_{12} s_{34}-s_{234} s_{34}\right)\nonumber\\
&&+\frac{1}{s_{24}s_{234}}\left(9 s_{12}+3 s_{13}-5 s_{134}+8 s_{34}\right)+\frac{2}{s_{34}s_{134}^2}\left(s_{13}^2+2 s_{12} s_{13}-2 s_{234} s_{13}\right)\nonumber\\
&&+\frac{1}{s_{134} s_{234}}\left(10 s_{12}+7 s_{13}-7 s_{24}+8 s_{34}\right)+\frac{9}{s_{13}}-\frac{17}{s_{134}}-\frac{17}{s_{234}}-\frac{9}{s_{24}}\nonumber\\
&&+\frac{1}{s_{134} s_{234} s_{24}}\left(-4 s_{12}^2-3 s_{13} s_{12}-6 s_{34} s_{12}-s_{13}^2-3 s_{34}^2-3 s_{13}s_{34}\right)\nonumber\\
&&+\frac{1}{s_{13} s_{24} s_{234}}\left(4 s_{12}^2-3 s_{134} s_{12}+6 s_{34} s_{12}+s_{134}^2+3 s_{34}^2-3 s_{134} s_{34}\right)\nonumber\\
&&+\frac{1}{s_{13} s_{24} s_{134}}\left(4 s_{12}^2-3 s_{234} s_{12}+6s_{34} s_{12}+s_{234}^2+3 s_{34}^2-3 s_{234} s_{34}\right)\nonumber\\
&&+\frac{1}{s_{13} s_{134} s_{234}}\left(4 s_{12}^2-3 s_{24} s_{12}+6 s_{34} s_{12}+s_{24}^2+3 s_{34}^2-3 s_{24}s_{34}\right)\nonumber\\
&&+\frac{1}{s_{13}s_{24} s_{134} s_{234}}\left(-2 s_{12}^3-4 s_{34} s_{12}^2-3 s_{34}^2 s_{12}-s_{34}^3\right)\nonumber\\
&&+\frac{2}{s_{13} s_{34}}\left(2 s_{12}-s_{134}-s_{234}-s_{24}\right)-\frac{1}{s_{34}s_{134}}\left(14 s_{12}+12 s_{13}-9 s_{234}-7 s_{24}\right)\nonumber\\
&&-\frac{1}{ s_{34}s_{234}}\left(14 s_{12}+7s_{13}-9 s_{134}-12 s_{24}\right)+\frac{2}{ s_{34}^2s_{134}^2} \left(s_{12} s_{13}^2-s_{13}^2s_{234}\right)\nonumber\\
&&+\frac{2}{s_{34}s_{234}^2}\left(s_{24}^2-2 s_{12} s_{24}+2 s_{134} s_{24}\right)+\frac{2}{ s_{34}^2s_{234}^2} \left(s_{12} s_{24}^2-s_{134} s_{24}^2\right)\nonumber\\
&&+\frac{1}{s_{13} s_{34} s_{234}}\left(-2 s_{12}^2+2 s_{134} s_{12}+2 s_{24} s_{12}-s_{134}^2-s_{24}^2-2 s_{134} s_{24}\right)\nonumber\\
&&+\frac{2}{s_{134} s_{234} s_{34}}\left(4 s_{12}^2+2 s_{13} s_{12}-2 s_{24} s_{12}+s_{13}^2+s_{24}^2-2s_{13} s_{24}\right)\nonumber\\
&&+\frac{1}{s_{13} s_{24} s_{34}}\left(2 s_{12}^2-2 s_{134} s_{12}-2 s_{234} s_{12}+s_{134}^2+s_{234}^2\right)\nonumber\\
&&+\frac{1}{s_{134} s_{24} s_{34}}\left(2 s_{12}^2+2 s_{13} s_{12}-2 s_{234} s_{12}+s_{13}^2+s_{234}^2-2 s_{13} s_{234}\right)\nonumber\\
&&+\frac{14}{s_{34}}+\frac{2}{s_{34}^2} \left(s_{12}+2 s_{13}-s_{134}-s_{234}-2 s_{24}\right)\nonumber\\
&&-\frac{4 s_{12} s_{13} s_{24}}{s_{134} s_{234}s_{34}^2}-\frac{2}{s_{134} s_{34}^2} \left(s_{13}^2+2 s_{12} s_{13}-2 s_{234} s_{13}-2 s_{24} s_{13}\right)\nonumber\\
&&+\frac{2}{s_{234} s_{34}^2} \left(-s_{24}^2+2 s_{12} s_{24}+2 s_{13}s_{24}-2 s_{134} s_{24}\right)\nonumber\\
&&+m_Q^2\bigg(\frac{1}{s_{13} s_{134} s_{24}}\left(-4 s_{12}+8 s_{234}-3 s_{34}\right)-\frac{1}{s_{13} s_{24} s_{134} s_{234}}\left(s_{12} s_{34}+s_{34}^2\right)\nonumber\\
&&+\frac{2}{s_{13}^2 s_{24}} \left(2 s_{12}-s_{134}-2 s_{234}+2 s_{34}\right)+\frac{4}{s_{13} s_{34}s_{134} } \left(s_{12}-s_{234}-s_{24}\right)\nonumber\\
&&+\frac{2}{s_{13}^2 s_{134}} \left(2 s_{12}-2s_{234}-s_{24}+2 s_{34}\right)+\frac{4}{s_{13}  s_{24} s_{34}s_{134}} \left(s_{12} s_{234}-s_{234}^2\right)\nonumber\\
&&-\frac{4}{s_{13}^2 s_{134} s_{24}} \left(-2 s_{12} s_{234}+2 s_{12}s_{34}+s_{12}^2-2 s_{234} s_{34}+s_{234}^2+s_{34}^2\right)+\frac{4}{s_{134}^2}\nonumber\\
&&+\frac{4s_{12} s_{24}}{s_{13} s_{134} s_{234} s_{34}}+\frac{s_{34}-4 s_{12}}{s_{13}s_{134} s_{234}}+\frac{4}{s_{13} s_{134}^2} \left(s_{12}-s_{234}+s_{34}\right)\nonumber\\
&&+\frac{4}{s_{134}^2 s_{34}} \left(s_{12}-s_{234}\right)-\frac{s_{34}}{s_{13}s_{234} s_{24}}-\frac{4 s_{24}}{s_{13} s_{234} s_{34}}-\frac{4 s_{234}}{s_{13} s_{24} s_{34}}+\frac{1}{s_{134} s_{24}}\nonumber\\
&&+\frac{5}{s_{13} s_{234}}+\frac{5}{s_{13} s_{24}}-\frac{4}{s_{13}s_{34}}-\frac{4}{s_{13}^2}-\frac{s_{34}}{s_{134} s_{234} s_{24}}+\frac{2}{s_{134} s_{234}}-\frac{4}{s_{134}s_{34}}\bigg)\nonumber\\
&&+m_Q^4\bigg(\frac{4}{s_{13}^2 s_{134}^2} \left(s_{12}-s_{234}+s_{34}\right)-\frac{4}{s_{13}^2 s_{134}}\bigg)\bigg]+\order{\e},
\eeqa
with $Q^2=-(p_1-p_2+p_3+p_4)^2$, $s_{134}=s_{13}+s_{14}+s_{34}$, and $s_{234}=-s_{23}-s_{24}+s_{34}$, using our convention $s_{ij}=2 p_i\dot p_j$

\bibliography{Main}

\providecommand{\href}[2]{#2}\begingroup\raggedright\begin{thebibliography}{10}

\bibitem{Aad:2012hg}
{ ATLAS Collaboration} , G.~Aad {\em et.~al.}, {\it {Measurements of top quark
  pair relative differential cross-sections with ATLAS in $pp$ collisions at
  $\sqrt{s}=7$ TeV}},  {\em Eur.Phys.J.} {\bf C73} (2013) 2261
  [\href{http://arXiv.org/abs/1207.5644}{{\tt 1207.5644}}].

\bibitem{CMS:fxa}
{ CMS Collaboration} , {\it {Measurement of differential top-quark pair
  production cross sections in the lepton+jets channel in pp collisions at 8
  TeV}}, .

\bibitem{CMS:cxa}
{ CMS Collaboration} , {\it {Measurement of the differential top-quark pair
  production cross section in the dilepton channel in pp collisions at p s = 8
  TeV}}, .

\bibitem{Abelof:2011ap}
G.~Abelof and A.~Gehrmann-De~Ridder, {\it {Double real radiation corrections to
  $t\bar{t}$ production at the LHC: the all-fermion processes}},  {\em JHEP}
  {\bf 1204} (2012) 076 [\href{http://arXiv.org/abs/1112.4736}{{\tt
  1112.4736}}].

\bibitem{Anastasiou:2008vd}
C.~Anastasiou and S.~M. Aybat, {\it {The One-loop gluon amplitude for
  heavy-quark production at NNLO}},  {\em Phys.Rev.} {\bf D78} (2008) 114006
  [\href{http://arXiv.org/abs/0809.1355}{{\tt 0809.1355}}].

\bibitem{Baernreuther:2012ws}
P.~Baernreuther, M.~Czakon and A.~Mitov, {\it {Percent level precision physics
  at the Tevatron: first genuine NNLO QCD corrections to $q \bar{q} \rightarrow
  t \bar{t} + X$}},  \href{http://arXiv.org/abs/1204.5201}{{\tt 1204.5201}}.

\bibitem{Bierenbaum:2011gg}
I.~Bierenbaum, M.~Czakon and A.~Mitov, {\it {The singular behavior of one-loop
  massive QCD amplitudes with one external soft gluon}},  {\em Nucl.Phys.} {\bf
  B856} (2012) 228--246 [\href{http://arXiv.org/abs/1107.4384}{{\tt
  1107.4384}}].

\bibitem{Bonciani:2008az}
R.~Bonciani, A.~Ferroglia, T.~Gehrmann, D.~Maitre and C.~Studerus, {\it
  {Two-Loop Fermionic Corrections to Heavy-Quark Pair Production: The
  Quark-Antiquark Channel}},  {\em JHEP} {\bf 0807} (2008) 129
  [\href{http://arXiv.org/abs/0806.2301}{{\tt 0806.2301}}].

\bibitem{Bonciani:2009nb}
R.~Bonciani, A.~Ferroglia, T.~Gehrmann and C.~Studerus, {\it {Two-Loop Planar
  Corrections to Heavy-Quark Pair Production in the Quark-Antiquark Channel}},
  {\em JHEP} {\bf 0908} (2009) 067 [\href{http://arXiv.org/abs/0906.3671}{{\tt
  0906.3671}}].

\bibitem{Bonciani:2010mn}
R.~Bonciani, A.~Ferroglia, T.~Gehrmann, A.~Manteuffel and C.~Studerus, {\it
  {Two-Loop Leading Color Corrections to Heavy-Quark Pair Production in the
  Gluon Fusion Channel}},  {\em JHEP} {\bf 1101} (2011) 102
  [\href{http://arXiv.org/abs/1011.6661}{{\tt 1011.6661}}].

\bibitem{Czakon:2008zk}
M.~Czakon, {\it {Tops from Light Quarks: Full Mass Dependence at Two-Loops in
  QCD}},  {\em Phys.Lett.} {\bf B664} (2008) 307--314
  [\href{http://arXiv.org/abs/0803.1400}{{\tt 0803.1400}}].

\bibitem{Czakon:2011ve}
M.~Czakon, {\it {Double-real radiation in hadronic top quark pair production as
  a proof of a certain concept}},  {\em Nucl.Phys.} {\bf B849} (2011) 250--295
  [\href{http://arXiv.org/abs/1101.0642}{{\tt 1101.0642}}].

\bibitem{Czakon:2012zr}
M.~Czakon and A.~Mitov, {\it {NNLO corrections to top-pair production at hadron
  colliders: the all-fermionic scattering channels}},
  \href{http://arXiv.org/abs/1207.0236}{{\tt 1207.0236}}.

\bibitem{Kniehl:2008fd}
B.~Kniehl, Z.~Merebashvili, J.~Korner and M.~Rogal, {\it {Heavy quark pair
  production in gluon fusion at next-to-next-to-leading $O(\alpha_s^{4})$
  order: One-loop squared contributions}},  {\em Phys.Rev.} {\bf D78} (2008)
  094013 [\href{http://arXiv.org/abs/0809.3980}{{\tt 0809.3980}}].

\bibitem{Korner:2008bn}
J.~Korner, Z.~Merebashvili and M.~Rogal, {\it {NNLO $O(\alpha_s^{4})$ results
  for heavy quark pair production in quark-antiquark collisions: The One-loop
  squared contributions}},  {\em Phys.Rev.} {\bf D77} (2008) 094011
  [\href{http://arXiv.org/abs/0802.0106}{{\tt 0802.0106}}].

\bibitem{Czakon:2013goa}
M.~Czakon, P.~Fiedler and A.~Mitov, {\it {The total top quark pair production
  cross-section at hadron colliders through ${\cal O} (\alpha_S^4)$}},
  \href{http://arXiv.org/abs/1303.6254}{{\tt 1303.6254}}.

\bibitem{Binoth:2000ps}
T.~Binoth and G.~Heinrich, {\it {An Automatized algorithm to compute infrared
  divergent multiloop integrals}},  {\em Nucl.Phys.} {\bf B585} (2000) 741--759
  [\href{http://arXiv.org/abs/hep-ph/0004013}{{\tt hep-ph/0004013}}].

\bibitem{Anastasiou:2003gr}
C.~Anastasiou, K.~Melnikov and F.~Petriello, {\it {A New method for real
  radiation at NNLO}},  {\em Phys.Rev.} {\bf D69} (2004) 076010
  [\href{http://arXiv.org/abs/hep-ph/0311311}{{\tt hep-ph/0311311}}].

\bibitem{Binoth:2004jv}
T.~Binoth and G.~Heinrich, {\it {Numerical evaluation of phase space integrals
  by sector decomposition}},  {\em Nucl.Phys.} {\bf B693} (2004) 134--148
  [\href{http://arXiv.org/abs/hep-ph/0402265}{{\tt hep-ph/0402265}}].

\bibitem{Anastasiou:2010pw}
C.~Anastasiou, F.~Herzog and A.~Lazopoulos, {\it {On the factorization of
  overlapping singularities at NNLO}},  {\em JHEP} {\bf 1103} (2011) 038
  [\href{http://arXiv.org/abs/1011.4867}{{\tt 1011.4867}}].

\bibitem{Catani:2007vq}
S.~Catani and M.~Grazzini, {\it {An NNLO subtraction formalism in hadron
  collisions and its application to Higgs boson production at the LHC}},  {\em
  Phys.Rev.Lett.} {\bf 98} (2007) 222002
  [\href{http://arXiv.org/abs/hep-ph/0703012}{{\tt hep-ph/0703012}}].

\bibitem{GehrmannDeRidder:2005cm}
A.~Gehrmann-De~Ridder, T.~Gehrmann and N.~Glover, {\it {Antenna subtraction at
  NNLO}},  {\em JHEP} {\bf 0509} (2005) 056
  [\href{http://arXiv.org/abs/hep-ph/0505111}{{\tt hep-ph/0505111}}].

\bibitem{Czakon:2010td}
M.~Czakon, {\it {A novel subtraction scheme for double-real radiation at
  NNLO}},  {\em Phys.Lett.} {\bf B693} (2010) 259--268
  [\href{http://arXiv.org/abs/1005.0274}{{\tt 1005.0274}}].

\bibitem{Boughezal:2011jf}
R.~Boughezal, K.~Melnikov and F.~Petriello, {\it {A subtraction scheme for NNLO
  computations}},  {\em Phys.Rev.} {\bf D85} (2012) 034025
  [\href{http://arXiv.org/abs/1111.7041}{{\tt 1111.7041}}].

\bibitem{Frixione:1997np}
S.~Frixione, {\it {A General approach to jet cross-sections in QCD}},  {\em
  Nucl.Phys.} {\bf B507} (1997) 295--314
  [\href{http://arXiv.org/abs/hep-ph/9706545}{{\tt hep-ph/9706545}}].

\bibitem{Abelof:2012rv}
G.~Abelof and A.~Gehrmann-De~Ridder, {\it {Double real radiation corrections to
  $t\bar{t}$ production at the LHC: the $gg\rightarrow t\bar{t}q\bar{q}$
  channel}},  {\em JHEP} {\bf 1211} (2012) 074
  [\href{http://arXiv.org/abs/1207.6546}{{\tt 1207.6546}}].

\bibitem{Abelof:2011jv}
G.~Abelof and A.~Gehrmann-De~Ridder, {\it {Antenna subtraction for the
  production of heavy particles at hadron colliders}},  {\em JHEP} {\bf 1104}
  (2011) 063 [\href{http://arXiv.org/abs/1102.2443}{{\tt 1102.2443}}].

\bibitem{Catani:2002hc}
S.~Catani, S.~Dittmaier, M.~H. Seymour and Z.~Trocsanyi, {\it {The Dipole
  formalism for next-to-leading order QCD calculations with massive partons}},
  {\em Nucl.Phys.} {\bf B627} (2002) 189--265
  [\href{http://arXiv.org/abs/hep-ph/0201036}{{\tt hep-ph/0201036}}].

\bibitem{Currie:2013dwa}
J.~Currie, A.~Gehrmann-De~Ridder, E.~Glover and J.~Pires, {\it {NNLO QCD
  corrections to jet production at hadron colliders from gluon scattering}},
  \href{http://arXiv.org/abs/1310.3993}{{\tt 1310.3993}}.

\bibitem{Currie:2013vh}
J.~Currie, E.~Glover and S.~Wells, {\it {Infrared Structure at NNLO Using
  Antenna Subtraction}},  {\em JHEP} {\bf 1304} (2013) 066
  [\href{http://arXiv.org/abs/1301.4693}{{\tt 1301.4693}}].

\bibitem{GehrmannDeRidder:2011aa}
A.~Gehrmann-De~Ridder, N.~Glover and J.~Pires, {\it {Real-Virtual corrections
  for gluon scattering at NNLO}},  {\em JHEP} {\bf 1202} (2012) 141
  [\href{http://arXiv.org/abs/1112.3613}{{\tt 1112.3613}}].

\bibitem{GehrmannDeRidder:2012dg}
A.~Gehrmann-De~Ridder, T.~Gehrmann, E.~Glover and J.~Pires, {\it {Double
  Virtual corrections for gluon scattering at NNLO}},  {\em JHEP} {\bf 1302}
  (2013) 026 [\href{http://arXiv.org/abs/1211.2710}{{\tt 1211.2710}}].

\bibitem{Bernreuther:2011jt}
W.~Bernreuther, C.~Bogner and O.~Dekkers, {\it {The real radiation antenna
  function for $S \to Q {\bar Q} q {\bar q}$ at NNLO QCD}},  {\em JHEP} {\bf
  1106} (2011) 032 [\href{http://arXiv.org/abs/1105.0530}{{\tt 1105.0530}}].

\bibitem{Bernreuther:2013uma}
W.~Bernreuther, C.~Bogner and O.~Dekkers, {\it {The real radiation antenna
  functions for $S\rightarrow Q\bar{Q}gg$ at NNLO QCD}},  {\em JHEP} {\bf 1310}
  (2013) 161 [\href{http://arXiv.org/abs/1309.6887}{{\tt 1309.6887}}].

\bibitem{Boughezal:2010mc}
R.~Boughezal, A.~Gehrmann-De~Ridder and M.~Ritzmann, {\it {Antenna subtraction
  at NNLO with hadronic initial states: double real radiation for
  initial-initial configurations with two quark flavours}},  {\em JHEP} {\bf
  1102} (2011) 098 [\href{http://arXiv.org/abs/1011.6631}{{\tt 1011.6631}}].

\bibitem{Daleo:2006xa}
A.~Daleo, T.~Gehrmann and D.~Maitre, {\it {Antenna subtraction with hadronic
  initial states}},  {\em JHEP} {\bf 0704} (2007) 016
  [\href{http://arXiv.org/abs/hep-ph/0612257}{{\tt hep-ph/0612257}}].

\bibitem{Daleo:2009yj}
A.~Daleo, A.~Gehrmann-De~Ridder, T.~Gehrmann and G.~Luisoni, {\it {Antenna
  subtraction at NNLO with hadronic initial states: initial-final
  configurations}},  {\em JHEP} {\bf 1001} (2010) 118
  [\href{http://arXiv.org/abs/0912.0374}{{\tt 0912.0374}}].

\bibitem{Gehrmann:2011wi}
T.~Gehrmann and P.~F. Monni, {\it {Antenna subtraction at NNLO with hadronic
  initial states: real-virtual initial-initial configurations}},  {\em JHEP}
  {\bf 1112} (2011) 049 [\href{http://arXiv.org/abs/1107.4037}{{\tt
  1107.4037}}].

\bibitem{GehrmannDeRidder:2007jk}
A.~Gehrmann-De~Ridder, T.~Gehrmann, N.~Glover and G.~Heinrich, {\it {Infrared
  structure of $e^+ e^- \rightarrow$ 3 jets at NNLO}},  {\em JHEP} {\bf 0711}
  (2007) 058 [\href{http://arXiv.org/abs/0710.0346}{{\tt 0710.0346}}].

\bibitem{GehrmannDeRidder:2009fz}
A.~Gehrmann-De~Ridder and M.~Ritzmann, {\it {NLO Antenna Subtraction with
  Massive Fermions}},  {\em JHEP} {\bf 0907} (2009) 041
  [\href{http://arXiv.org/abs/0904.3297}{{\tt 0904.3297}}].

\bibitem{GehrmannDeRidder:2012ja}
A.~Gehrmann-De~Ridder, T.~Gehrmann and M.~Ritzmann, {\it {Antenna subtraction
  at NNLO with hadronic initial states: double real initial-initial
  configurations}},  {\em JHEP} {\bf 1210} (2012) 047
  [\href{http://arXiv.org/abs/1207.5779}{{\tt 1207.5779}}].

\bibitem{Glover:2010im}
N.~Glover and J.~Pires, {\it {Antenna subtraction for gluon scattering at
  NNLO}},  {\em JHEP} {\bf 1006} (2010) 096
  [\href{http://arXiv.org/abs/1003.2824}{{\tt 1003.2824}}].

\bibitem{GehrmannDeRidder:2013mf}
A.~Gehrmann-De~Ridder, T.~Gehrmann, E.~Glover and J.~Pires, {\it {Second order
  QCD corrections to jet production at hadron colliders: the all-gluon
  contribution}},  \href{http://arXiv.org/abs/1301.7310}{{\tt 1301.7310}}.

\bibitem{Abelof:2012he}
G.~Abelof, O.~Dekkers and A.~Gehrmann-De~Ridder, {\it {Antenna subtraction with
  massive fermions at NNLO: Double real initial-final configurations}},  {\em
  JHEP} {\bf 1212} (2012) 107 [\href{http://arXiv.org/abs/1210.5059}{{\tt
  1210.5059}}].

\bibitem{Cascioli:2011va}
F.~Cascioli, P.~Maierhofer and S.~Pozzorini, {\it {Scattering Amplitudes with
  Open Loops}},  {\em Phys.Rev.Lett.} {\bf 108} (2012) 111601
  [\href{http://arXiv.org/abs/1111.5206}{{\tt 1111.5206}}].

\bibitem{Ossola:2007ax}
G.~Ossola, C.~G. Papadopoulos and R.~Pittau, {\it {CutTools: A Program
  implementing the OPP reduction method to compute one-loop amplitudes}},  {\em
  JHEP} {\bf 0803} (2008) 042 [\href{http://arXiv.org/abs/0711.3596}{{\tt
  0711.3596}}].

\bibitem{Badger:2011yu}
S.~Badger, R.~Sattler and V.~Yundin, {\it {One-Loop Helicity Amplitudes for
  $t\bar{t}$ Production at Hadron Colliders}},  {\em Phys.Rev.} {\bf D83}
  (2011) 074020 [\href{http://arXiv.org/abs/1101.5947}{{\tt 1101.5947}}].

\bibitem{Beenakker:1988bq}
W.~Beenakker, H.~Kuijf, W.~van Neerven and J.~Smith, {\it {QCD Corrections to
  Heavy Quark Production in p anti-p Collisions}},  {\em Phys.Rev.} {\bf D40}
  (1989) 54--82.

\bibitem{Nason:1989zy}
P.~Nason, S.~Dawson and R.~K. Ellis, {\it {The One Particle Inclusive
  Differential Cross-Section for Heavy Quark Production in Hadronic
  Collisions}},  {\em Nucl.Phys.} {\bf B327} (1989) 49--92.

\bibitem{Catani:2000ef}
S.~Catani, S.~Dittmaier and Z.~Trocsanyi, {\it {One loop singular behavior of
  QCD and SUSY QCD amplitudes with massive partons}},  {\em Phys.Lett.} {\bf
  B500} (2001) 149--160 [\href{http://arXiv.org/abs/hep-ph/0011222}{{\tt
  hep-ph/0011222}}].

\bibitem{Altarelli:1977zs}
G.~Altarelli and G.~Parisi, {\it {Asymptotic Freedom in Parton Language}},
  {\em Nucl.Phys.} {\bf B126} (1977) 298.

\bibitem{Berends:1988zn}
F.~A. Berends and W.~Giele, {\it {Multiple Soft Gluon Radiation in Parton
  Processes}},  {\em Nucl.Phys.} {\bf B313} (1989) 595.

\bibitem{Campbell:1997hg}
J.~M. Campbell and N.~Glover, {\it {Double unresolved approximations to
  multiparton scattering amplitudes}},  {\em Nucl.Phys.} {\bf B527} (1998)
  264--288 [\href{http://arXiv.org/abs/hep-ph/9710255}{{\tt hep-ph/9710255}}].

\bibitem{deFlorian:2001zd}
D.~de~Florian and M.~Grazzini, {\it {The Structure of large logarithmic
  corrections at small transverse momentum in hadronic collisions}},  {\em
  Nucl.Phys.} {\bf B616} (2001) 247--285
  [\href{http://arXiv.org/abs/hep-ph/0108273}{{\tt hep-ph/0108273}}].

\bibitem{Weinzierl:2003ra}
S.~Weinzierl, {\it {Subtraction terms for one loop amplitudes with one
  unresolved parton}},  {\em JHEP} {\bf 0307} (2003) 052
  [\href{http://arXiv.org/abs/hep-ph/0306248}{{\tt hep-ph/0306248}}].

\bibitem{Bern:1994zx}
Z.~Bern, L.~J. Dixon, D.~C. Dunbar and D.~A. Kosower, {\it {One loop n point
  gauge theory amplitudes, unitarity and collinear limits}},  {\em Nucl.Phys.}
  {\bf B425} (1994) 217--260 [\href{http://arXiv.org/abs/hep-ph/9403226}{{\tt
  hep-ph/9403226}}].

\bibitem{Bern:1998sc}
Z.~Bern, V.~Del~Duca and C.~R. Schmidt, {\it {The Infrared behavior of one loop
  gluon amplitudes at next-to-next-to-leading order}},  {\em Phys.Lett.} {\bf
  B445} (1998) 168--177 [\href{http://arXiv.org/abs/hep-ph/9810409}{{\tt
  hep-ph/9810409}}].

\bibitem{Kosower:1999xi}
D.~A. Kosower, {\it {All order collinear behavior in gauge theories}},  {\em
  Nucl.Phys.} {\bf B552} (1999) 319--336
  [\href{http://arXiv.org/abs/hep-ph/9901201}{{\tt hep-ph/9901201}}].

\bibitem{Kosower:1999rx}
D.~A. Kosower and P.~Uwer, {\it {One loop splitting amplitudes in gauge
  theory}},  {\em Nucl.Phys.} {\bf B563} (1999) 477--505
  [\href{http://arXiv.org/abs/hep-ph/9903515}{{\tt hep-ph/9903515}}].

\bibitem{Bern:1999ry}
Z.~Bern, V.~Del~Duca, W.~B. Kilgore and C.~R. Schmidt, {\it {The Infrared
  behavior of one loop QCD amplitudes at next-to-next-to leading order}},  {\em
  Phys.Rev.} {\bf D60} (1999) 116001
  [\href{http://arXiv.org/abs/hep-ph/9903516}{{\tt hep-ph/9903516}}].

\bibitem{Catani:2000pi}
S.~Catani and M.~Grazzini, {\it {The soft gluon current at one loop order}},
  {\em Nucl.Phys.} {\bf B591} (2000) 435--454
  [\href{http://arXiv.org/abs/hep-ph/0007142}{{\tt hep-ph/0007142}}].

\bibitem{Kosower:2002su}
D.~A. Kosower, {\it {Multiple singular emission in gauge theories}},  {\em
  Phys.Rev.} {\bf D67} (2003) 116003
  [\href{http://arXiv.org/abs/hep-ph/0212097}{{\tt hep-ph/0212097}}].

\bibitem{Kosower:2003cz}
D.~A. Kosower, {\it {All orders singular emission in gauge theories}},  {\em
  Phys.Rev.Lett.} {\bf 91} (2003) 061602
  [\href{http://arXiv.org/abs/hep-ph/0301069}{{\tt hep-ph/0301069}}].

\bibitem{Catani:2003vu}
S.~Catani, D.~de~Florian and G.~Rodrigo, {\it {The Triple collinear limit of
  one loop QCD amplitudes}},  {\em Phys.Lett.} {\bf B586} (2004) 323--331
  [\href{http://arXiv.org/abs/hep-ph/0312067}{{\tt hep-ph/0312067}}].

\bibitem{Bern:2004cz}
Z.~Bern, L.~J. Dixon and D.~A. Kosower, {\it {Two-loop $g\rightarrow gg$
  splitting amplitudes in QCD}},  {\em JHEP} {\bf 0408} (2004) 012
  [\href{http://arXiv.org/abs/hep-ph/0404293}{{\tt hep-ph/0404293}}].

\bibitem{Badger:2004uk}
S.~Badger and E.~N. Glover, {\it {Two loop splitting functions in QCD}},  {\em
  JHEP} {\bf 0407} (2004) 040 [\href{http://arXiv.org/abs/hep-ph/0405236}{{\tt
  hep-ph/0405236}}].

\bibitem{Denner:2002ii}
A.~Denner and S.~Dittmaier, {\it {Reduction of one loop tensor five point
  integrals}},  {\em Nucl.Phys.} {\bf B658} (2003) 175--202
  [\href{http://arXiv.org/abs/hep-ph/0212259}{{\tt hep-ph/0212259}}].

\bibitem{Denner:2005nn}
A.~Denner and S.~Dittmaier, {\it {Reduction schemes for one-loop tensor
  integrals}},  {\em Nucl.Phys.} {\bf B734} (2006) 62--115
  [\href{http://arXiv.org/abs/hep-ph/0509141}{{\tt hep-ph/0509141}}].

\bibitem{Ossola:2006us}
G.~Ossola, C.~G. Papadopoulos and R.~Pittau, {\it {Reducing full one-loop
  amplitudes to scalar integrals at the integrand level}},  {\em Nucl.Phys.}
  {\bf B763} (2007) 147--169 [\href{http://arXiv.org/abs/hep-ph/0609007}{{\tt
  hep-ph/0609007}}].

\bibitem{Mastrolia:2010nb}
P.~Mastrolia, G.~Ossola, T.~Reiter and F.~Tramontano, {\it {Scattering
  AMplitudes from Unitarity-based Reduction Algorithm at the Integrand-level}},
   {\em JHEP} {\bf 1008} (2010) 080 [\href{http://arXiv.org/abs/1006.0710}{{\tt
  1006.0710}}].

\bibitem{Grazzini:2013bna}
M.~Grazzini, S.~Kallweit, D.~Rathlev and A.~Torre, {\it {$Z\gamma$ production
  at hadron colliders in NNLO QCD}},  {\em Phys.Lett.} {\bf B731} (2014)
  204--207 [\href{http://arXiv.org/abs/1309.7000}{{\tt 1309.7000}}].

\bibitem{Denner:2014gla}
A.~Denner, S.~Dittmaier and L.~Hofer, {\it {COLLIER -- A fortran-library for
  one-loop integrals}},  \href{http://arXiv.org/abs/1407.0087}{{\tt
  1407.0087}}.

\bibitem{Denner:2010tr}
A.~Denner and S.~Dittmaier, {\it {Scalar one-loop 4-point integrals}},  {\em
  Nucl.Phys.} {\bf B844} (2011) 199--242
  [\href{http://arXiv.org/abs/1005.2076}{{\tt 1005.2076}}].

\bibitem{Hoeche:2014qda}
S.~Hoeche, F.~Krauss, P.~Maierhoefer, S.~Pozzorini, M.~Schonherr {\em et.~al.},
  {\it {Next-to-leading order QCD predictions for top-quark pair production
  with up to two jets merged with a parton shower}},
  \href{http://arXiv.org/abs/1402.6293}{{\tt 1402.6293}}.

\bibitem{Draggiotis:2009yb}
P.~Draggiotis, M.~Garzelli, C.~Papadopoulos and R.~Pittau, {\it {Feynman Rules
  for the Rational Part of the QCD 1-loop amplitudes}},  {\em JHEP} {\bf 0904}
  (2009) 072 [\href{http://arXiv.org/abs/0903.0356}{{\tt 0903.0356}}].

\bibitem{Kleiss:1985gy}
R.~Kleiss, W.~J. Stirling and S.~Ellis, {\it {A new Monte Carlo treatment of
  multiparticle phase space at high-energies}},  {\em Comput.Phys.Commun.} {\bf
  40} (1986) 359.

\bibitem{Nogueira:1991ex}
P.~Nogueira, {\it {Automatic Feynman graph generation}},  {\em J.Comput.Phys.}
  {\bf 105} (1993) 279--289.

\bibitem{Alwall:2011uj}
J.~Alwall, M.~Herquet, F.~Maltoni, O.~Mattelaer and T.~Stelzer, {\it {MadGraph
  5 : Going Beyond}},  {\em JHEP} {\bf 1106} (2011) 128
  [\href{http://arXiv.org/abs/1106.0522}{{\tt 1106.0522}}].

\end{thebibliography}\endgroup

\end{document}